\documentclass[journal]{IEEEtran}

\usepackage{amsmath}
\usepackage{graphicx}
\usepackage[dvipsnames]{xcolor}
\usepackage{amssymb,amsfonts}
\usepackage{caption}
\usepackage{subcaption}
\usepackage{algorithmic}
\usepackage{algorithm}
\usepackage{cite}
\usepackage[export]{adjustbox}
\usepackage[footnote]{acronym}

\usepackage{array}
\usepackage{multirow}
\usepackage{hhline}
\makeatletter
\newcommand{\thickhline}{%
	\noalign {\ifnum 0=`}\fi \hrule height 1.1pt
	\futurelet \reserved@a \@xhline
}
\newcolumntype{"}{@{\hskip\tabcolsep\vrule width 1.1pt\hskip\tabcolsep}}
\makeatother

\usepackage{tikz}

\usepackage[hidelinks]{hyperref}

\usepackage[
    acronym,
    nopostdot,
    nonumberlist,
    nogroupskip,
    style=super
]{glossaries}

\newglossary*{tensor}{Matrix \& Tensor Decomposition}
\newglossary*{hsi}{Hyperspectral Imaging}
\newglossary*{mm}{Mathematical Morphology}

\makeglossaries

\newacronym[type=tensor]{td}{TD}{tensor decomposition}
\newacronym[type=tensor]{als}{ALS}{alternating least squares}
\newacronym[type=tensor]{asc}{ASC}{abundance sum-to-one constraint}
\newacronym[type=tensor]{admm}{ADMM}{alternating direction method of multipliers}
\newacronym[type=tensor]{ao}{AO}{alternating optimization}
\newacronym[type=tensor]{aoadmm}{AO-ADMM}{alternating optimization ADMM}
\newacronym[type=tensor]{bss}{BSS}{blind source separation}
\newacronym[type=tensor]{nmf}{NMF}{nonnegative matrix factorization}
\newacronym[type=tensor]{btd}{BTD}{block term decomposition}
\newacronym[type=tensor]{cpd}{CPD}{canonical polyadic decomposition}
\newacronym[type=tensor]{procoals}{ProCo-ALS}{projected compressed ALS}

\newacronym[type=hsi]{hsi}{HSI}{hyperspectral image}
\newacronym[type=hsi]{hu}{HU}{hyperspectral unmixing}
\newacronym[type=hsi]{multihutd}{MultiHU-TD}{multi-feature hyperspectral unmixing based on tensor decomposition}
\newacronym[type=hsi]{am}{AM}{abundance map}
\newacronym[type=hsi]{em}{EM}{endmember}
\newacronym[type=hsi]{lmm}{LMM}{linear mixing model}
\newacronym[type=hsi]{elmm}{ELMM}{extended linear mixing model}
\newacronym[type=hsi, longplural={spectral variabilities}]{sv}{SV}{spectral variability}
\newacronym[type=hsi]{gt}{GT}{ground-truth}
\newacronym[type=hsi]{rmse}{RMSE}{root mean squared error}
\newacronym[type=hsi]{sad}{SAD}{spectral angular distance}
\newacronym[type=hsi]{lidar}{LiDAR}{light detection and ranging}
\newacronym[type=hsi]{snr}{SNR}{signal-to-noise ratio}

\newacronym[type=mm]{mm}{MM}{mathematical morphology}
\newacronym[type=mm, longplural={closings by reconstruction}]{cbr}{CbR}{closing by reconstruction}
\newacronym[type=mm, longplural={openings by reconstruction}]{obr}{ObR}{opening by reconstruction}
\newacronym[type=mm]{se}{SE}{structuring element}
\newacronym[type=mm]{mp}{MP}{morphological profile}
\newacronym[type=mm]{emp}{EMP}{extended morphological profile}

\input{Inputs/new_commands.tex}

\begin{document}

\title{MultiHU-TD: Multi-feature Hyperspectral Unmixing Based on Tensor Decomposition}

\author{
	Mohamad~Jouni,~\IEEEmembership{Member,~IEEE,}
	Mauro~Dalla~Mura,~\IEEEmembership{Senior Member,~IEEE,}
	Lucas~Drumetz,~\IEEEmembership{Member,~IEEE,}
	and~Pierre~Comon,~\IEEEmembership{Fellow,~IEEE}
	\thanks{
	    This work is partly supported by grant ANR FuMultiSPOC (ANR-20-ASTR-0006),
	    and partly by Région Auvergne-Rhône-Alpes grant ``Pack Ambition International 2021'' (21-007356-01FONC, 21-007356-02INV).
	}
	\thanks{
	    M. Jouni, M. Dalla Mura, and P. Comon are with
		Univ. Grenoble Alpes, CNRS, Inria, Grenoble INP, GIPSA-lab,
		38000 Grenoble, France
		(e-mail: mohamad.jouni@grenoble-inp.fr; mauro.dalla-mura@grenoble-inp.fr; pierre.comon@grenoble-inp.fr).
		M. Dalla Mura is also with Institut Universitaire de France (IUF), France.
	}
	\thanks{
	    L. Drumetz is with {IMT Atlantique, Lab-STICC, UMR CNRS 6285, F-29238, Brest, France} (e-mail: lucas.drumetz@imt-atlantique.fr).
	}
	\thanks{Manuscript received \dots; revised \dots.}
}

\markboth{Journal of \LaTeX\ Class Files,~Vol.~, No.~, \dots}%
{Shell \MakeLowercase{\textit{et al.}}: Bare Demo of IEEEtran.cls for Journals}

\maketitle

\begin{abstract}
	\Acrlong{hu} allows to represent mixed pixels as a set of pure materials weighted by their abundances.
	Spectral features alone are often insufficient, so it is common to rely on other features of the scene.
    Matrix models become insufficient when the \acrlong{hsi} is represented as a high-order tensor with additional features in a multimodal, multi-feature framework.
    Tensor models such as \Acrlong{cpd} allow for this kind of unmixing, but lack a general framework and interpretability of the results.
    In this paper, we propose an interpretable methodological framework for low-rank \Acrfull{multihutd} which incorporates the \acrlong{asc} in the \Acrlong{aoadmm} algorithm, and provide in-depth mathematical, physical and graphical interpretation and connections with the \acrlong{elmm}.
    As additional features, we propose to incorporate \acrlong{mm} and reframe a previous work on neighborhood patches within \acrshort{multihutd}.
	Experiments on real hyperspectral images showcase the interpretability of the model and the analysis of the results.
	\texttt{Python} and \texttt{MATLAB} implementations are made available on GitHub.
\end{abstract}


%
\IEEEpeerreviewmaketitle

\section{Introduction}
\label{sec:Introduction}

\IEEEPARstart{H}{yperspectral imaging} refers to the acquisition of images of a scene over a wide and almost continuous spectrum.
A \gls{hsi} contains pixels that can cover areas of pure or mixed materials and amounts to a high spectral feature diversity \cite{ma2013signal,Amig19:book}.
These characteristics allow to perform \gls{bss} \cite{ComoJ10:book, ChabKMSTY14:spm, VincGF06:taslp} on the observed spectral signatures to blindly extract those of pure materials (\textit{sources}), also called \textit{\glspl{em}}, and their per-pixel (\textit{per-sample}) abundances.
This case of \gls{bss} is known as \gls{hu}, which is an active research topic with several applications like remote sensing, chemometrics, biomedical imagery, etc \cite{BajjZDD19:whispers, ImbiBB18:ssp, HongYCZ18:tip, BorsIB19:icassp, AzarMBR21:sp, borsoi2021spectral, DrumVHPCJ16:tip, DrumDTF20:icassp, VegaCFUDCC16:eusipco, Joun21:ugaEn}. \gls{hu} allows to understand and quantify the physical components of a scene.

A significant part of research in \gls{bss} and \gls{hu} relies on matrix factorization with additional constraints that aim at modeling the context of the problem. Consequently, an observed data matrix $\matr{M} \in \RR{I}{J}$ (i.e., with $I$ pixel samples and $J$ spectral features) is decomposed into two factor matrices $\matr{A} \in \RR{I}{R}$ and $\matr{B} \in \RR{J}{R}$ such that:
\begin{equation}
	\matr{M} =
	\matr{A} \matr{B}^{\T} =
	\sum_{r=1}^{R} \vect{a}_r \vect{b}_r^{\T}
	\label{eq:matrixfactorization}
\end{equation}
where $R$ is the number of latent components to be estimated, and $\vect{a}_r$ and $\vect{b}_r$ are the columns of $\vect{A}$ and $\vect{B}$, respectively, $\forall\, r \in \{1,\dots,R\}$.
As such, the columns of $\matr{B}$ represent the estimated source signals, and the rows of $\matr{A}$ represent the per-sample abundances of the sources.
The decomposition is often carried out by minimizing the generic cost function \cite{WangZ12:tkde}:
\begin{equation}
    \argmin{\matr{A},\matr{B}}{\|\matr{M} - \matr{A}\matr{B}^{\T}\|_F^2}
    + r(\matr{A}) + r(\matr{B})
    \label{eq:matrixfactorization_cost}
\end{equation}
where $r(\matr{\cdot})$ encodes the imposed constraints and/or regularizations to enforce desirable properties on the solutions.

In the case of \gls{hu}, a classical approach is the \gls{nmf}, which relies on the \gls{lmm} of the observed \gls{hsi} matrix (see Fig. \ref{fig:LMM_NMF_Illust}).
Hence, $\matr{A}$ and $\matr{B}$ are element-wise \textit{nonnegative}, which applies also in most domains of \gls{bss} (other than \gls{hu}) where the interpretability of the factor matrices is important.
Moreover, the rows of $\matr{A}$ are subject to the \textit{\gls{asc}}, which means that each row sums to $1$:
\begin{equation}
	\sum_{r=1}^R a_{ir} = 1
	\;\;\;
	\forall\, i \in \{1,\dots,I\}
\end{equation}
which applies to domains where the coefficients of the decomposition are proportions.

When only few materials concur in the mixture for each pixel, sparsity is imposed on the abundances \cite{YangZXDYZ10:tgrs}.
Finally, real \glspl{hsi} often contain \glspl{sv} in the sources, e.g., variations in the \glspl{em} due to local physico-chemical variations, illumination changes or topographic effects.
In order to account to these \glspl{sv}, the \gls{elmm} was proposed to extend the \gls{lmm} to account to said \glspl{sv}, which
is an active topic that has seen a lot of progress recently \cite{borsoi2021spectral, DrumVHPCJ16:tip, DrumDTF20:icassp, VegaCFUDCC16:eusipco}.

\subsection{Tensor Analysis of \glsentryshortpl{hsi}}

An \gls{hsi} can be treated as a data cube \cite{XionQZT18:tgrs, TanLXL19:igarss, HeYLYZ19:cvpr, GattDKFJ21:esa, ZareHKS21:rs, YaoHXMCX21:tgrs} (i.e., a third-order tensor with two spatial and one spectral dimensions).
However, sometimes the \gls{hsi} does not come alone but is associated with additional modalities such as:
\begin{itemize}
	\item A time series or multi-angular data of \gls{hsi} images \cite{VegaCFCC16:tgrs}.
	
	\item The \gls{hsi} is combined with images acquired by different sensors (e.g., panchromatic, multispectral and \acrshort{lidar} fusion) \cite{KanaFSM18:tsp, PrevUCB20:tsp, UtoDC18:whispers, XueYZD19:rs}.
	
	\item Some spatial features are extracted from the \gls{hsi} (such as in spectral-spatial classification problems \cite{JounDC19:ismm, JounDC19:igarss, JounDC20:mmta, GuLL19:tgrs}).
\end{itemize}
Such scenarios have recently also concerned other areas of \gls{bss} such as multi-channel signal processing \cite{mitsufuji2020multichannel, niknazar2014blind, virta2017blind} and multidimensional biomedical signal and image processing \cite{becker2015brain, becker2012multi, sole2018brain, zhang2019tensor, mishra2021recent}.
In the aforementioned scenarios, the data are represented natively as
\textit{tensors}\footnote{A tensor can be represented as a multidimensional array. The order of a tensor refers to the number of its array's indices, which is also the number of its \textit{modes}. For example, a tensor of dimensions $I \times J \times K$ is said to have three \textit{modes}, and is called a \textit{third-order tensor}. Data sets with order $3$ or above are described as \textit{high-order} tensors.}
\cite{Como14:spmag}, and the challenge usually boils down to the proper modeling of a joint factorization of multivariate representations without losing the multimodal structure, and hence its interpretation in terms of \gls{bss}.

\begin{figure}[t]
	\centering
	
	\begin{minipage}[b]{\linewidth}
		\centering
		\includegraphics[width=\textwidth]{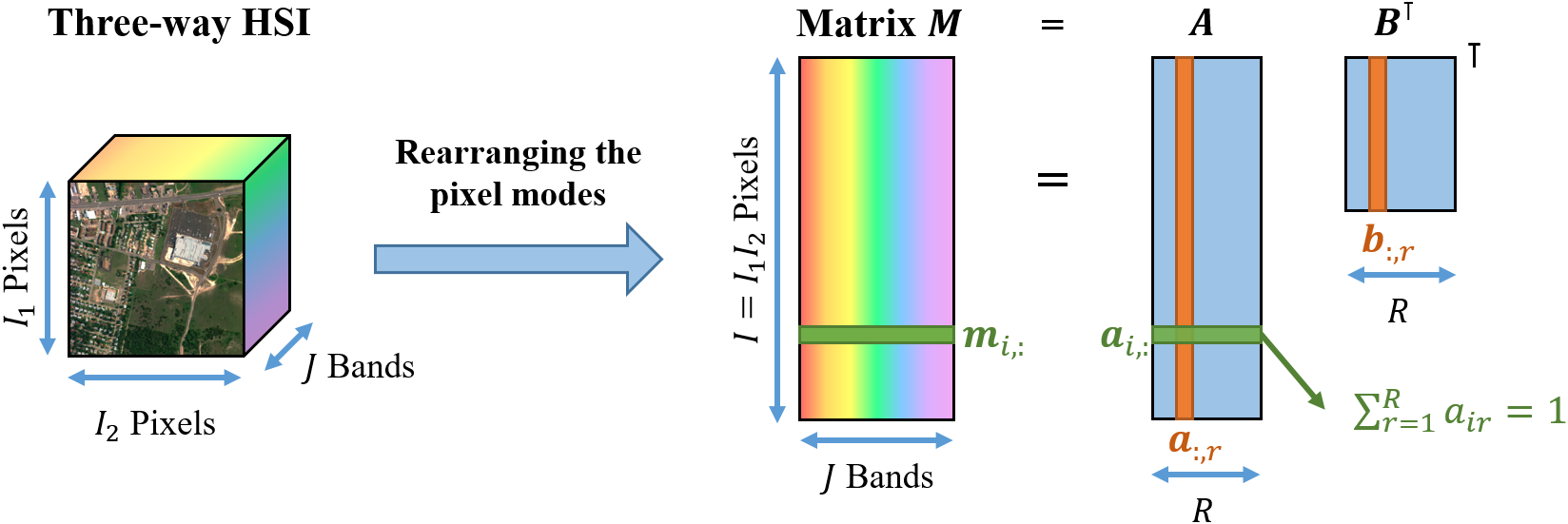}
		\caption{
			Classical matrix-based \gls{hu} using \gls{nmf} (i.e., \gls{lmm})
		}
		\label{fig:LMM_NMF_Illust}
	\end{minipage}
	
	\vspace{1mm}
	
	\begin{minipage}[b]{\linewidth}
		\centering
		\includegraphics[width=\textwidth]{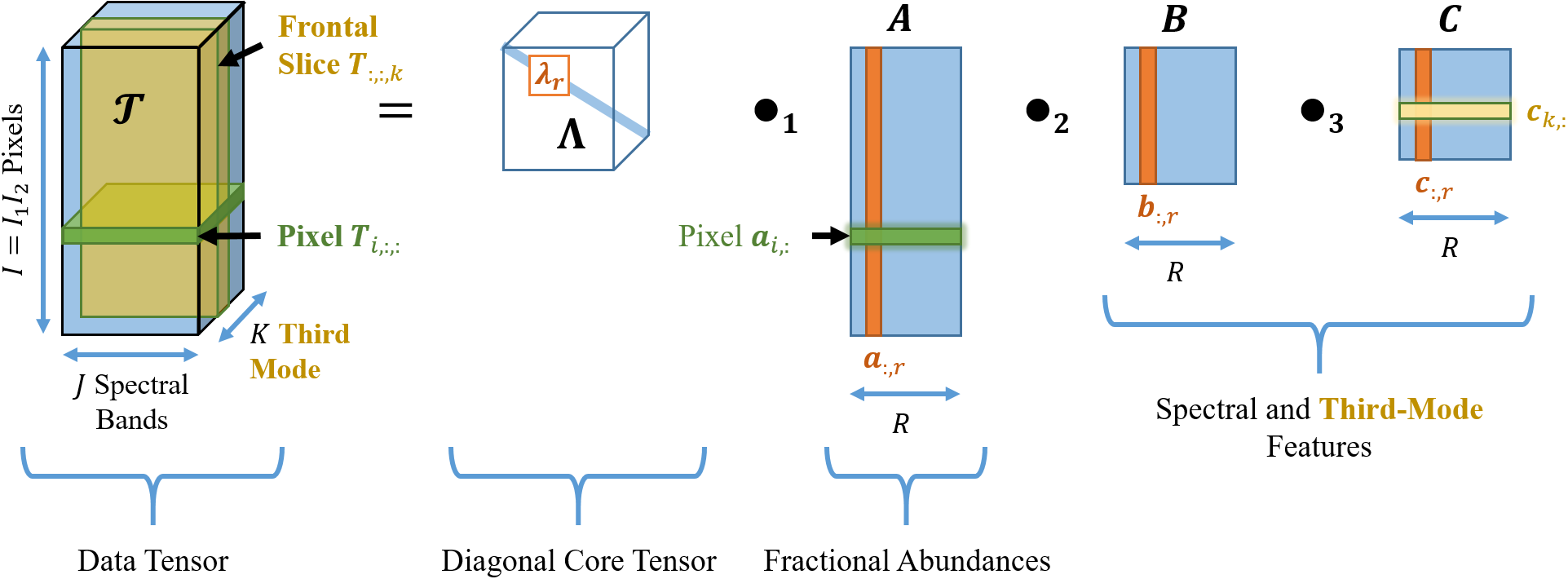}
		\caption{
			\Acrlong{cpd} of a third-order tensor. The tensor is formed of $K$ matricized \glspl{hsi} that are stacked along the third mode.
			A frontal slice $\matr{T}_{:,:,k}$ (in yellow) represents a matricized \gls{hsi} and is associated with one row of $\matr{C}$.
			A horizontal (pixel) slice $\matr{T}_{i,:,:}$ (in green) represents a matrix of features and is associated with one row of $\matr{A}$.
		}
		\label{fig:CPD_Illust}
	\end{minipage}
\end{figure}

Among these scenarios,
we focus on the case of \gls{hu} where the \gls{hsi} is associated with an additional set of features in the form of a new tensor mode, which we coin as \textit{Multi-feature \gls{hu} based on Tensor Decomposition} \glsunset{multihutd}(\gls{multihutd}).
For example, we consider a set of spatial features extracted from the image itself that can be considered as new modes.
Moreover, the pixels are rearranged in lexicographic order, promoting \textit{low-rank} tensor decomposition\footnote{
    Pixels form only one mode, even if images are often seen as 2D objects.
    This suppresses the high-rankness introduced by the complex spatial features of the whole scene \cite{Joun21:ugaEn}, which is an inconvenience for \gls{bss} methods
    \cite{PrevUCB20:tsp}.
}.
In this scenario, there are some challenging questions to answer such as: 
how can we jointly perform a constrained factorization in such settings? And how can we interpret the extracted factors?
An analysis of the literature shows that there are works that perform \gls{nmf} with additional constraints \cite{QianJZR11:tgrs, ZhuWXFP14:isprs, XuLLP19:tgrs}, and others that consider the case of multimodal inputs with coupled \gls{nmf} \cite{YokoYI11:tgrs,HenrCJ16:tip}, but this is different from considering data as tensors in our case.

As the native structure of our data is a tensor, we consider the problem in terms of tensor decomposition \cite{Como14:spmag}, which is the natural framework for processing multimodal data in the signal and image processing community \cite{CoheCC15:spl,JounDC18:lvaica,HuanSL16:tsp}.
There are many types of decomposition, such as tucker decomposition, \gls{btd}, \gls{cpd}, etc \cite{KoldB09:siam}.
However, we choose \gls{cpd}
thanks to the diagonality of its core tensor which allows the interaction only between similarly indexed columns in the factor matrices,
which \textit{naturally} promotes \textit{source separation} and a straightforward \textit{interpretability} of the extracted components in unmixing, both of which are core aspects of our work.

In the third-order case, as illustrated in Fig. \ref{fig:CPD_Illust}, \gls{cpd} decomposes a data tensor $\tens{T} \in \RRR{I}{J}{K}$ into
a diagonal core tensor $\vect{\Lambda} \in \RRR{R}{R}{R}$ and
$3$ factor matrices $\{\matr{A} \in \mathbb{R}^{I\times R}, \matr{B} \in \mathbb{R}^{J\times R}, \matr{C} \in \mathbb{R}^{K\times R}\}$,
each representing one of the $3$ modes of $\tens{T}$ respectively, such that:
\begin{equation}
	\tens{T} =
	\vect{\Lambda}
	\con_1 \matr{A} \con_2 \matr{B} \con_3 \matr{C}
\end{equation}
where $\con_d$ denotes the mode-$d$ product (product along the $d$-th mode), further described in expressions (\ref{eq:mode_product}), (\ref{eq:CPDp_1}), and (\ref{eq:CPDp_2}).

\gls{cpd} extends \gls{nmf} to high-order data and can adopt all of its features, especially that of imposing constraints.
\gls{cpd} is often computed by minimizing the cost function \cite{Como14:spmag}:
\begin{gather}
    \argmin{\matr{A},\matr{B},\matr{C}}
    {\|\tens{T} - \vect{\Lambda}
    \con_1 \matr{A} \con_2 \matr{B} \con_3 \matr{C}\|_F^2
    + r(\matr{\matr{A},\matr{B},\matr{C}})}
    \label{eq:tensorfactorization_cpd_cost}
\end{gather}
where $r(\matr{\cdot})$ encodes the imposed constraints. 
Note that the nonnegative constraint ensures the existence of a minimum; in fact, without an appropriate regularisation term $r(\matr{\cdot})$, the above cost function could admit only an infimum, which may not be reachable \cite{QiCL16:tit}. On the other hand, with an appropriate regularisation, Problem (\ref{eq:tensorfactorization_cpd_cost}) is well posed.

\subsection{Related Works and Limitations}
\label{subsec:related_works_limitations}

In the context of \gls{multihutd},
\gls{cpd} has been used with multitemporal/angular \glspl{hsi} \cite{VegaCFCC16:tgrs} as well as with \glspl{hsi} having an additional diversity of extracted neighborhood patches \cite{VegaCFUDCC16:eusipco} (see Fig. \ref{fig:patches_illustration} for a $5\times5$ patch-\gls{hsi} tensor).
Moreover, some works \cite{JounDC19:ismm,JounDC19:igarss,JounDC20:mmta} jointly considered \glspl{hsi} with spatial features extracted by \gls{mm} filters \cite{NajmT13:jws,MarpPDPBB12:ijidf,DallBWB10:ijrs,DallBWB10:tgrs} in the framework of multi-feature scene classification (see Fig. \ref{fig:morpho_illustration}).
These works show that \gls{cpd} is a suitable approach for joint decomposition.
However, they present some limitations.

\subsubsection{Algorithmic perspective}
The nonnegative constraint is implemented in \cite{CichZA09} by projection onto the nonnegative orthant, which exhibits some computational issues \cite{JounDC18:lvaica}.
In \cite{VegaCFUDCC16:eusipco}, nonnegative \gls{als} is used where \gls{asc} is also naively implemented by projecting the abundances on the unit simplex, contrary to the common practice in the matrix case \cite{YangZXDYZ10:tgrs} where \gls{asc} is embedded in the updates.
In \cite{VegaCFCC16:tgrs}, the nonnegative \gls{cpd} is computed using the \gls{procoals} algorithm, which is considerably fast \cite{CoheCC15:spl} but not so flexible with additional constraints.
Finally, in \cite{JounDC19:ismm,JounDC19:igarss,JounDC20:mmta}, an alternative algorithm is proposed based on \acrlong{ao}\glsunset{ao} \acrlong{admm}\glsunset{admm} (\acrshort{aoadmm})\glsunset{aoadmm} \cite{HuanSL16:tsp} with compression and nonnegative constraints, which is flexible and stable with large datasets, but has not yet addressed \gls{multihutd} which requires further modeling (i.e., sparsity, \gls{asc}).

\subsubsection{Interpretability}
The work of \cite{VegaCFCC16:tgrs} faced a certain challenge in interpreting the third-mode factors, perhaps due to the naive employment of \gls{cpd}.
In \cite{VegaCFUDCC16:eusipco}, a link was established between \gls{cpd} and \gls{elmm} but was not deeply investigated as it was restricted to the case of patches and tested only with synthetic data. Moreover, it faced another challenge in interpreting the factors, which poses an ambiguity on its performance and the meaning of the extracted features.
Finally, \cite{VegaCFUDCC16:eusipco} and \cite{JounDC19:ismm,JounDC19:igarss,JounDC20:mmta} perform tensor decomposition with spatial features. However, the used spatial features can allow limited flexibility (e.g., patches), or the interpretation of the factor matrices was not addressed, noting here that incorporating them with \gls{cpd} showed improvement in supervised classification (e.g., \gls{mm}).

In this paper, we wish to consider such operations in a \gls{bss} framework from the lens of \gls{multihutd} with in-depth interpretability.
This presents us with two main challenges:
\begin{itemize}
	\item Tuning \gls{aoadmm} to incorporate \gls{asc}, which is challenging due to the multilinear structure of \gls{cpd}, particularly in modeling the samples as a convex combination of the spectral sources in a multimodal setting.
	
	\item Exploring the meaning of the extracted features in these conditions.
\end{itemize}

\subsection{Contributions}
\label{subsec:contributions}

To our knowledge, imposing \gls{asc} in \gls{cpd} or \gls{aoadmm} as a natural extension of \gls{nmf} in \cite{YangZXDYZ10:tgrs} has not been done. Furthermore, we are interested in finding a generalized framework for \gls{multihutd} favouring the interpretation of its results under any third-mode diversity.
More precisely, our contributions to jointly deal with these limitations are the following:
\begin{itemize}
    \item We propose a methodological framework for dealing with \gls{multihutd} based on \gls{aoadmm} by Huang \cite{HuanSL16:tsp}, and expand it to incorporate \gls{asc} with joint nonnegativity and sparsity.
    The proposed \gls{aoadmm}-\gls{asc} is a general algorithm that can be applied in other domains of \gls{bss} where convex combinations of sources apply.
    
    \item We establish a unified framework for the interpretability of \gls{multihutd}. In particular, the link between \gls{elmm} and \gls{cpd} \cite{VegaCFUDCC16:eusipco} is expanded by providing in-depth physical and graphical insights for better interpretability of the \gls{cpd} model and its factors.
    
    \item We propose to include \gls{mm} as spatial features to perform a spectral-spatial \gls{hu} and demonstrate the aforementioned points. We also revise \cite{VegaCFUDCC16:eusipco} and provide detailed interpretations on the cases of patches and \gls{mm}, which has not been addressed in any of the previous works \cite{VegaCFUDCC16:eusipco,JounDC19:ismm,JounDC19:igarss,JounDC20:mmta}.
    This analysis also shows that \gls{mm} is better suited since it embeds physically meaningful features (scale and brightness of objects) into \gls{hu} unlike patches.
\end{itemize}
That said, note that our main goal is to reason about the interpretability of this factorization and to describe this framework rather than to propose yet another \gls{hu} algorithm.

The remainder is organized as follows.
In section \ref{sec:Background}, we introduce some background.
In section \ref{sec:MultimodalHU}, we  detail the proposed framework.
In section \ref{sec:Unmixing_Experiments}, we present our experiments and results.
Finally, we draw out some conclusions in \ref{sec:Conclusion}.

\section{Notations and Definitions}
\label{sec:notations_and_definitions}

Table \ref{tab:array_notations} shows a list of notations for the different types of objects used throughout the paper: scalars, vectors, matrices, tensors, and array dimensions and indices.
Table \ref{tab:data_notations} denotes the types of observed data in the paper, their dimensions and different ways of indexing.

\begin{table}[h]
	\centering
	
	\begin{tabular}{c"c|c}
		\textbf{Type} & \textbf{Font style} & \textbf{Example} \\
		\thickhline
		\textbf{Scalars} & unformatted lowercase & $a$, $b$, $c$, $t$ \\
		\textbf{Vectors} & bold lowercase & $\vect{a}$, $\vect{b}$, $\vect{c}$, $\vect{t}$ \\
		\textbf{Matrices} & bold uppercase & $\matr{A}$, $\matr{B}$, $\matr{C}$, $\matr{T}$ \\
		\textbf{Tensors} & bold calligraphic & $\tens{T}$ \\
		\thickhline
		\textbf{Dimension} & unformatted uppercase & $I$, $J$, $K$, $R$ \\
		\textbf{Indices} &
		\begin{tabular}{c}
			lowercase version of \\
			the spanned dimension
		\end{tabular} &
		$i$, $j$, $k$, $r$ \\
	\end{tabular}
	
	\caption{Array notations}
	\label{tab:array_notations}
\end{table}

Table \ref{tab:tensor_slicing_unfolding} denotes the different ways to slice and unfold a third-order tensor.
The \emph{mode unfolding} (or \textit{matricization}) of a tensor means to reshape it into a matrix by fixing the targeted mode and rearrange the others in lexicographic order.

Table \ref{tab:cpd_factor_matrices} denotes the factor matrices of an \gls{nmf} (matrix case) or \gls{cpd} (tensor case). Mode-$1$, Mode-$2$ and Mode-$3$ correspond to the modes of \textit{pixels}, spectral \textit{bands}, and set of extracted spatial features (\textit{transforms}) respectively.

\begin{table*}[ht]
	
	\begin{minipage}[b]{\textwidth}
		\centering
		
		\begin{tabular}{c"c c c c c c}
			\textbf{Type} &
			\textbf{Symbol} &
			\begin{tabular}{c}
				\textbf{Dimensions} \\
				\textbf{(pixel $\times$ band $\times$ transform)}
			\end{tabular} &
			\begin{tabular}{c}
				$i$\textbf{-th} \\
				\textbf{pixel}
			\end{tabular} &
			\begin{tabular}{c}
				$j$\textbf{-th} \\
				\textbf{band}
			\end{tabular} &
			\begin{tabular}{c}
				$k$\textbf{-th} \\
				\textbf{transform}
			\end{tabular} &
			\begin{tabular}{c}
				$(i,j,k)$\textbf{-th} \\
				\textbf{element}
			\end{tabular} \\
			\thickhline
			\textbf{\gls{hsi} matrix} & $\matr{M}$ & $I \times J$ & $\vect{m}_{i,:}$ & $\vect{m}_{:,j}$ & - & $m_{i,j}$ \\
			\textbf{\gls{hsi} tensor} & $\tens{T}$ & $I \times J \times K$ & $\matr{T}_{i,:,:}$ & $\matr{T}_{:,j,:}$ & $\matr{T}_{:,:,k}$ & $t_{i,j,k}$ \\
		\end{tabular}
		
		\caption{
			The pixels are rearranged in \textit{lexicographic order} spanning the first mode, so $I$ is the total number of pixels.
			The symbol ``$:$'' in the index indicates a span of the whole mode.
			For example, $\vect{m}_{i,:}$ and $\vect{m}_{:,j}$ represent the $i$-th row and $j$-th column \textit{vectors} of $\matr{M}$ respectively (see Fig. \ref{fig:LMM_NMF_Illust}),
			and $\matr{T}_{:,:,k}$ represents the $k$-th frontal \textit{matrix} slice of $\tens{T}$ (see Fig. \ref{fig:CPD_Illust}).
		}
		\label{tab:data_notations}
	\end{minipage}
	
	\vspace{3mm}
	
	\begin{minipage}[b]{0.37\textwidth}
		\centering
		
		\begin{tabular}{c"cc}
			\textbf{Variable} &
			\textbf{Symbol} &
			\begin{tabular}{c}
				\textbf{Dimensions}
			\end{tabular} \\
			\thickhline
			\begin{tabular}{c}
				\textbf{Horizontal slice}
			\end{tabular} &
			$\matr{T}_{i,:,:}$ &
			$J \times K$ \\
			\begin{tabular}{c}
				\textbf{Lateral slice}
			\end{tabular} &
			$\matr{T}_{:,j,:}$ &
			$I \times K$ \\
			\begin{tabular}{c}
				\textbf{Frontal slice}
			\end{tabular} &
			$\matr{T}_{:,:,k}$ &
			$I \times J$ \\
			\thickhline
			\begin{tabular}{c}
				\textbf{Mode-$1$ unfolding}
			\end{tabular} &
			$\matr{T}_{(1)}$ & $JK \times I$ \\
			\begin{tabular}{c}
				\textbf{Mode-$2$ unfolding}
			\end{tabular} &
			$\matr{T}_{(2)}$ & $IK \times J$ \\
			\begin{tabular}{c}
				\textbf{Mode-$3$ unfolding}
			\end{tabular} &
			$\matr{T}_{(3)}$ & $IJ \times K$ \\
		\end{tabular}
		
		\caption{Tensor slicing and mode-unfolding.}
		\label{tab:tensor_slicing_unfolding}
	\end{minipage}
	\hspace{3mm}
	\begin{minipage}[b]{0.6\textwidth}
		\centering
		\begin{tabular}{c"ccccc}
			\begin{tabular}{c}
				\textbf{Factor Mat.}
			\end{tabular} &
			\textbf{Symbol} &
			\begin{tabular}{c}
				\textbf{Dimensions}
			\end{tabular} &
			\begin{tabular}{c}
				\textbf{Row} \\
				\textbf{index}
			\end{tabular} &
			\begin{tabular}{c}
				\textbf{Col.} \\
				\textbf{index}
			\end{tabular} &
			\begin{tabular}{c}
				\textbf{Element} \\
				\textbf{index}
			\end{tabular} \\
			\thickhline
			\begin{tabular}{c}
				\textbf{Mode}-$1$
			\end{tabular} &
			$\matr{A}$ &
			$I \times R$ &
			$\vect{a}_{i,:}$ &
			$\vect{a}_{:,r}$ &
			$a_{i,r}$ \\
			\begin{tabular}{c}
				\textbf{Mode}-$2$
			\end{tabular} &
			$\matr{B}$ &
			$J \times R$ &
			$\vect{b}_{j,:}$ &
			$\vect{b}_{:,r}$ &
			$b_{j,r}$ \\
			\begin{tabular}{c}
				\textbf{Mode}-$3$
			\end{tabular} &
			$\matr{C}$ &
			$K \times R$ &
			$\vect{c}_{k,:}$ &
			$\vect{c}_{:,r}$ &
			$c_{k,r}$ \\
		\end{tabular}
		
		\caption{The factor matrices, each corresponding to one of the matrix or tensor modes. $R$ is the number of rank-$1$ additive terms in the decomposition.}
		\label{tab:cpd_factor_matrices}
	\end{minipage}
\end{table*}

We use the notation ``$\Diag{\vect{v}}$'' to refer to the diagonal matrix whose entries are the elements of any vector $\vect{v}$.

The \emph{outer product} of two vectors $\vect{a} \in \R{I}$ and $\vect{b} \in \R{J}$ results in a matrix $\matr{M} \in \RR{I}{J}$ as follows:
\begin{gather*}
		\matr{M} =
		\vect{a} \out \vect{b} =
		\vect{a} \vect{b}^{\T}
		\Longleftrightarrow
		m_{i,j} = a_i b_j
		\\
		\forall\, i \in \{1,\dots,I\},
		\;\;
		\forall\, j \in \{1,\dots,J\}
\end{gather*}
The \emph{outer product} of three vectors $\vect{a} \in \R{I}$, $\vect{b} \in \R{J}$ and $\vect{c} \in \R{K}$ results in a third-order tensor $\tens{T} \in \RRR{I}{J}{K}$ as follows:
\begin{gather*}
		\tens{T} =
		\vect{a} \out \vect{b} \out \vect{c}
		\Longleftrightarrow
		t_{i,j,k} = a_i b_j c_k
		\\
		\forall\, i \in \{1,\dots,I\},
		\;\;
		\forall\, j \in \{1,\dots,J\},
		\;\;
		\forall\, k \in \{1,\dots,K\}
\end{gather*}

The \emph{mode-$d$ product} $\con_d$ represents the product of a tensor by a matrix along the $d$-th mode.
For example, assuming that we have
$\tens{G} \in \RRR{L}{M}{N}$,
$\matr{A} \in \RR{I}{L}$ and
$\matr{B} \in \RR{J}{M}$,
the mode-$1$ and mode-$2$ product of $\tens{G}$ by $\matr{A}$ and $\matr{B}$ respectively results in a tensor $\tens{T} \in \RRR{I}{J}{N}$ defined as:
\begin{equation}
	\tens{T} = \tens{G} \con_1 \matr{A} \con_2 \matr{B}
	\Longleftrightarrow
	t_{ijn} = \sum_{l=1}^{L} \sum_{m=1}^{M} G_{lmn} \, a_{il} b_{jm}
	\label{eq:mode_product}
\end{equation}

\section{Background}
\label{sec:Background}

In this section, we briefly  review the existing notions in the literature upon which we base our algorithm and generalized interpretation of the \gls{multihutd} framework.
First, we explain how \gls{asc} is applied in \gls{nmf} \cite{YangZXDYZ10:tgrs} as the proposed framework extends this for tensor decomposition.
Then, we give a brief account on \gls{elmm},  including graphical and visual interpretations, which will be the  basis for the proposed interpretation.
Finally, we  discuss the link between \gls{cpd} and \gls{elmm}  preliminarily presented in \cite{VegaCFUDCC16:eusipco}.

\subsection{\glsentryshort{nmf} with \glsentryshort{asc}, Nonnegativity, and Sparsity}

In the \gls{nmf} case \cite{YangZXDYZ10:tgrs}, when sparsity ($\ell_1$ norm) and \gls{asc} are imposed on the abundances, (\ref{eq:matrixfactorization})  becomes:
\begin{equation}
    \begin{aligned}
    \label{eq:cost_sparse_NMF_original}
        \argmin{\matr{A},\matr{B}}
        {\frac{1}{2}\|\matr{M}-\matr{A}\matr{B}^{\T}\|_F^2 +
        \alpha \|\matr{A}\|_1} \\
        \textrm{s.t.}
        \;
        \matr{A}\succeq0,
        \;
        \matr{B}\succeq0,
        \;
        \sum_{r=1}^R a_{i,r} = 1
        \;
        |_{\forall i \in \{1,\dots,I\}}
    \end{aligned}
\end{equation}
where $\alpha>0$, and $\succeq$  denotes element-wise nonnegativity.
A simple strategy to embed \gls{asc} goes by stacking a row vector in $\matr{B}$ and a column vector in $\matr{M}$ such that \cite{YangZXDYZ10:tgrs}:
\begin{equation}
    \Tilde{\matr{M}} = 
    \begin{bmatrix}
        \matr{M}
        \; | \;
        \delta \matr{1}_{I\times 1}
    \end{bmatrix},
    \;
    \Tilde{\matr{B}} = 
    \begin{bmatrix}
        \matr{B} \\
        \delta \matr{1}_{1\times R}
    \end{bmatrix},
\end{equation}
where $\delta$ is a constant that is usually set as the mean of $\matr{M}$, and the last row of $\Tilde{\matr{B}}$ is reset to $\delta$ after each iteration.
This operation ensures that \gls{asc} is softly embedded in \gls{nmf} since $\forall i \in \{1,\dots,I\}$ we have:
\begin{equation}
    m_{i,J+1}
    = \sum_{r=1}^{R} a_{i,r} b_{J+1,r}
    = \sum_{r=1}^{R} a_{i,r} \delta
    = \delta
\end{equation}
corresponding to $\sum_{r=1}^{R} a_{i,r}$ = $1$. Then, (\ref{eq:cost_sparse_NMF_original}) becomes:
\begin{equation}
    \begin{aligned}
    \label{eq:cost_sparse_NMF}
        \argmin{\matr{A},\Tilde{\matr{B}}}
        {\frac{1}{2}
        \|\Tilde{\matr{M}}-\matr{A}\Tilde{\matr{B}}^{\T}\|_F^2
        + \alpha \|\matr{A}\|_1}
        \textrm{ s.t. } \matr{A}\succeq0,\Tilde{\matr{B}}\succeq0
    \end{aligned}
\end{equation}

There are many algorithms proposed in the literature that deal with sparse \gls{nmf} and \gls{asc}, which  are out of the scope of this work \cite{QianJZR11:tgrs,ZhuWXFP14:isprs}. In our case, we extend \gls{nmf} within the \gls{aoadmm} framework for \gls{cpd}. \gls{nmf} then becomes a special case for order-$2$ tensors. This tensor extension, proposed in Section \ref{subsec:ASC-AOADMM} is referred to as \gls{aoadmm}-\gls{asc}.

\subsection{\glsentryshort{elmm}}
\label{subsec:Unmixing_ELMM_Classical}

While \gls{lmm} is seen as a direct approach for \gls{hu}, it cannot model \glspl{sv} represented by nonlinear effects or illumination conditions.
One way to account to said effects is through \gls{elmm} \cite{DrumVHPCJ16:tip}, which in general assumes additional degrees of freedom that account to said \glspl{sv} \textit{at the pixel level} by introducing a \textit{pixel-dependent} \gls{sv} function $\vect{f}_{i} : \R{J} \rightarrow \R{J}$ $\forall\, i \in \{1, \dots, I\}$, which maps each \gls{em} $\vect{b}_{:,r}$ $\forall\, r \in \{1, \dots, R\}$ to a new spectral signature $\vect{b}^{(i)}_{:,r}$ that best reflects the targeted \glspl{sv}:
\begin{equation}
    \label{eq:ELMM_pixel_}
    \vect{m}_i = \sum_{r=1}^R a_{ir} \vect{f}_{i}(\vect{b}_{:,r})
    = \sum_{r=1}^R a_{ir} \vect{b}^{(i)}_{:,r}.
\end{equation}
For example, in the case of different illumination conditions, this can be represented as a scaling factor for each pixel on the \glspl{em}. In the following, we present the parts that are at the basis of the interpretability of our proposed framework.

\begin{figure}[b]
    \centering
    
    \begin{subfigure}[b]{0.34\linewidth}
        \centering
        \includegraphics[width=\textwidth]{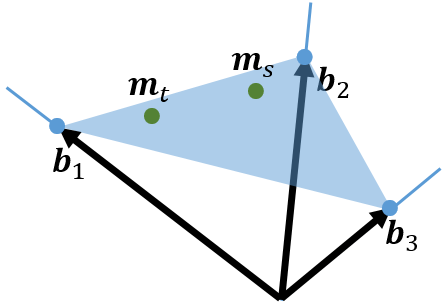}
        \caption{\gls{lmm}}
        \label{subfig:LMM_graph}
    \end{subfigure}
    \begin{subfigure}[b]{0.62\linewidth}
        \centering
        \includegraphics[width=\textwidth]{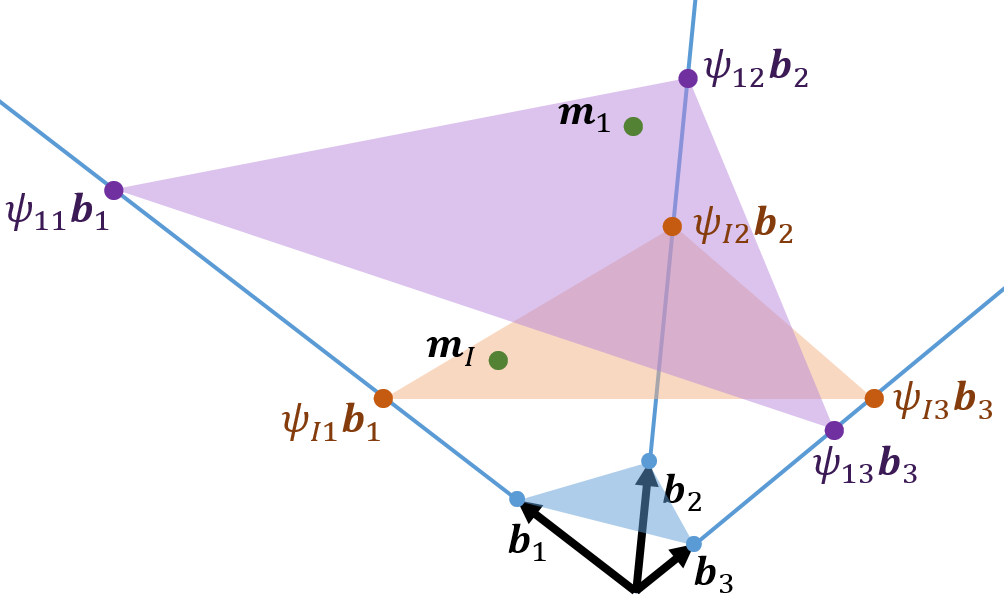}
        \caption{\gls{elmm} (scalar factors)}
        \label{subfig:ELMM_CLS_graph}
    \end{subfigure}
    
    \caption{
    Graphical comparison between \gls{lmm} and \gls{elmm} (scalar factors) in the case of three spectral signatures $\{\vect{b}_{:,1}$, $\vect{b}_{:,2}$, $\vect{b}_{:,3}\}$ and two example pixels $\{\vect{m}_{s,:}$, $\vect{m}_{t,:}\}$.
    The relative coordinates in the simplices are conserved.
    }
    \label{fig:graph_LMM_ELMM_CLS}
\end{figure}

When \gls{asc} is imposed in \gls{lmm}, all the pixels will lie on  the convex hull of the set of estimated \glspl{em}
(i.e., the columns of $\matr{B}$\footnote{Here, we note that unless the \glspl{em} are not affinely independent, which is unlikely, and $R \leq J+1$, then the convex hull is a simplex.}),
and the fractional abundances in each row of $\matr{A}$ define the coordinates of each pixel on the convex hull, which is illustrated in Fig. \ref{subfig:LMM_graph}.
With the introduction of \gls{elmm}, the pixels will not lie on the same simplex anymore as each pixel is mapped to a new set of \glspl{em}, which is illustrated in Fig. \ref{subfig:ELMM_CLS_graph} in the case where the \gls{sv} is modeled by a scaling factor $\psi_{ir}$ that is pixel- and \gls{em}-dependent such that
$\vect{b}^{(i)}_r = \psi_{ir} \vect{b}_{:,r}$ \cite{DrumVHPCJ16:tip}.
Accordingly, the new spectral signatures $\vect{b}^{(i)}_r$ are allowed to move only along the directions of $\vect{b}_{:,r}$.

In the case where the \glspl{sv} are modeled with scaling factors, we denote by $\matr{\Psi} \in \RR{I}{R}$ the matrix whose entries are the scaling factors $\psi_{ir}$ $\forall\, i \in \{1, \dots, I\}$ and $\forall\, r \in \{1, \dots, R\}$.
Then, the following are equivalent:
\begin{subequations}
    \begin{align}
        & \matr{M} = (\matr{A} \hadam \matr{\Psi}) \matr{B}^{\T}
        \label{eq:ELMM_CLS_matrix} \Longleftrightarrow
        \\
        & \vect{m}_{i,:}
        = \sum_{r=1}^R a_{ir} \psi_{ir} \vect{b}_{:,r}^{\T}
        = \vect{a}_{i,:} \Diag{\matr{\Psi}_{i,:}} \matr{B}^{\T} = \vect{a}_{i,:} \matr{\Psi}_{(i)} \matr{B}^{\T}
        \label{eq:ELMM_CLS_pixel3}
    \end{align}
\end{subequations}
where $\hadam$ is the Hadamard product.
$\matr{\Psi}_{i,:} \in \R{R}$ is the $i$-th row of $\matr{\Psi}$, and $\matr{\Psi}_{(i)} \in \RR{R}{R}$ represents the diagonal matrix formed from $\matr{\Psi}_{i,:}$.
These mathematical, graphical, and visual relationships are the key to elaborate and interpret the case of \gls{cpd} in Section \ref{subsec:elmminterpretation}.

\subsection{\glsentryshort{cpd} and \glsentryshort{elmm}}
\label{subsec:CPDandELMM}

In the work of \cite{VegaCFUDCC16:eusipco}, a representation of \gls{cpd} in terms of \gls{elmm} has been presented, which will be reported here.
\gls{cpd} decomposes a third-order tensor $\tens{T} \in \RRR{I}{J}{K}$ such that:
\begin{subequations}
    \begin{align}
        \tens{T} = \vect{\Lambda}
        \con_1 \matr{A}
        \con_2 \matr{B}
        \con_3 \matr{C}
        & \Leftrightarrow
        \tens{T} = \sum_{r=1}^{R}
        \lambda_{r,r,r} \;
        \vect{a}_{:,r} \out
        \vect{b}_{:,r} \out
        \vect{c}_{:,r}
        \label{eq:CPDp_0}
        \\
        & \Leftrightarrow
        \tens{T} = \sum_{r=1}^{R}
        \vect{a}_{:,r} \out
        \vect{b}_{:,r} \out
        \vect{\psi}_{:,r}
        \label{eq:CPDp_1}
        \\
        & \Leftrightarrow
        t_{i,j,k} = \sum_{r=1}^{R}
        \; a_{i,r}
        \; b_{j,r}
        \; \psi_{k,r}
        \label{eq:CPDp_2}
    \end{align}
\end{subequations}
where $t_{i,j,k}$ is a (scalar) entry of $\tens{T}$, $R$ is the number of estimated sources, and
$\vect{\Lambda}$ $\in$ $\RRR{R}{R}{R}$ is a diagonal tensor that absorbs the $\ell_2$-norms of the columns of the factor matrices
\cite{Como14:spmag,JounDC18:lvaica}.
In (\ref{eq:CPDp_1}) and (\ref{eq:CPDp_2}), we suppress the expression of $\vect{\Lambda}$ by absorbing its entries in the columns of $\matr{C}$, resulting in $\matr{\Psi} \in \RR{K}{R}$ whose columns are the scaled version of those of $\matr{C}$ such that $\vect{\psi}_{:,r} = \lambda_{r,r,r} \vect{c}_{:,r}$ $\forall\; r$.

As shown in \cite{VegaCFUDCC16:eusipco},
given that $\matr{T}_{:,:,k}$ denotes the $k$-th frontal slice of $\tens{T}$,  $\forall\, k \in \{1, \dots, K\}$,
and assuming that $\matr{\Psi}_{(k)} \in \RR{R}{R}$ represents the diagonal matrix formed from the row $\vect{\psi}_{k,:} \in \R{R}$ of the factor $\matr{\Psi}$,
one can write:
\begin{subequations}
    \begin{eqnarray}
        \matr{T}_{:,:,k}
        = \matr{A} \Diag{\vect{\psi}_{k,:}} \matr{B}^{\T}
        = \matr{A} \matr{\Psi}_{(k)} \matr{B}^{\T}
        = \matr{A} \Tilde{\matr{f}}_{k}(\matr{B})^{\T}
        \label{eq:RegELMM_frontalslice_intro}
        \\
        \iff
        \vect{t}_{i,:,k}
        = \sum_{r=1}^R a_{i,r} \; (\vect{b}_{:,r}  \psi_{k,r})
        = \sum_{r=1}^R a_{i,r} \vect{f}_{k}(\vect{b}_{:,r})
        \label{eq:RegELMM_pixel_intro}
    \end{eqnarray}
\end{subequations}
where $\vect{t}_{i,:,k}$ is the $i$-th pixel row of $\matr{T}_{:,:,k}$.
From the perspective of each frontal slice, \gls{cpd} resembles a regularized coupled \gls{nmf} of the frontal slices where $\matr{A}$ is a common factor:
\begin{eqnarray}
    \argmin{\matr{A},\matr{B}^{(k)}}
    {\|\matr{T}_{:,:,k} - \matr{A} \matr{B}^{(k)\T}\|_F^2}
    \textrm{ s.t. } \matr{A}\succeq0, \matr{B}^{(k)}\succeq0
    \label{eq:RegELMM_frontalslice_intro_}
\end{eqnarray}
where $\matr{B}^{(k)} = \Tilde{\matr{f}}_{k}(\matr{B})|_{\forall\, k \in \{1,\dots,K\}}$, which has an analogous expression to the case of \gls{elmm}. The latter was used in \cite{VegaCFUDCC16:eusipco} as a way to impose a spatial smoothing (\ref{eq:RegELMM_frontalslice_intro_}) on the abundances within a patch of neighboring pixels.
This link between \gls{cpd} and \gls{elmm} was preliminary presented and restricted to an application of patches with only simulated data, and the meaning of the \glspl{sv} was not explored.

In Section \ref{subsec:elmminterpretation}, we propose a generalized in-depth interpretation of \gls{multihutd},
and in Section \ref{subsec:Spatialfeatures} we  present how to incorporate spatial features (e.g., patches and \gls{mm}) in such a setting.

\section{Proposed \glsentryshort{multihutd} Framework}
\label{sec:MultimodalHU}

In this section, we present the proposed \gls{multihutd} framework based on \gls{cpd}.
First, we talk about the implementation of \gls{aoadmm}-\gls{asc}.
Then, we provide an interpretation of \gls{multihutd} including the physical meaning of \gls{asc} and the \gls{elmm} model in the tensor case. Finally, we propose to include spatial features as examples of the third modality.

\subsection{\glsentryshort{aoadmm}-\glsentryshort{asc} with Nonnegativity and Sparsity}
\label{subsec:ASC-AOADMM}

In \gls{cpd}, after imposing nonnegativity on the factor matrices, and sparsity
and \gls{asc} on the abundances, (\ref{eq:tensorfactorization_cpd_cost}) becomes:
\begin{equation}
    \begin{aligned}
    \label{eq:cost_sparse_CPD_original}
        \argmin{\matr{A},\matr{B},\matr{C}}
        {\|\tens{T} - \vect{\Lambda}
        \con_1 \matr{A} \con_2 \matr{B} \con_3 \matr{C}\|_F^2
        +
        \alpha \|\matr{A}\|_1} \\
        \textrm{s.t.}
        \;
        \matr{A}\succeq0,
        \;
        \matr{B}\succeq0,
        \;
        \matr{C}\succeq0,
        \;
        \sum_{r=1}^R a_{i,r} = 1
        \;
        |_{\forall i \in \{1,\dots,I\}}
    \end{aligned}
\end{equation}
In (\ref{eq:cost_sparse_CPD_original}), since $\matr{A}$ adopts the sum-to-one constraint on its rows,
it is hence enough to normalize only the columns of $\matr{B}$ and $\matr{C}$. In principle, these $\ell_2$-norms are absorbed in $\matr{\Lambda}$, but for the sake of consistency, we use the variable $\matr{\Psi}=\matr{\Lambda}\matr{C}$ instead of $\matr{\Lambda}$ and $\matr{C}$, as explained in Section \ref{subsec:CPDandELMM}.
In order to solve (\ref{eq:cost_sparse_CPD_original}), we propose an algorithm inspired by \gls{aoadmm} \cite{HuanSL16:tsp}, where the factor matrices are updated in an alternating way and where each update of a factor matrix is optimized as an \gls{admm} subproblem.

\subsubsection{ASC Solution}
In order to model the problem as close as possible to \gls{lmm},
we need a tensor decomposition algorithm embedding both the \gls{asc} and the non-negativity of factors as constraints.
To this end, the strategy we follow is to extend the concept from \gls{nmf} to \gls{cpd} by stacking:
\begin{itemize}
	\item a row vector to $\matr{B}$ (i.e., $\vect{b}_{J+1,:} \in \R{R}$)
	
	\item a lateral slice to $\tens{T}$ (i.e., $\matr{T}_{:,J+1,:} \in \RRR{I}{1}{K}$)
\end{itemize}
such that $\sum_{r=1}^R a_{i,r} = 1 |_{\forall i \in \{1,\dots,I\}}$ is ensured.

In general, $\matr{T}_{:,J+1,:}$ can be constructed such that $\forall i\in\{1,\dots,I\}$ and $\forall k\in\{1,\dots,K\}$:
\begin{equation}
	t_{i,J+1,k}
	=
	\sum_{r=1}^R a_{i,r} b_{J+1,r} \psi_{k,r}
	\label{eq:extra_slice}
\end{equation}
So if we set:
\begin{itemize}
	\item $\matr{T}_{:,J+1,K} = \delta \matr{1}_{I}$, i.e., $t_{i,J+1,K} = \delta$ $\forall i\in\{1,\dots,I\}$
	
	\item $b_{J+1,r}=\delta \psi_{K,r}^{-1} \;\; \forall r\in\{1,\dots,R\}$,
\end{itemize}
where $\delta$ is the mean of $\tens{T}$, by substituting the expressions in (\ref{eq:extra_slice}) for $k=K$, we have:
\begin{subequations}
    \begin{eqnarray}
	t_{i,J+1,K}
	&=&
	\sum_{r=1}^R a_{i,r} b_{J+1,r} \psi_{K,r}
	\\
	\implies
	\delta
	&=&
	\delta \sum_{r=1}^R a_{i,r} \psi_{K,r}^{-1} \psi_{K,r}
	=
	\delta \sum_{r=1}^R a_{i,r}
    \end{eqnarray}
\end{subequations}
which implies that $\sum_{r=1}^R a_{i,r} = 1$.

At the end of each \gls{aoadmm} iteration, $\tens{T}$ and $\matr{B}$ have to be updated, which boils down to a matrix and a vector updates after each iteration. We denote by $\Tilde{\tens{T}}$ and $\Tilde{\matr{B}}$ the extensions of $\tens{T}$ and $\matr{B}$ with the additional lateral slice and row vector respectively, roughly described as follows:
\begin{equation}
    \Tilde{\tens{T}} = 
    \begin{bmatrix}
        \tens{T}
        \; | \;
        \matr{T}_{:,J+1,:}
    \end{bmatrix},
    \;
    \Tilde{\matr{B}} = 
    \begin{bmatrix}
        \matr{B}
        \\
        \vect{b}_{J+1,:}
    \end{bmatrix},
    \label{eq:T_B_updates}
\end{equation}
Then, expression (\ref{eq:cost_sparse_CPD_original}) becomes:
\begin{equation}
    \begin{aligned}
    \label{eq:cost_sparse_CPD_original_}
        \argmin{\matr{A},\Tilde{\matr{B}},\matr{\Psi}}
        {\|\Tilde{\tens{T}} - \tens{I}
        \con_1 \matr{A} \con_2 \Tilde{\matr{B}} \con_3 \matr{\Psi}\|_F^2
        +
        \alpha \|\matr{A}\|_1} \\
        \textrm{s.t.}
        \;
        \matr{A}\succeq0,
        \;
        \Tilde{\matr{B}}\succeq0,
        \;
        \matr{\Psi}\succeq0
    \end{aligned}
\end{equation}
where $\tens{I}$ is a diagonal tensor of ones.

\subsubsection{\glsentryshort{admm} Updates}
At this stage, solving (\ref{eq:cost_sparse_CPD_original_}) with \gls{aoadmm} becomes simple.
We demonstrate the \gls{admm} sub-problem updates for each factor matrix starting with $\matr{A}$.

Supposing that $\Tilde{\matr{T}}_{(1)}$ represents the mode-$1$ unfolding of $\Tilde{\tens{T}}$, we can write the sub-problem of $\matr{A}$ as follows:
\begin{equation}
    \begin{aligned}
    \label{eq:cost_NCPD_A_1}
        \matr{A} = \argmin{\matr{A}}
        {\frac{1}{2}\|\Tilde{\matr{T}}_{(1)} - \Tilde{\matr{W}}_{(A)}\matr{A}^{\T}\|_F^2
        +
        \alpha \|\matr{A}\|_1}
        \\
        \textrm{ s.t. } \matr{A}\succeq0
    \end{aligned}
\end{equation}
where $\Tilde{\matr{W}}_{(A)} \in \RR{(J+1)K}{I} = \Tilde{\matr{B}} \khatri \matr{\Psi}$ represents the Khatri-Rao product \cite{Como14:spmag}.
By introducing the splitting variable $\Bar{\matr{A}} = \matr{A}^{\T}$, expression (\ref{eq:cost_NCPD_A_1}) becomes:
\begin{equation}
    \begin{aligned}
    \label{eq:cost_NCPD_A_2}
        \argmin{\matr{A},\Bar{\matr{A}}}
        {\frac{1}{2}\|\Tilde{\matr{T}}_{(1)} - \Tilde{\matr{W}}_{(A)}\Bar{\matr{A}}\|_F^2
        +
        \alpha \|\matr{A}\|_1}
        \\
        \textrm{ s.t. } \Bar{\matr{A}} = \matr{A}^{\T}
        \textrm{ and } \matr{A}\succeq0
    \end{aligned}
\end{equation}
Adopting \gls{admm} for (\ref{eq:cost_NCPD_A_2}), the updates of $\Bar{\matr{A}}$ and $\matr{A}$ become:
\begin{equation}
	\begin{aligned}
        \label{eq:cost_ADMM_A}
        &\Bar{\matr{A}} \leftarrow (\Tilde{\matr{W}}_{(A)}^{\T} \Tilde{\matr{W}}_{(A)} + \rho \matr{I})^{-1}
        (\Tilde{\matr{W}}^{\T}_{(A)} \Tilde{\matr{T}}_{(1)} + \rho(\matr{A}+\matr{U}_{(A)})^{\T})
        \\
        &\matr{A} \leftarrow \max(0,\Bar{\matr{A}}^{\T}-\matr{U}_{(A)} - \frac{\alpha}{\rho})
        \\
        &\matr{U}_{(A)} \leftarrow \matr{U}_{(A)} + \matr{A} -\Bar{\matr{A}}^{\T}
	\end{aligned}
\end{equation}
where $\matr{U}_{(A)} \in \RR{I}{R}$ is called the dual variable.

Similarly, the updates of $\Tilde{\matr{B}}$ and $\matr{\Psi}$ become:
\begin{equation}
	\begin{aligned}
		\label{eq:cost_ADMM_B}
		&\Bar{\matr{B}} \leftarrow (\Tilde{\matr{W}}_{(B)}^{\T} \Tilde{\matr{W}}_{(B)} + \rho \matr{I})^{-1}
		(\Tilde{\matr{W}}^{\T}_{(B)} \Tilde{\matr{T}}_{(2)} + \rho(\matr{B}+\matr{U}_{(B)})^{\T})
		\\
		&\Tilde{\matr{B}} \leftarrow \max(0,\Bar{\matr{B}}^{\T}-\matr{U}_{(B)})
		\\
		&\matr{U}_{(B)} \leftarrow \matr{U}_{(B)} + \Tilde{\matr{B}} -\Bar{\matr{B}}^{\T}
	\end{aligned}
\end{equation}
\begin{equation}
    \begin{aligned}
    	\label{eq:cost_ADMM_C}
    	&\Bar{\matr{\Psi}} \leftarrow (\Tilde{\matr{W}}_{(\Psi)}^{\T} \Tilde{\matr{W}}_{(\Psi)} + \rho \matr{I})^{-1}
    	(\Tilde{\matr{W}}^{\T}_{(\Psi)} \Tilde{\matr{T}}_{(3)} + \rho(\matr{\Psi}+\matr{U}_{(\Psi)})^{\T})
    	\\
    	&\matr{\Psi} \leftarrow \max(0,\Bar{\matr{\Psi}}^{\T}-\matr{U}_{(\Psi)})
    	\\
    	&\matr{U}_{(\Psi)} \leftarrow \matr{U}_{(\Psi)} + \matr{\Psi} -\Bar{\matr{\Psi}}^{\T}
    \end{aligned}
\end{equation}
where
$\Tilde{\matr{T}}_{(2)}$ and $\Tilde{\matr{T}}_{(3)}$ are the mode-$2$ and mode-$3$ unfoldings of $\Tilde{\tens{T}}$,
$\Tilde{\matr{W}}_{(B)} = \matr{A} \khatri \matr{\Psi}$ and $\Tilde{\matr{W}}_{(\Psi)} = \matr{A} \khatri \Tilde{\matr{B}}$ are the Khatri-Rao products, and
$\matr{U}_{(B)}$ and $\matr{U}_{(\Psi)}$ are the dual variables.

Finally, for order-$2$ tensors, this model becomes equivalent to solving \gls{nmf} (\ref{eq:cost_sparse_NMF}).
The implementation of \gls{aoadmm}-\gls{asc} is summarized in Algorithm \ref{algo:AOADMMASC}. The code is available on GitHub in \texttt{Python}\footnote{
    \href{https://github.com/mhmdjouni/MultiHU-TD-Python}{\texttt{https://github.com/mhmdjouni/MultiHU-TD-Python}}
} and \texttt{MATLAB}\footnote{
    \href{https://github.com/mhmdjouni/MultiHU-TD-MATLAB}{\texttt{https://github.com/mhmdjouni/MultiHU-TD-MATLAB}}
}.

\begin{algorithm}[ht]
\caption{\gls{aoadmm}-\gls{asc}}
\begin{algorithmic}\label{algo:AOADMMASC}
\REQUIRE $\tens{T}$, $\matr{A},\matr{B},\matr{\Psi}$, $\matr{U}_{(A)},\matr{U}_{(B)},\matr{U}_{(\Psi)}$, $\alpha$
    \STATE Initialize $\matr{A},\matr{B},\matr{\Psi}$;
    \STATE Initialize $\matr{U}_{(A)},\matr{U}_{(B)},\matr{U}_{(\Psi)}$ to zero;
    
    \REPEAT
        \STATE Set $\Tilde{\matr{B}}$ and $\Tilde{\tens{T}}$ based on (\ref{eq:T_B_updates});
        \STATE Set $\matr{B} \leftarrow \Tilde{\matr{B}}$ and
        $\tens{T} \leftarrow \Tilde{\tens{T}}$;
        
        \FOR{$\forall \matr{D} \in \{\matr{A},\matr{B},\matr{\Psi}\}$}
            \STATE $\Tilde{\matr{W}}_{(D)} = \khatri_{\matr{J}\neq \matr{D}} \matr{J}$;
            \STATE $\rho = \textrm{trace} (\Tilde{\matr{W}}_{(D)}^{\T} \Tilde{\matr{W}}_{(D)})/R$; \cite{HuanSL16:tsp}
            \STATE Update $\matr{D}$ with either (\ref{eq:cost_ADMM_A}), (\ref{eq:cost_ADMM_B}), or (\ref{eq:cost_ADMM_C});
        \ENDFOR
        
        \STATE
        Absorb the column-wise $\ell_2$-norms of $\matr{B}$ into $\matr{\Psi}$
        such that $\vect{\psi}_{:,r} \leftarrow
        \vect{\psi}_{:,r} ~ \|\vect{b}_{:,r}\|_2 ~~{\forall r\in\{1,\dots,R\}}$
        \STATE
        Normalize the columns of $\matr{B}$
    \UNTIL{Termination criterion (e.g., number of iterations)}
    
\RETURN $\matr{A},\matr{B},\matr{\Psi}$
\end{algorithmic}
\end{algorithm}

\subsection{Interpreting Tensor-based unmixing, \glsentryshort{asc}, and \glsentryshort{elmm}}
\label{subsec:elmminterpretation}

\begin{figure*}[t]
    \begin{minipage}{\textwidth}
        \centering
        
        \begin{minipage}[b]{0.487\textwidth}
            \centering
            \includegraphics[width=0.80\textwidth]{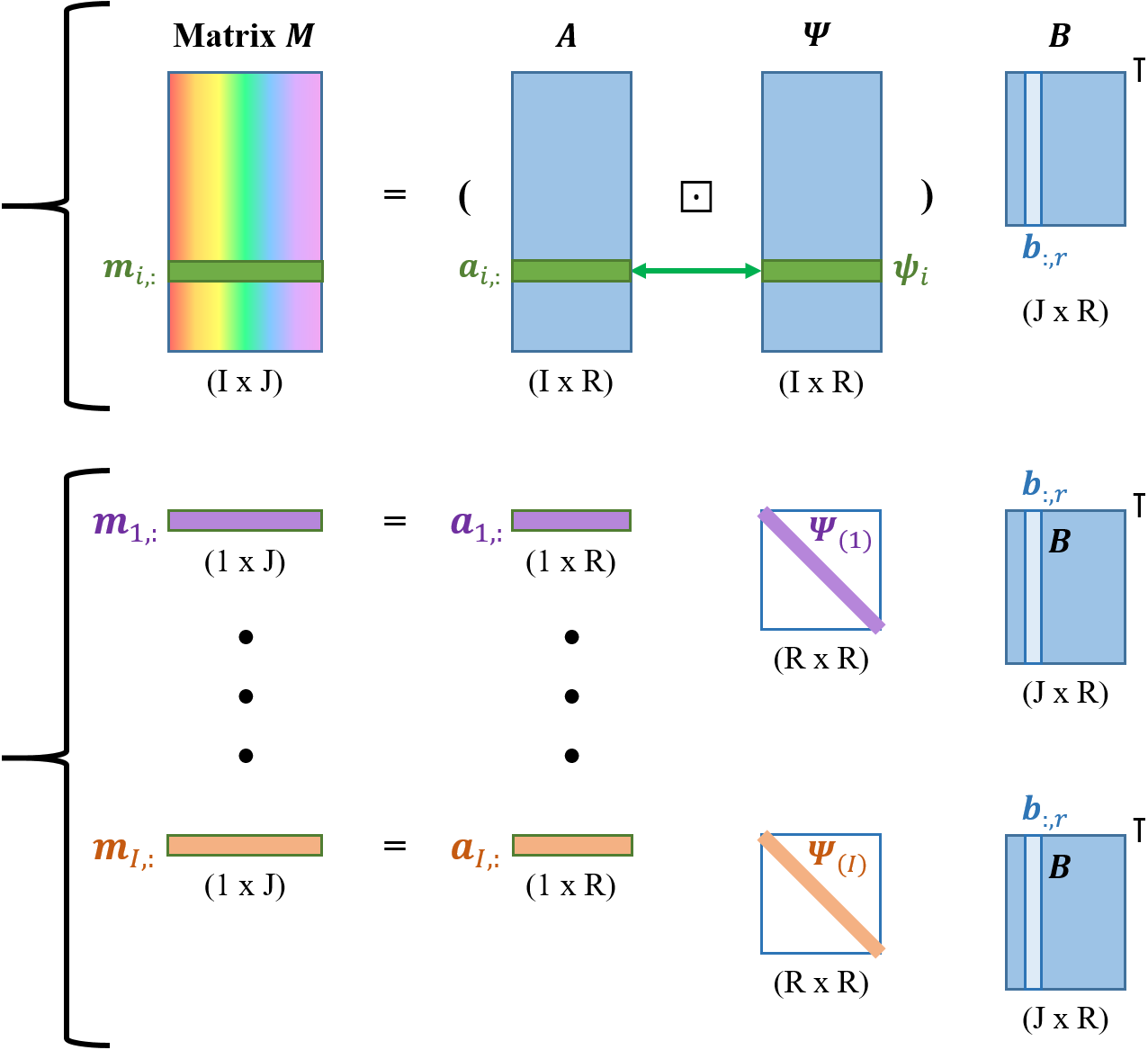}
            \caption{
            Our visualization of equations (\ref{eq:ELMM_CLS_matrix}) (Hadamard product) and (\ref{eq:ELMM_CLS_pixel3}) (matrix product). The color code of the bottom part follows that of Fig. \ref{subfig:ELMM_CLS_graph}. We have $\matr{\Psi}_{(i)} = \Diag{\matr{\Psi}_{i,:}}$.}
            \label{fig:ELMM_CLS_visual}
        \end{minipage}
        ~~
        \begin{minipage}[b]{0.487\textwidth}
            \centering
            \includegraphics[width=0.80\textwidth]{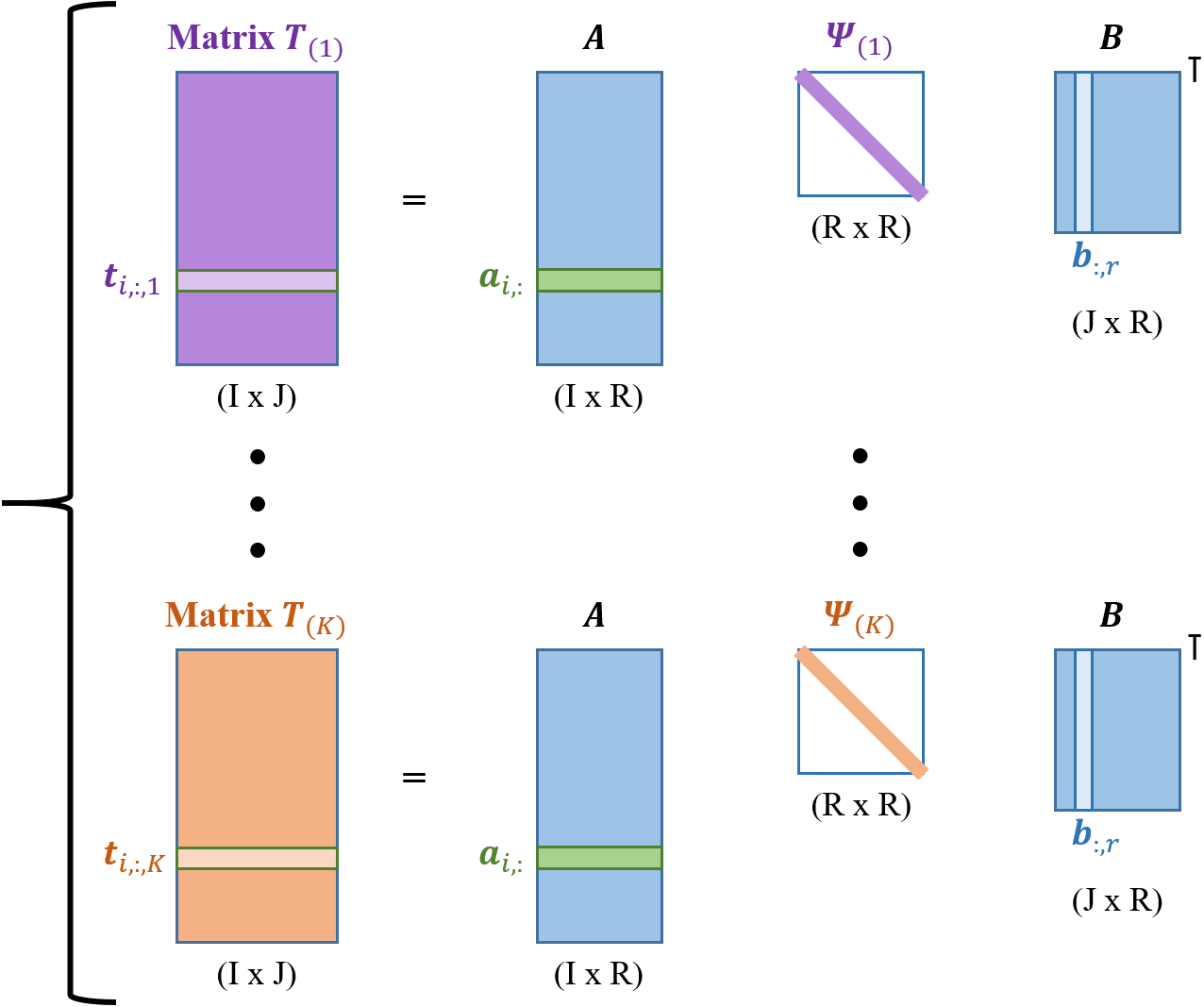}
            \caption{
            Our visualization of equation (\ref{eq:RegELMM_frontalslice_intro}). The color code is made analogous to that of Fig. \ref{fig:CPD_Illust} and follows that of Fig. \ref{subfig:RegELMM_graph}. We have $\matr{\Psi}_{(k)} = \Diag{\vect{\psi}_{k,:}}$.}
            \label{fig:RegELMM_visual}
        \end{minipage}
    \end{minipage}
    
    \vspace{3mm}
    
    \begin{minipage}{\textwidth}
        \centering
        
        \begin{subfigure}[b]{0.285\textwidth}
            \centering
            \includegraphics[width=\textwidth]{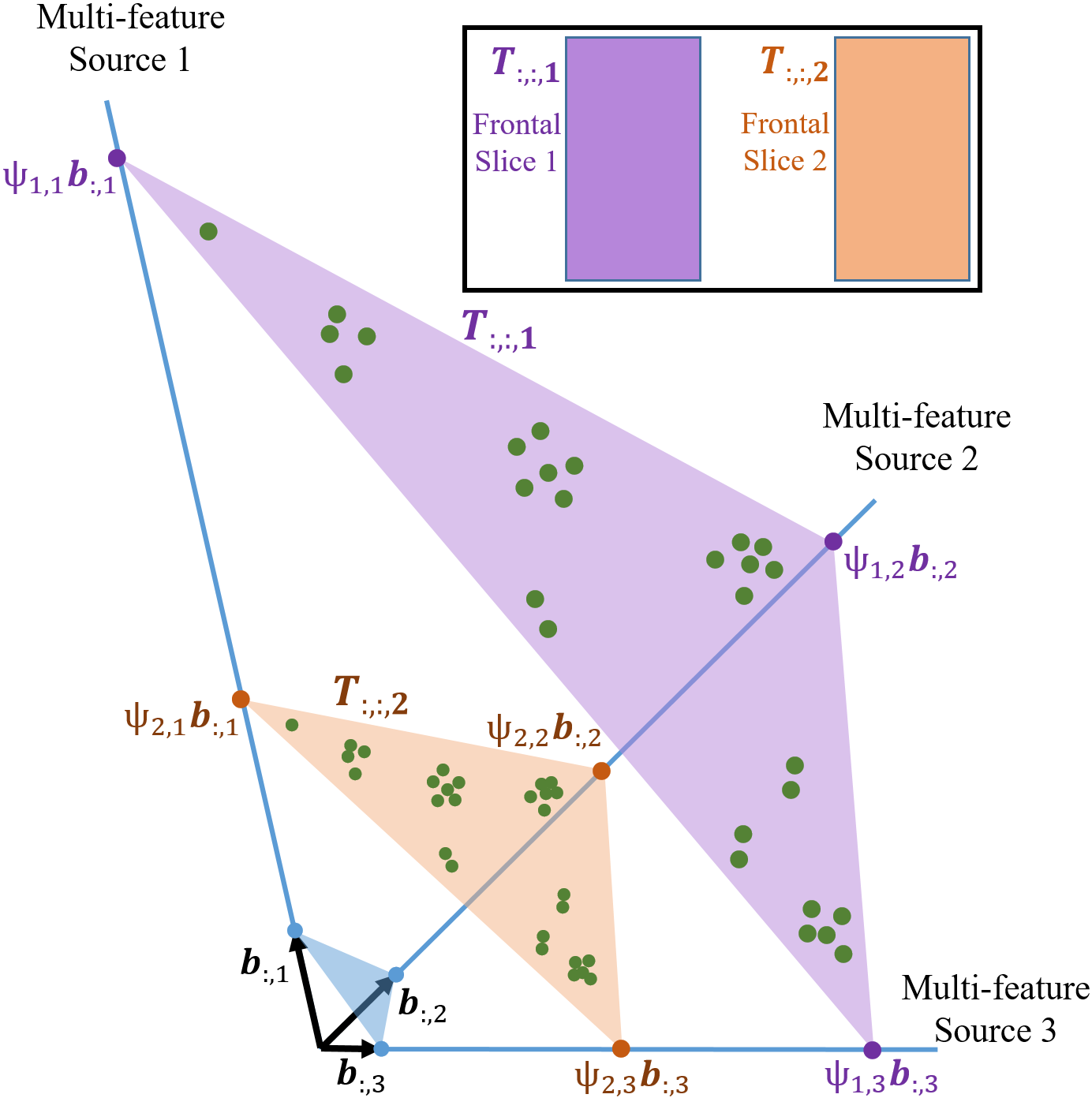}
            \caption{\gls{elmm}-\gls{cpd} (General)}
            \label{subfig:RegELMM_graph}
        \end{subfigure}
        ~
        \begin{subfigure}[b]{0.335\textwidth}
            \centering
            \includegraphics[width=\textwidth]{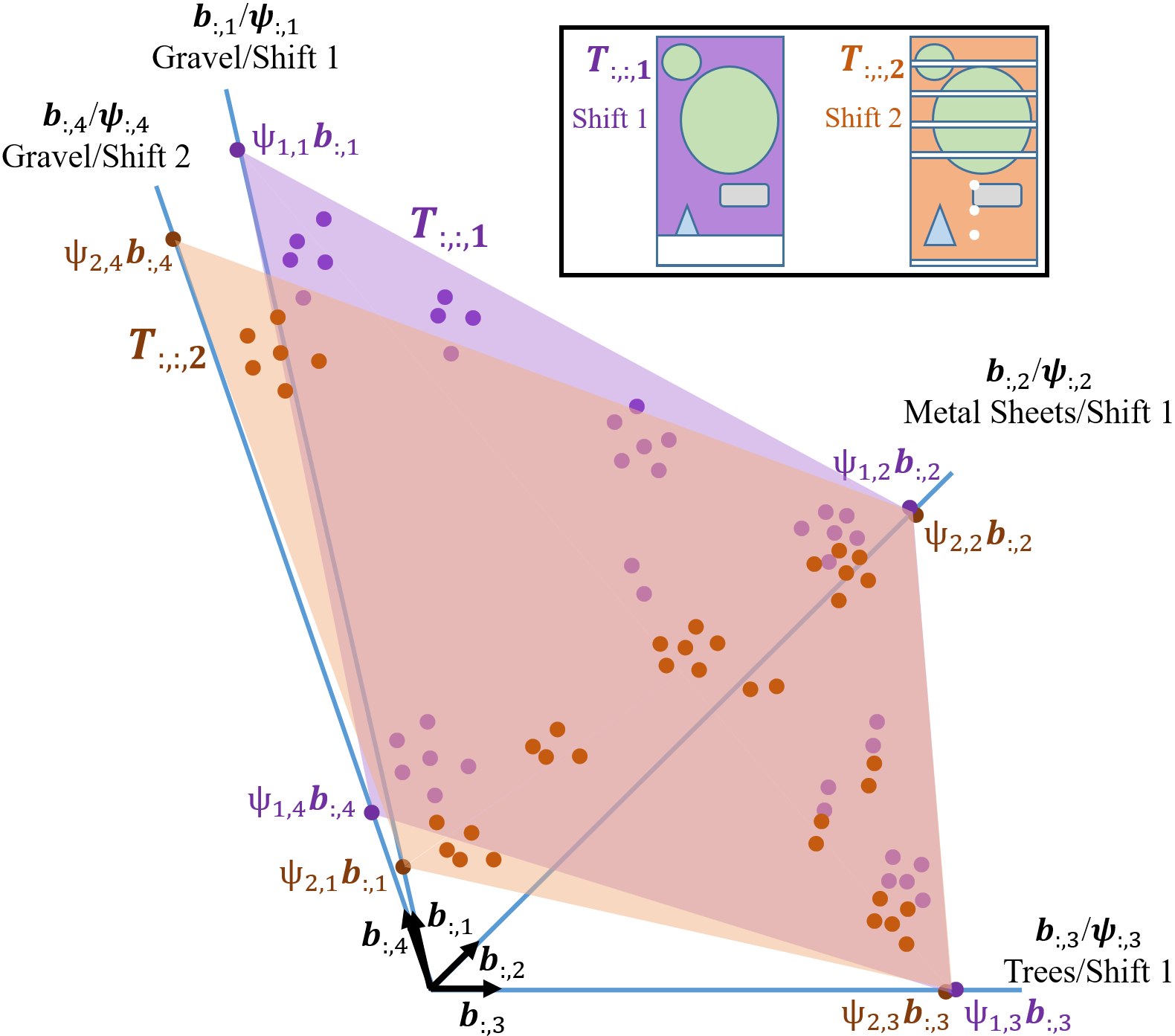}
            \caption{\gls{elmm}-\gls{cpd} (Patches)}
            \label{subfig:RegELMM_graph_patches}
        \end{subfigure}
        ~
        \begin{subfigure}[b]{0.335\textwidth}
            \centering
            \includegraphics[width=\textwidth]{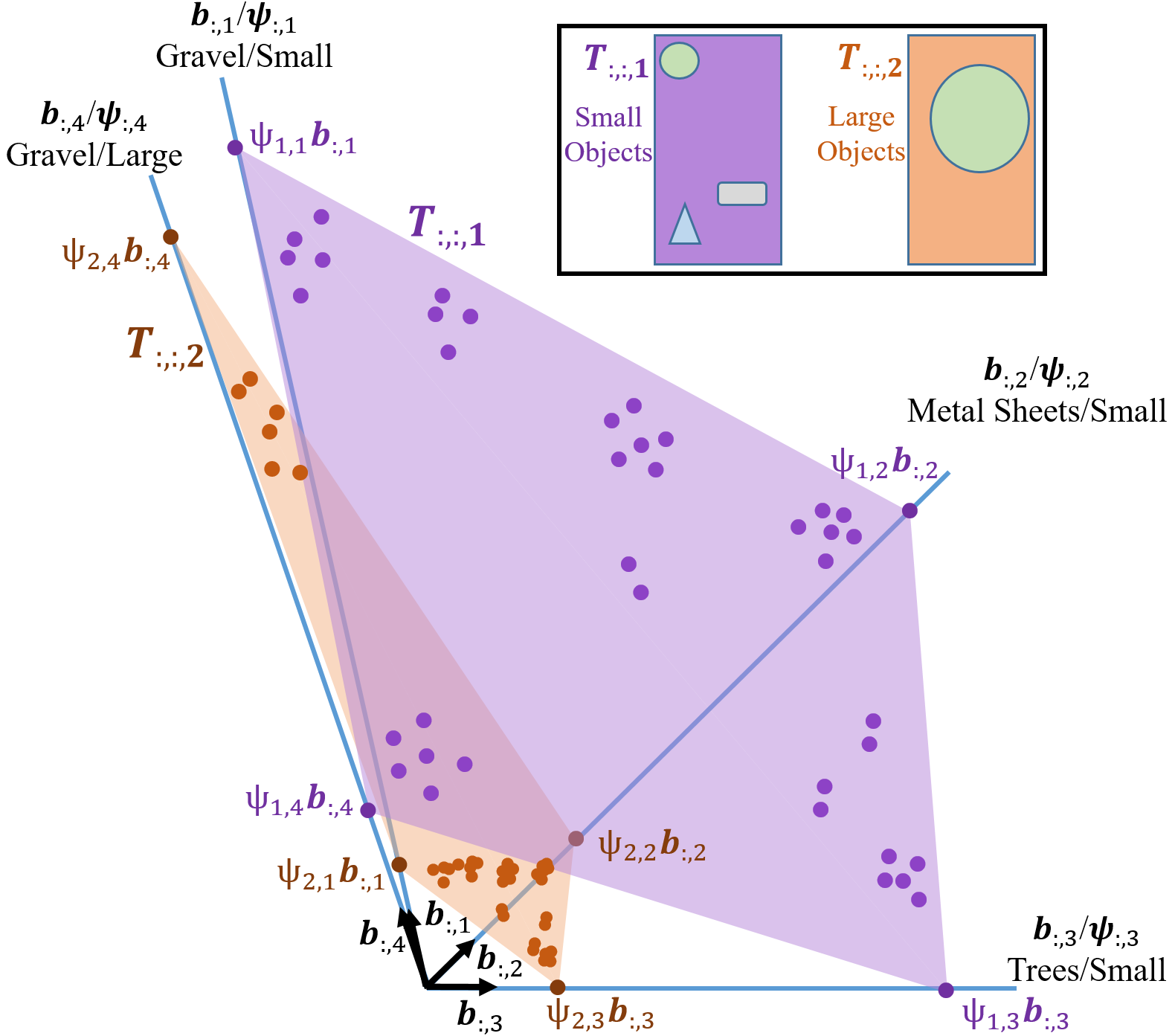}
            \caption{\gls{elmm}-\gls{cpd} (\gls{mm})}
            \label{subfig:RegELMM_graph_MM}
        \end{subfigure}
        
        \caption{
            Graphical representations of (a) \gls{cpd} with $R=3$ components in the case of three spectral signatures $\{\vect{b}_{:,1}, \vect{b}_{:,2}, \vect{b}_{:,3}\}$ and two frontal slices $\{\matr{T}_{:,:,1},\matr{T}_{:,:,K}\}$, and (b) patch-\gls{cpd} and (c) \gls{mm}-\gls{cpd} with $R=4$ components in the case of four spectral signatures $\{(\vect{b}_{:,1},\vect{\psi}_{:,1}), \dots, (\vect{b}_{:,4},\vect{\psi}_{:,4})\}$
            and two frontal slices $\{\matr{T}_{:,:,1},\matr{T}_{:,:,2}\}$.
            Very correlated vectors can be seen as ``spectral bundles'' with different third-modality characteristics.
            The relative coordinates of the pixels in the convex hulls must be the same since $\matr{A}$ is common for all the frontal slices.
        }
        \label{fig:graph_RegELMM_Patches_MM}
    \end{minipage}
\end{figure*}

Here, we build upon what has been presented in Sections \ref{subsec:Unmixing_ELMM_Classical} and \ref{subsec:CPDandELMM} as methodological, physical, and graphical bases for the \gls{multihutd} interpretation.
For that, we first draw the analogies between the expressions of Sections \ref{sec:Background} and \ref{sec:MultimodalHU} by starting from the interpretation of the matrix case and elaborating that of the tensor case. In the process, we break down the physical meaning of \gls{asc} and that of the so-called \gls{sv} function of \gls{elmm} in \gls{multihutd} at the base of the composition of the frontal slices of the tensor.
Then, we visualize the expressions in order to interpret \gls{multihutd} through graphical representations of subspaces while commenting on the physical role of the extracted factors and the number of latent components $R$.

\subsubsection{Interpretation of ASC}

In the matrix case, assuming a matricized \gls{hsi} $\matr{M}$ (that is, after reordering the two pixel modes into one mode in lexicographic order) such that:
\begin{equation}
	\matr{M} = \matr{A} \, \matr{B}^{\T}
\end{equation}
where $\matr{A}$ and $\matr{B}$ represent the estimated abundances and endmembers respectively,
the physical meaning of \gls{asc} is that it constrains the columns of $\matr{B}$ to form a simplex. Then, the rows of $\matr{A}$ (which sum to one) represent the position of the pixels on said simplex.
This is visualized in \figurename~\ref{fig:graph_LMM_ELMM_CLS}.

In the tensor case, we assume a tensor $\tens{T}$ whose \gls{cpd} is expressed as:
\begin{equation}
	\tens{T} = \tens{I} \con_1 \matr{A} \con_2 \matr{B} \con_3 \matr{\Psi}
\end{equation}
where $\tens{I}$ is a diagonal tensor of ones, $\matr{A}$ and $\matr{B}$ represent the estimated abundances and endmembers respectively, and $\matr{\Psi}$ represents the factor matrix of the third modality.

Here, each slice of the tensor $\matr{T}_{:,:,k}$ $\forall k \in \{1, \dots, K\}$ represents a matricized \gls{hsi} similar to $\matr{M}$ (e.g., corresponding to acquisitions at different dates in a time series, at different angles in multi-angular acquisitions or at different scales in a multi-scale decomposition), and we have the following linear relationship:
\begin{equation}
	\matr{T}_{:,:,k}
	= \matr{A} \, (\diag{\Psi_{k, :}} \, \matr{B}^{\T})
	= \matr{A} \, \matr{B}^{(k)\,\T}
	\, \forall k,
\end{equation}
where $\Psi_{k,:}$ represents the $k$-th row of $\matr{\Psi}$.
First, let us look at the properties of the aforementioned expression:
\begin{itemize}
	\item If we look at each slice separately, the physical meaning of applying \gls{asc} is similar to that of the matrix case where the simplex is formed out of the columns of $\matr{B}^{(k)}$.
	In fact, we have that $\matr{B}^{(k)} = \matr{B} \, \diag{\Psi_{k,:}}$, which means that the columns of $\matr{B}^{(k)}$ are only scaled versions of those of $\matr{B}$ such that:
	\begin{equation}
		\vect{b}^{(k)}_{:,r} = \psi_{k,r} \vect{b}_{:,r}
		\,\,\,
		\forall r \in \{1, \dots, R\}
		\label{eq:scaled_endmembers}
	\end{equation}
	
	\item If we consider all the slices together, we notice that the abundance matrix $\matr{A}$ is common to all of them.
	Moreover, we notice that the factor matrix of estimated endmembers $\matr{B}$ (obtained through \gls{cpd}) is at the base of their estimated endmembers, influenced only by the corresponding scaling factors in $\Psi_{k,:}$, which encode the corresponding third-mode features.
	
	\item Given that $\matr{A}$ and $\matr{B}$ factorize the physical and data structures along the pixel and spectral mode respectively, they are indepedent of the third-mode differences in the hyperspectral scene between the slices.
	One could even construct a matrix $\matr{M}^{(\textrm{CPD})}$ from the first two factor matrices $\matr{A}$ and $\matr{B}$ (obtained through \gls{cpd}) such that:
	\begin{equation}
		\matr{M}^{(\textrm{CPD})} = \matr{A} \, \matr{B}^{\T}
	\end{equation}
	where $\matr{B}$ is independent of the spectral variabilities present along the third mode.
	
	\item As $\matr{A}$ and $\matr{B}$ factorize the pixel and spectral information, the third-mode factor matrix $\matr{\Psi}$ encodes the changes between the slices along the third mode where the $k$-th row $\Psi_{k,:}$ is associated to the $k$-th slice.
	Intuitively speaking, this allows some degrees of freedom to express the nonlinearities along the third mode in a linear sense, which is reflected on the level of each slice by scaling the columns of $\matr{B}$ as expressed in \eqref{eq:scaled_endmembers}.
\end{itemize}

Finally, assuming that we have $K$ slices in the tensor, imposing \gls{asc} is the equivalent of having $K$ simplices whose edges, defined by the columns of $\matr{B}^{(k)}$ $\forall k \in \{1, \dots, K\}$, can move only along the directions of the columns of $\matr{B}$ such that:
\begin{itemize}
	\item The positions of the edges of each simplex is defined by the corresponding scaling factor $\psi_{k,r}$, which encodes the third-mode physical property (whether it is time, morphological properties such as scale and brightness, neighborhood pixels, etc) of the $r$-th estimated endmember $\vect{b}_{:,r}$ in the $k$-th tensor slice $\matr{T}_{:,:,k}$.
	
	\item For instance, an estimated endmember $\vect{b}_{:,r}$ can be relevant in a given slice $\matr{T}_{:,:,1}$ due to a high factor $\psi_{1,r}$, such that $\vect{b}^{(1)}_{:,r} = \psi_{1,r} \vect{b}_{:,r}$, but also have a low contribution in another slice $\matr{T}_{:,:,2}$ due to a low factor $\psi_{2,r}$, such that $\vect{b}^{(2)}_{:,r} = \psi_{2,r} \vect{b}_{:,r}$.
	We can see that $\vect{b}_{:,r}$ remains independent of the physical entity that the third mode represents, but also that $\vect{b}^{(1)}_{:,r}$ and $\vect{b}^{(2)}_{:,r}$ move along the direction of $\vect{b}_{:,r}$ (due to the scaling factors $\psi_{1,r}$ and $\psi_{2,r}$) based on the effect that the the third-mode physical entity applies on $\vect{b}_{:,r}$ in the given slice.
	
	\item Since $\matr{A}$ is common to all the slices, the positions of the pixels are relatively fixed to each of the $K$ simplices.
\end{itemize}
This is demonstrated in \figurename~\ref{fig:RegELMM_visual} and \ref{fig:graph_RegELMM_Patches_MM} in the manuscript.

\subsubsection{Interpretation of tensor-based ELMM}

First, we note that expressions (\ref{eq:RegELMM_frontalslice_intro}) and (\ref{eq:RegELMM_pixel_intro}) are analogous to the \gls{elmm} expression (\ref{eq:ELMM_CLS_pixel3}). The major difference between the two cases is that in \gls{cpd}, the scaling factors are \textit{frontal slice-dependent} ($\psi_{k,r}$), while in \gls{elmm}, they are pixel-dependent ($\psi_{i,r}$).
Second, we visualize (\ref{eq:ELMM_CLS_matrix}) and (\ref{eq:ELMM_CLS_pixel3}) in Fig. \ref{fig:ELMM_CLS_visual},
and (\ref{eq:RegELMM_frontalslice_intro}) and (\ref{eq:RegELMM_pixel_intro}) in Fig. \ref{fig:RegELMM_visual}.
Looking at (\ref{eq:RegELMM_frontalslice_intro}), the frontal slices $\matr{T}_{:,:,k}$ and the physical meaning that they represent have a direct influence on the \gls{sv} function $\matr{f}_k$ and the interpretation of the \glspl{sv}, which is simply reflected as scaling factors in each row of $\matr{\Psi}$, i.e. $\vect{\psi}_{k,:}$ (or $\matr{\Psi}_{(k)} = \Diag{\vect{\psi}_{k,:}}$).

As a result, since the spatial and spectral information are factorized and represented by $\matr{A}$ and $\matr{B}$ respectively, and since $\matr{A}$ and $\matr{B}$ are shared by all the frontal slices,
then each frontal slice $\matr{T}_{:,:,k}$ is inherently differentiated through a set of $R$ scaling factors $\{\psi_{k,1}, \dots, \psi_{k,R}\}$.
Consequently, the spectral information in each frontal slice $\matr{T}_{:,:,k}$ can be seen as the set of scaled sources $\{\vect{b}^{(k)}_{:,r} = \psi_{k,r} \vect{b}_{:,r}\} |_ {\forall\; r \in \{1,\dots,R\}}$,
where $\{\vect{b}_{:,r}\}$ are the columns of $\matr{B}$ (independent of the slices), $\vect{b}^{(k)}_{:,r}$ are their spectral variations per frontal slice, and $\vect{\psi}_{k,:}$ encodes the scaling factors of these variations.

This also means that $R$ is a major parameter that represents the degrees of freedom especially through the scaling factors of $\matr{\Psi}$, which then jointly encodes:
\begin{itemize}
	\item the mode-$3$ evolution of the extracted components of $\matr{A}$ and $\matr{B}$ in its columns $\vect{\psi}_{:,r}$
	
	\item the per-slice modeling of the \glspl{sv} in its rows $\vect{\psi}_{k,:}$.
\end{itemize}
Intuitively, when obtaining an augmented \gls{hsi} tensor, one can say that the physical representations of any applied transformations (e.g., scale, illumination) \cite{JounDC20:mmta} and any natural evolution of a scene (e.g., time series) \cite{VegaCFCC16:tgrs}, and resulting in $\matr{T}_{:,:,k}$, are reflected and observed through the matrix $\matr{\Psi}$ of the decomposition.
Moreover, we point out the following:
\begin{itemize}
    \item In \gls{cpd}, there are as many \gls{sv} functions (and simplices) as the frontal slices of the tensor, which is significantly lower than the number of pixels ($K \lll I$).
    
    \item In \gls{cpd}, one row of $R$ scaling factors in $\matr{\Psi}$ corresponds to a full frontal slice and is shared by all the pixel rows of $\matr{A}$, while in classical \gls{elmm}, each row of scaling factors in $\matr{\Psi}$ corresponds to one pixel of $\matr{M}$ and interacts with only one row of $\matr{A}$. This clearly appears when we compare (\ref{eq:ELMM_pixel_}) and (\ref{eq:ELMM_CLS_pixel3}) to (\ref{eq:RegELMM_frontalslice_intro}) and (\ref{eq:RegELMM_pixel_intro}), and Fig. \ref{fig:ELMM_CLS_visual} to Fig. \ref{fig:RegELMM_visual}.
    
    \item On a graphical representation, when \gls{asc} is imposed, \gls{cpd} suggests that each simplex contains $I$ pixels as illustrated in Fig. \ref{subfig:RegELMM_graph}, such that the relative coordinates of the pixels inside each convex hull are the same since each row of $\matr{\Psi}$ interacts with all of $\matr{A}$.
\end{itemize}
In summary, having a third mode in \gls{hsi} produces scaling factors in \gls{elmm} that absorb the \glspl{sv} based on the physical meaning of the frontal slices along the third mode (e.g., time, patches, \gls{mm}), which balances the extracted factors in $\matr{A}$ and $\matr{B}$ independently of said \glspl{sv}.
Moreover, the imposed value of $R$ represents the number of extracted sources and scaling factors and has a major effect on the results and the \gls{sv} interpretation.
Since the extraction of materials is also driven by the third mode diversity, it is possible to expect a few more latent components than the number of pure ``\glspl{em}'' existing in the scene, with some spectral correlations as roughly demonstrated in \figurename~\ref{fig:graph_RegELMM_Patches_MM}.
As $R$ decreases, we tend towards having fewer degrees of freedom, where \gls{cpd} tends towards extracting the \glspl{em} while applying a regularization on the observed pixels influenced by the physical meaning of the information across the third mode.
As $R$ increases, we tend towards having more degrees of freedom, where \gls{cpd} tends towards extracting factors with multi-feature separability of the sources. We note that $R$ should not be too high in order to avoid
over-fitting\footnote{
    There is no exact value of the tensor rank, and finding a good estimate is an open challenge (as for matrix factorization problems), which is out of the scope of this paper.
	As a rule of thumb, $R$ could be chosen by looking at the reconstruction error in the factorization, e.g. by choosing the smallest value of $R$ providing an acceptable reconstruction, or by observing the elbow of the plot of singular values of the mode-$1$ unfolding of the data.
} (given the relative low-rankness of the data) and to ensure uniqueness of the \gls{cpd} \cite{Como14:spmag, QiCL16:tit}.

\begin{figure*}[!ht]
	\centering
	\begin{minipage}{\textwidth}
		\centering
		\includegraphics[width=0.65\textwidth]{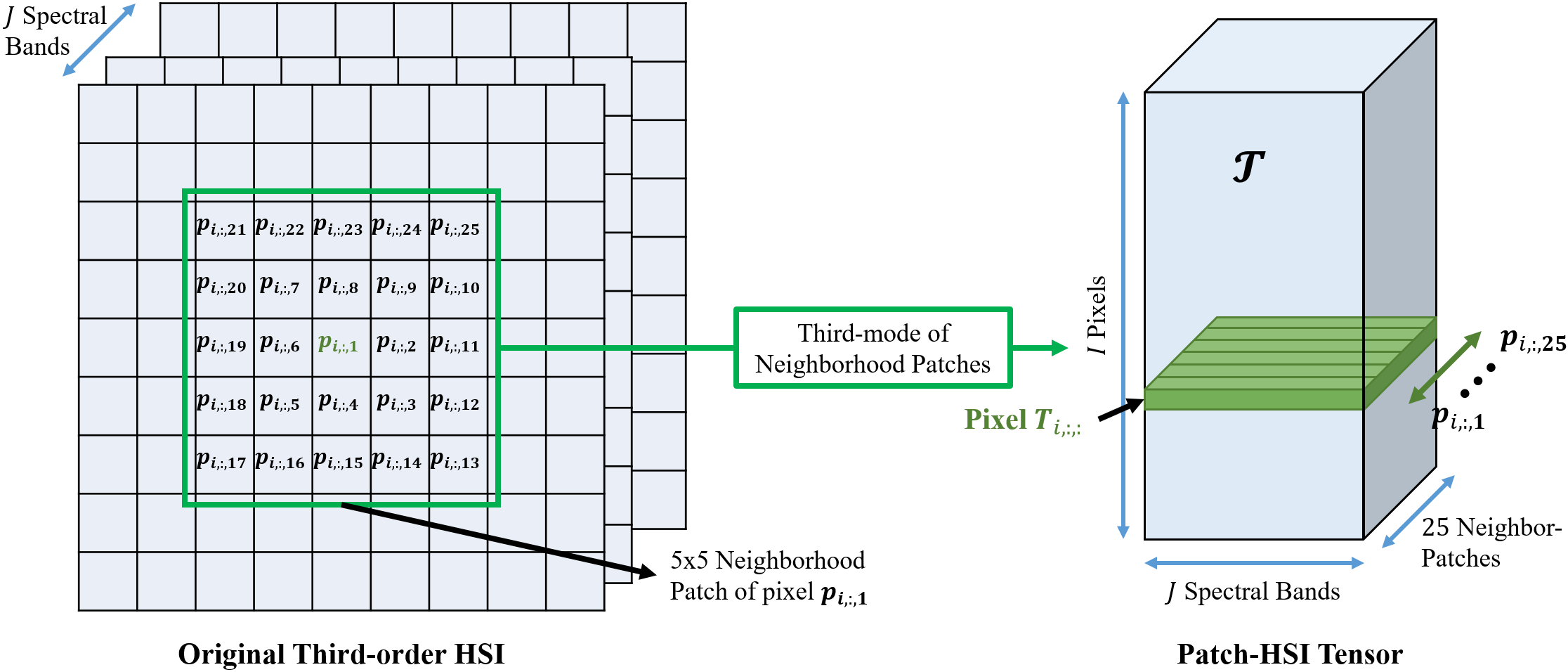}
		\caption{An illustration of constructing a $5\times5$ Patch-\gls{hsi} tensors based on \cite{VegaCFUDCC16:eusipco}.}
		\label{fig:patches_illustration}
	\end{minipage}
	
	\vspace{3mm}
	
	\begin{minipage}{\textwidth}
		\centering
		\includegraphics[width=0.9\textwidth]{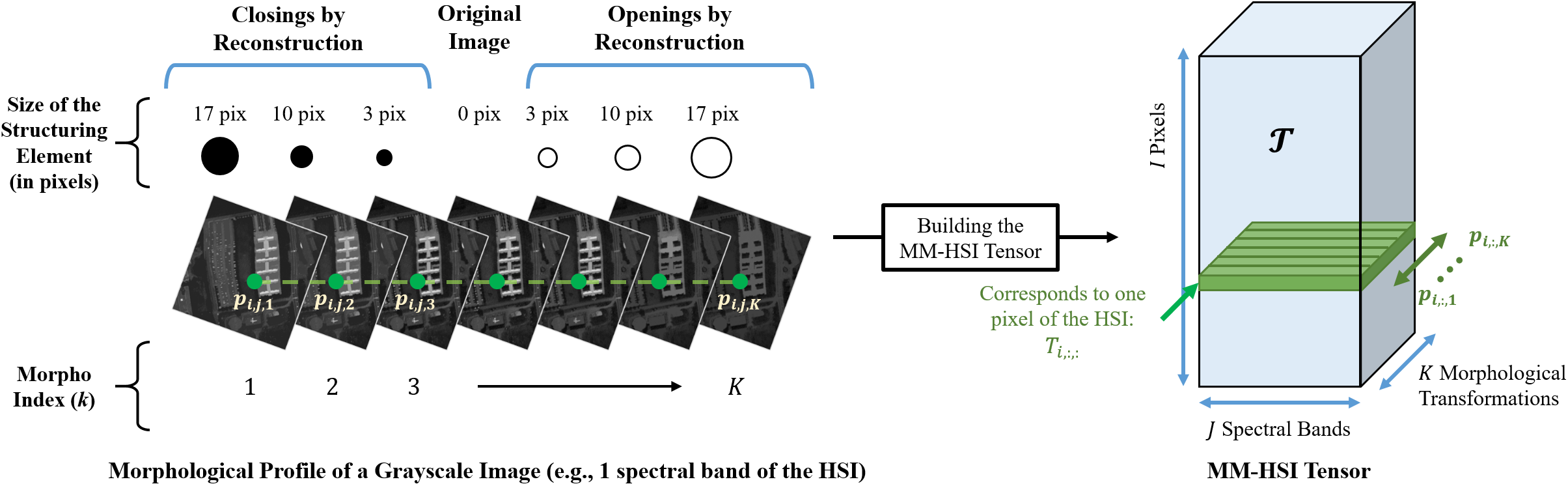}
		\caption{
			Example of a sequential morphological filtering of a grayscale image (corresponding to one spectral band of the \gls{hsi} of Pavia University) with Openings and Closings by Reconstruction using successive sizes of the \acrlong{se}, which is a disk in this case. The stacking of the transformations, with the original image corresponding to $0$ size being placed in the middle, is referred to as the \acrlong{mp} of the image.
			The green spots mark the same pixel positioning in each of the transformations. Then, to create the \gls{mm}-\gls{hsi} tensor \cite{JounDC20:mmta}, the pixel positioning modes are reordered into lexicographic order along the first mode, and the yellow spots are stored along the third mode, in order to create the tensor.
		}
		\label{fig:morpho_illustration}
	\end{minipage}
\end{figure*}

\subsection{Examples of third-mode features: Spatial Features (patches / \glsentrylong{mm})}
\label{subsec:Spatialfeatures}

In this section, for the sake of comparison and illustration, we consider two examples of spatial features that augment a \gls{hsi} into a third-order tensor for \gls{multihutd}: neighborhood patches and \gls{mm}.
Having these two types of features allow for a more comprehensive comparison in terms of the properties of the \gls{multihutd} framework and its links to previous works on said features.
We revisit the case of patches with additional insights, and introduce \gls{mm}.
Consequently, this helps demonstrate the interpretability of the model especially in terms of physical significance and the variation of $R$.
We often refer to Fig. \ref{fig:RegELMM_visual} and \ref{fig:graph_RegELMM_Patches_MM} for illustration.

\subsubsection{Patches}
\label{subsubsec:Unmixing_ELMM_Patches}

We recall that the motivation for adding patches as features is to perform a spatial regularization by considering the spatial correlation of neighboring pixels \cite{VegaCFUDCC16:eusipco}.
Constructing a \gls{hsi} tensor from neighborhood patches (coined as ``Patch-\gls{hsi} tensor'') is illustrated in Fig. \ref{fig:patches_illustration}.
In short, each pixel tube in the original \gls{hsi} cube is taken with a patch of its neighboring pixels (of predetermined size), then the pixel and its neighbors are stacked as a horizontal slice $\matr{T}_{i,:,:}$ in the third-order tensor. That said, the first frontal slice $\matr{T}_{:,:,1}$ of said tensor is usually the matricized \gls{hsi} (where the pixels represent one mode).

Here, we add that a Patch-\gls{hsi} tensor has an inherently low-rank structure which is equal to that of the \gls{nmf} of the matricized \gls{hsi}, that is, the information contained along the third mode by each of the frontal slices of the tensor is almost essentially the same. In fact, the frontal slices are just \textit{spatially-shifted versions} of the original image $\matr{T}_{:,:,1}$, and the values of these shifts correspond to a \textit{small spatial kernel}, usually around $3\times3$ or $5\times5$.
However, what is different in the tensor case is that when this \textit{shifting} information is stacked along the third mode and \gls{cpd} is imposed with such a low value of $R$, the model automatically applies an implicit smoothing of the pixels that belong to the same patch (i.e., the same horizontal slice of $\tens{T}$). This is because the frontal slices are jointly factorized with the degrees of freedom of a single one of them, while also sharing the information of $\matr{A}$ and $\matr{B}$.

Therefore, a main advantage over \gls{nmf} is that one expects to extract the same sources with a patch-local smoothing of the \glspl{sv} of the estimated \glspl{em}, where the \glspl{sv} are balanced out in the form of scaling factors stored in the rows of $\matr{\Psi}$.
An important note here is that the scaling factors stored in $\matr{\Psi}$ may not have a significant physical meaning.

Now, what happens when $R$ increases? Since the information across the frontal slices are essentially the same (implying redundancy), the sources and abundances are expected to replicate, and we expect to observe slightly spatially-shifted versions of the abundance maps (i.e., in the columns of $\matr{A}$).
In this case, the scaling factors in $\matr{\Psi}$ only indicate whether an estimated \gls{em} in $\matr{B}$ corresponds to a certain spatial shifting or another.
This point is roughly illustrated in Fig. \ref{subfig:RegELMM_graph_patches} (inspired by Fig. \ref{fig:pavia_pat_NNCPD_R4_Sparse_ASC}) where we have three spectral sources: Gravel, Metal Sheets, and Trees, but \gls{cpd} is carried out with $R=4$. Here, the convex hull of $\matr{T}_{:,:,1}$ gives a high scaling factor at $\vect{b}_1$ and a low factor at $\vect{b}_4$, while that of $\matr{T}_{:,:,2}$ gives the opposite with almost the same quantity.
This is due to the fact that the materials are present with almost the same quantity in both frontal slices.
In other words, there may be a problem of redundancy if some components account for the same material with patches, which does happen in practice.

This problem does not occur when the third mode represents a physical meaning such as the case of \gls{mm}.

\subsubsection{Mathematical morphology}
\label{subsubsec:Unmixing_ELMM_MM}

While using patches is efficient, it still ignores the physical properties of connected pixels, and the \glspl{sv} are regularized indifferently among pixels belonging to different types of materials.
On the other hand, morphological features \cite{GhamDB14:tgrs, DallBCB11:ors} take into account physical properties such as scale and brightness of objects and promote dealing with \gls{sv} among pixels sharing these properties.
Constructing a \gls{hsi} tensor using \gls{mm} (coined as ``\gls{mm}-\gls{hsi} tensor'') is illustrated in Fig. \ref{fig:morpho_illustration}. In short, the matricized versions of the original \gls{hsi} and the results of its morphological transformations are stacked as the frontal slices of the
tensor\footnote{
	The details of constructing \gls{mm}-\gls{hsi} tensors are out of the scope of this paper, but can be found in Section 3 of \cite{JounDC20:mmta}.
}.

Through \gls{mm}, we emphasize the role of incorporating spatial diversities that add physical significance to the objects of the scene.
As such, one expects that a \gls{mm}-\gls{hsi} tensor has a more complicated structure than that of a Patch-\gls{hsi} tensor since its frontal slices contain additional context on the materials, such as their sizes and brightness levels.
As such, imposing a low $R$ promotes spectral smoothing of the \glspl{sv} based on a morphological regularization of the abundances, while imposing a sufficiently high $R$ promotes a distinctive spectral-morphological multi-feature separation of the materials, unlike Patch-\gls{hsi} tensors.

Since each frontal slice is seen as a characteristic of spatial scale (i.e., size of objects in the scene) and / or brightness (which is particularly relevant for \gls{elmm} because scaling factors can be directly linked to brightness), then the scaling factors represented by $\matr{\Psi}$ indicate the quantitative correspondence of an extracted material to the aforementioned physical properties per frontal slice.
This point is roughly illustrated in Fig. \ref{subfig:RegELMM_graph_MM} (inspired by Fig. \ref{fig:pavia_emp_NNCPD_R4_Sparse_ASC}) where $\matr{T}_{:,:,1}$ and $\matr{T}_{:,:,2}$ characterize small and large objects respectively.
Here, the convex hull of $\matr{T}_{:,:,1}$ gives high scaling factors for $\vect{b}_1$, $\vect{b}_2$, and $\vect{b}_3$ (corresponding to small objects), and a low scaling factor at $\vect{b}_4$ (corresponds to large objects), while that of $\matr{T}_{:,:,2}$ gives the opposite.

\section{Experiments and Results}
\label{sec:Unmixing_Experiments}

In this section, we discuss the experiments and results of \gls{multihutd} on real \glspl{hsi} in terms of \gls{aoadmm}-\gls{asc} (compared to Naive \gls{asc} \cite{VegaCFUDCC16:eusipco}), extracted factors, \gls{elmm} interpretability, and qualitative comparisons between Patch-\gls{hsi} and \gls{mm}-\gls{hsi} tensors with low and high values of the number of latent components.
In each experiment, among 30 random initializations of the factor matrices, the result with the minimum \gls{rmse} is chosen based on \eqref{eq:rmse_root_mean_squared_error}. The estimated \glspl{em} of $\matr{B}$ are identified based on their minimum \gls{sad}, in degrees, with respect to the reference \glspl{em} based on \eqref{eq:sad_spectral_angular_distance}.

\begin{subequations}
    \begin{align}
        \textrm{RMSE}\left(\tens{T}, \hat{\tens{T}}\right) = 
        \frac
        {\|\tens{T} - \hat{\tens{T}}\|_F^2}
        {\|\tens{T}\|_F^2}
        \label{eq:rmse_root_mean_squared_error}
        \\
        \textrm{SAD}(\vect{e}, \vect{b}) = \arccos \left(
        \frac
        {\vect{e} \cdot \vect{b}}
        {\|\vect{e}\|_2 \cdot \|\vect{b}\|_2}
        \right)
        \label{eq:sad_spectral_angular_distance}
    \end{align}
\end{subequations}

The maps and plots shown in the experiments represent the \textit{columns} of the factor matrices.
Above each \gls{am}, we show the material that corresponds to it with its minimum \gls{sad} value. We recall that a set of similarly indexed columns, e.g., $\{\matr{A}_1$, $\matr{B}_1$, $\matr{\Psi}_1\}$, represent the abundance, spectral source signal, and third-mode source pattern (e.g., morphological print, shifting print) of one extracted material respectively.
Since $\matr{\Psi}$ plays a crucial role in the interpretability of \gls{elmm} and \gls{multihutd}, we also highlight the relevance of its row components.
Each fixed index $k$ in the plot corresponds to a row of $\matr{\Psi}$ and thus to a frontal slice in $\tens{T}$, and the vertical grouping of points at said index, as indicated in Fig. \ref{subfig:pavia_pat_NNCPD_R4_Sparse_ASC_morpho_normalized}, 
\ref{subfig:pavia_emp_NNCPD_R4_Sparse_ASC_morpho_normalized}, 
\ref{subfig:pavia_pat_NNCPD_R8_Sparse_ASC_morpho_normalized}, 
and \ref{subfig:pavia_NNCPD_Sparse_ASC_morpho_normalized}, represents the scalars in that row. This also means that when projected on Fig. \ref{fig:graph_RegELMM_Patches_MM}, in Patch-\gls{cpd}, $k = 1$ contains the \gls{sv} scaling factors that balance the convex hull of the original \gls{hsi} and are responsible for its reconstruction, while in \gls{mm}-\gls{cpd}, the middle index does that. This will eventually show how \gls{mm} accounts to physical effects in the scene, while patches do not.

That said, we note that we carried part of the experiments using sparse \gls{nmf} with \gls{asc} \cite{YangZXDYZ10:tgrs} for the sake of qualitative comparison of the abundance maps and spectral sources obtained from the original \gls{hsi} (i.e., the \gls{hsi} matrix without additional filtering).
Due to the difference in the type of information contained between the matrix and tensor cases, and since \gls{nmf} does not apply in the framework of \gls{multihutd}, these results will serve only as a reference for the extracted components of \gls{cpd} as they do not serve the main aim and message of this work. For that reason, we include them in Appendix \ref{appendix:NMF_results} with further reasoning and explanation of the \gls{nmf} case analysis and its relevance to this work.

\begin{figure}[t]
	\centering
	\begin{subfigure}[b]{0.35\columnwidth}
		\centering
		\includegraphics[width=\textwidth]{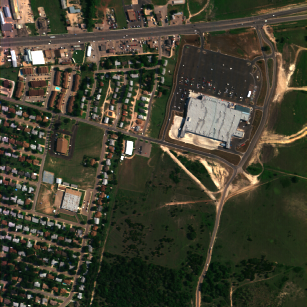}
		\caption{}
		\label{subfig:urban_hsi_color}
	\end{subfigure}
	\begin{subfigure}[b]{0.45\columnwidth}
		\centering
		\includegraphics[width=\textwidth]{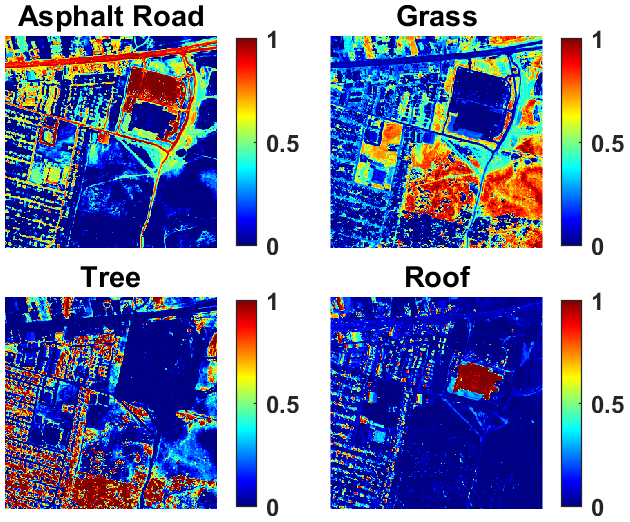}
		\caption{}
		\label{subfig:urban_gt_abundancemaps}
	\end{subfigure}
	
	\begin{subfigure}[b]{0.55\columnwidth}
		\centering
		\includegraphics[width=\textwidth]{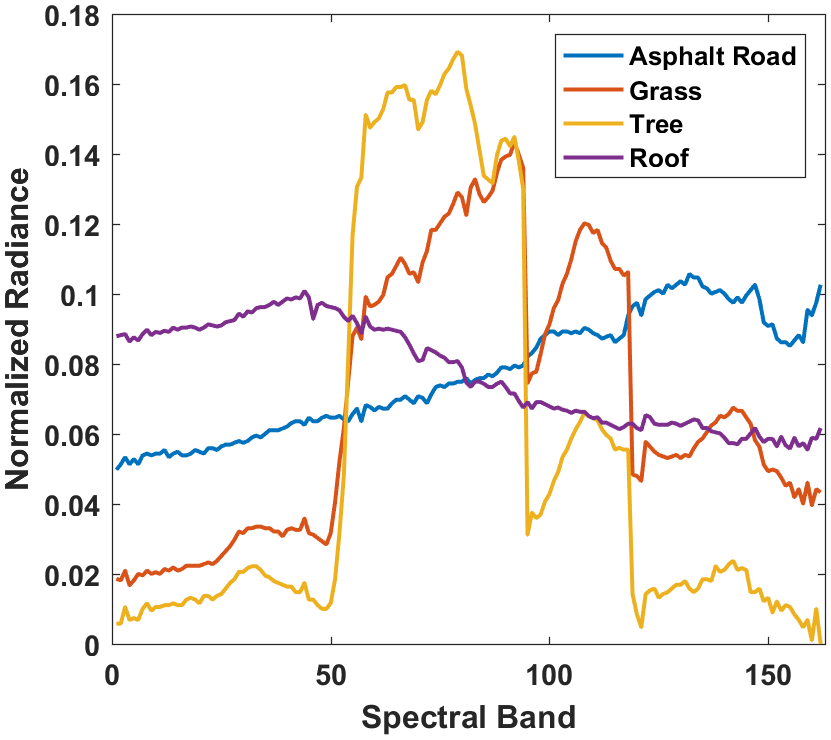}
		\caption{}
		\label{subfig:urban_spectralreference}
	\end{subfigure}
	
	\caption{
		\ref{subfig:urban_hsi_color}) Urban in false colors.
		\ref{subfig:urban_gt_abundancemaps}) Urban's spatial reference.
		\ref{subfig:urban_spectralreference}) Urban's spectral reference.
	}
	\label{fig:Urban_spectralreference}
\end{figure}

In each case analysis, we look into the components of $\matr{A}$ and $\matr{B}$ first, which visually and spectrally identify the materials, then we explain their correspondence to those of $\matr{\Psi}$, where we are interested in the significance of the third-mode patterns then their relevance to the original \gls{hsi}.
We note that quantitative validation of the \glspl{am} and \glspl{em} is usually not evident,
especially in the case of the Urban \gls{hsi} in Fig. \ref{subfig:urban_gt_abundancemaps} where the spatial \gls{gt} is not a real \gls{gt} but actually just a reference, and is not usable for quantitative comparison.
Moreover, there is neither a quantitative nor a qualitative reference for third-mode patterns in the literature, so highlight an in-depth qualitative analysis.

\begin{figure}[t]
	\centering
	\begin{subfigure}[b]{0.34\columnwidth}
		\centering
		\includegraphics[width=\textwidth]{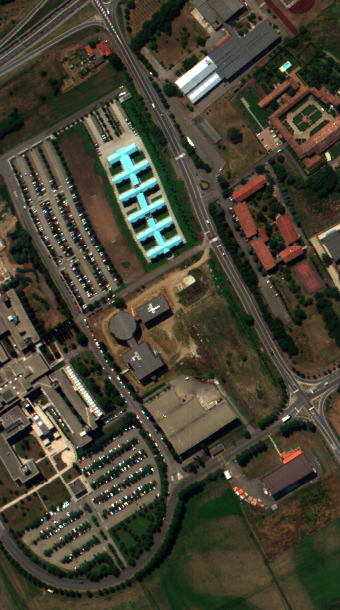}
		\caption{}
		\label{subfig:pavia_hsi_color}
	\end{subfigure}
	\begin{subfigure}[b]{0.30\columnwidth}
		\centering
		\includegraphics[width=\textwidth]{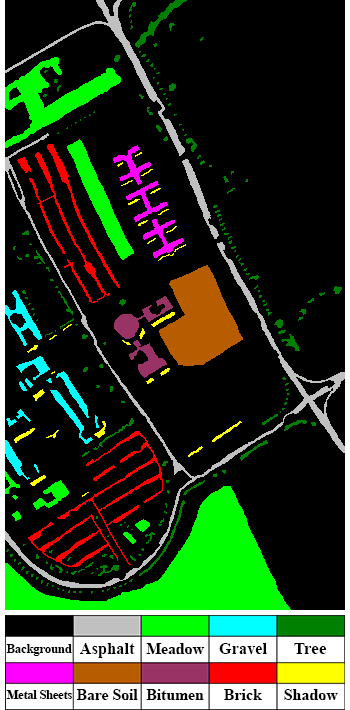}
		\caption{}
		\label{subfig:pavia_gt_classes}
	\end{subfigure}
	
	\begin{subfigure}[b]{0.55\columnwidth}
		\centering
		\includegraphics[width=\textwidth]{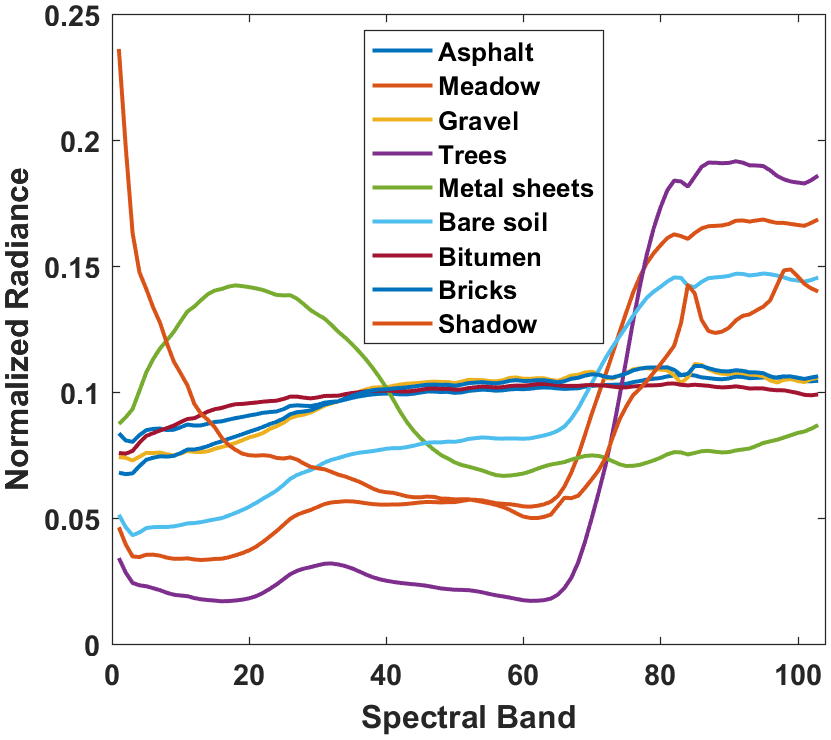}
		\caption{}
		\label{subfig:pavia_spectralreference}
	\end{subfigure}
	
	\caption{
		\ref{subfig:pavia_hsi_color}) Pavia in false colors.
		\ref{subfig:pavia_gt_classes}) Pavia's spatial reference. 
		\ref{subfig:pavia_spectralreference}) Pavia's spectral reference extracted by averaging each class of the spatial reference.
	}
	\label{fig:Pavia_spectralreference}
	\vspace{-2mm}
\end{figure}

We want to consider \glspl{hsi} which show objects with spatial features of different scale and brightness levels, for which urban areas are good candidates. For that, we choose two real \glspl{hsi}\footnote{
    The data sets with detailed information are available on the website:
    \\
    \texttt{http://lesun.weebly.com/hyperspectral-data-set.html}
}:
\textit{Pavia University} and \textit{Urban},
shown in \figurename~\ref{fig:Pavia_spectralreference} and \ref{fig:Urban_spectralreference} respectively in false colors with their spatial and spectral references.
It is worth noting for the sake of interpretation that the \gls{mm}-\gls{hsi} tensors are built following the \gls{emp} technique used in \cite{JounDC20:mmta} with \glspl{obr} and \glspl{cbr}, which correspond to bright and dark objects respectively, with varying sizes of the \gls{se}, which in turns corresponds to the scales of objects.
In the following experiments, $4$ sizes of the \gls{se} are used, corresponding to $8$ \gls{obr} and \gls{cbr} transformations, then the dimension of the third mode is $K=9$ where the original image (corresponding to scale $0$) is placed in the middle as roughly shown in Fig. \ref{fig:morpho_illustration}.
The Patch-\gls{hsi} tensors are built following \cite{VegaCFUDCC16:eusipco} as shown in Fig. \ref{fig:patches_illustration} with $3\times 3$ patches.
This means that the dimension of the third mode is $K=3 \times 3=9$ where the original image is placed at $k=1$.
In addition to the real \glspl{hsi}, we demonstrate the performance of the proposed framework under different levels of noise through a synthetic \gls{hsi} tensor in Appendix \ref{appendix:synthetic_hsi}.
We run our experiments with Intel® Core™ i7-1185G7, $32$GB RAM $3200$MHz LPDDR4.

\subsection{Results Discussion - Pavia University}
\label{subsec:Unmixing_Experiments_Pavia}

In this section, we present the experiments of the \gls{hsi} of Pavia,
but first, we note that the reference for this dataset is originally composed of training and testing sets, where each pixel is manually associated with one of 9 labeled classes as seen in \figurename~\ref{subfig:pavia_gt_classes}.
The spectral reference in Fig. \ref{subfig:pavia_spectralreference} is extracted by averaging the spectral signatures of each subset of pixels belonging to one class.
In Fig. \ref{subfig:pavia_spectralreference}, some classes have very similar spectral signatures, so, in the following, sometimes we refer to \textit{Trees} and \textit{Meadows} as \textit{vegetation}, and to \textit{Asphalt}, \textit{Bitumen}, \textit{Gravel}, and \textit{Bricks} as \textit{roads} or \textit{roofs}, while \textit{Bare Soil} may belong to either of both groups.

We start by comparing \gls{aoadmm}-\gls{asc} and Naive \gls{asc}. After that, we focus on \gls{cpd} and the \gls{elmm} analysis of the factors while interpreting the cases of patches and \gls{mm}.
For the \gls{mm}-\gls{hsi} tensor, our \glspl{se} are disks with the successive radii: $\{2,7,12,17\}$ pixels.
Both Patch- and \gls{mm}-\gls{hsi} tensors then have $K = 9$ frontal slices and dimensions $207400\times 103\times 9$.
Finally, we find that $R = 4$ and $R = 8$ are the best for low and high values of the number of latent components respectively.

\subsubsection{\glsentryshort{aoadmm}-\glsentryshort{asc}}

Here, we compare the \gls{rmse} results of \gls{mm}-\gls{cpd} between \gls{aoadmm}-\gls{asc} and Naive \gls{asc} \cite{VegaCFUDCC16:eusipco}. The results are shown in Table \ref{tab:pavia_Algo}, where we see that with \gls{aoadmm}-\gls{asc} we gain in \gls{rmse}, which corresponds to a better estimation of the factors with respect to the observed tensor with a small difference in the execution time.
\begin{table}[ht]
\begin{center}\begin{tabular}{c"c|c|c}
Algorithm & $R$ & \gls{rmse} $\%$ & Time (s) \\
\thickhline
Naive \gls{asc} \cite{VegaCFUDCC16:eusipco}  & 8 & 7.07 & \underline{231} \\
\gls{aoadmm}-\gls{asc} & 8 & \underline{6.34} & 384 \\
\end{tabular}
\caption{Pavia. The results of \gls{aoadmm}-\gls{asc} and Naive \gls{asc} in terms of \gls{rmse} and execution time: $R$ indicates the number of latent components. The results of the minimum \gls{rmse} are shown.} \label{tab:pavia_Algo}
\end{center}
\end{table}

\begin{figure*}[!t]
    \centering
    
    \begin{minipage}[b]{0.46\textwidth}
        \centering
        \begin{subfigure}[b]{\textwidth}
            \includegraphics[width=\textwidth]{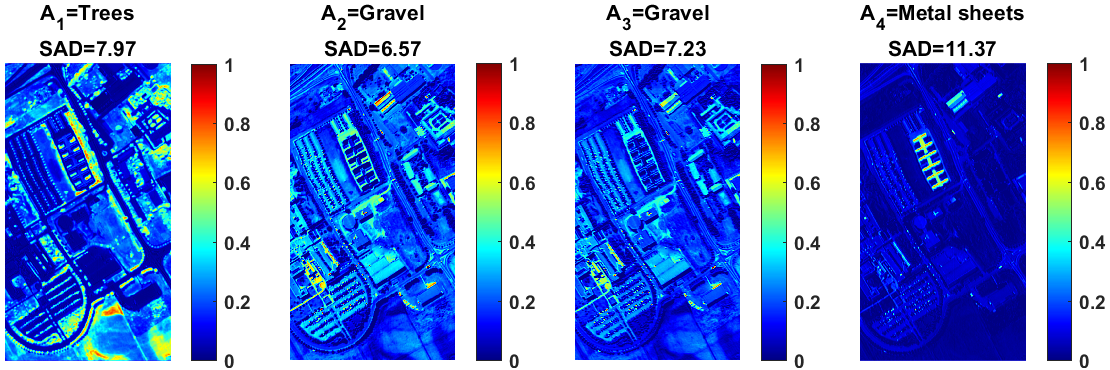}
            \caption{Components of $\matr{A}$}
            \label{subfig:pavia_pat_NNCPD_R4_Sparse_ASC_spatial_ensemble}
        \end{subfigure}
        
        \begin{subfigure}[b]{0.46\textwidth}
            \includegraphics[width=\textwidth]{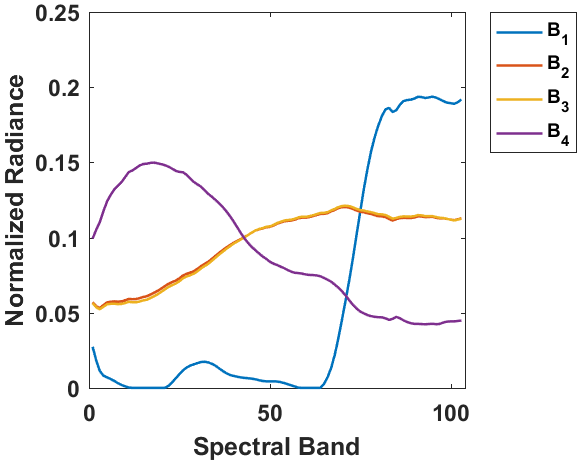}
            \caption{Components of $\matr{B}$}
            \label{subfig:pavia_pat_NNCPD_R4_Sparse_ASC_spectra_normalized}
        \end{subfigure}
        ~
        \begin{subfigure}[b]{0.46\textwidth}
            \includegraphics[width=\textwidth]{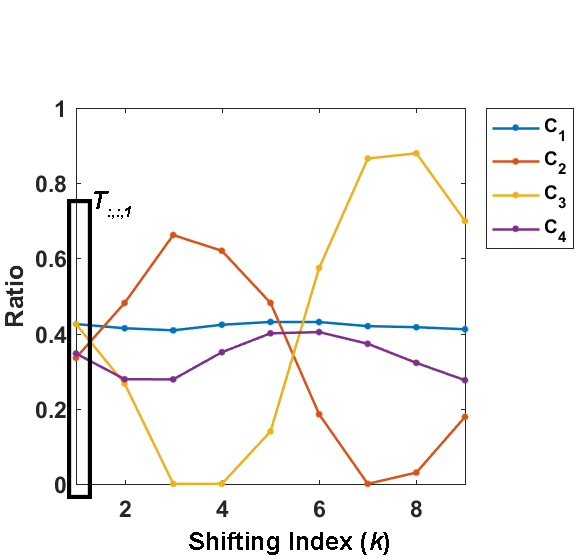}
            \caption{Components of $\matr{\Psi}$}
            \label{subfig:pavia_pat_NNCPD_R4_Sparse_ASC_morpho_normalized}
        \end{subfigure}
        \caption{Pavia. \gls{cpd} results of the Patch-tensor for $R$=$4$}
        \label{fig:pavia_pat_NNCPD_R4_Sparse_ASC}
    \end{minipage}
    ~~~~~~
    \begin{minipage}[b]{0.46\textwidth}
        \centering
        \begin{subfigure}[b]{\textwidth}
            \includegraphics[width=\textwidth]{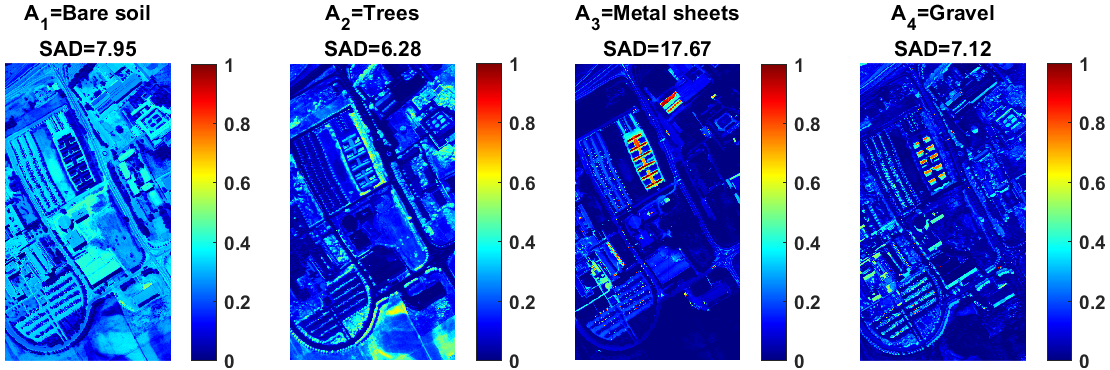}
            \caption{Components of $\matr{A}$}
            \label{subfig:pavia_emp_NNCPD_R4_Sparse_ASC_spatial_ensemble}
        \end{subfigure}
    
        \begin{subfigure}[b]{0.46\textwidth}
            \includegraphics[width=\textwidth]{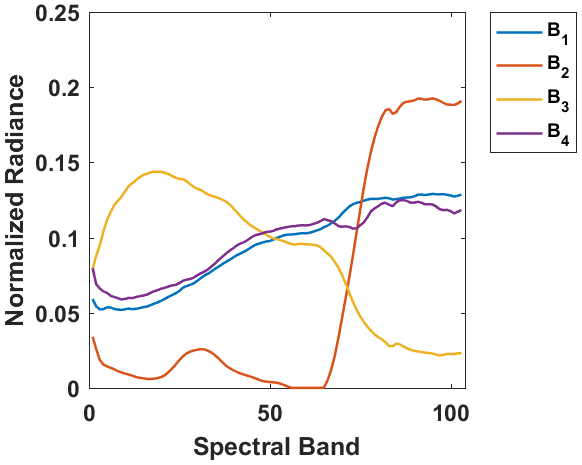}
            \caption{Components of $\matr{B}$}
            \label{subfig:pavia_emp_NNCPD_R4_Sparse_ASC_spectra_normalized}
        \end{subfigure}
        ~
        \begin{subfigure}[b]{0.46\textwidth}
            \includegraphics[width=\textwidth]{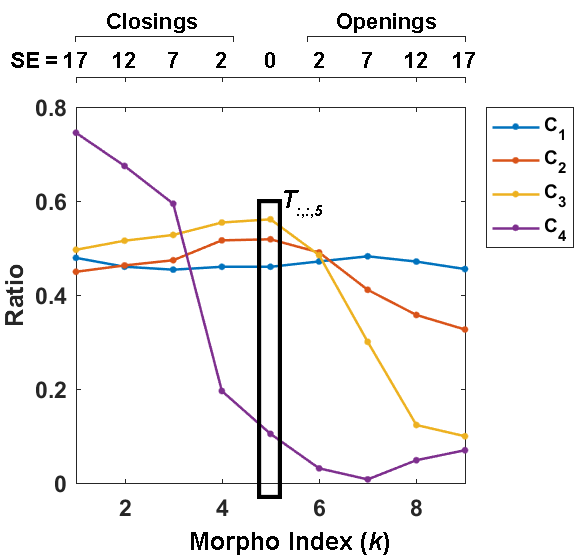}
            \caption{Components of $\matr{\Psi}$}
            \label{subfig:pavia_emp_NNCPD_R4_Sparse_ASC_morpho_normalized}
        \end{subfigure}
        \caption{Pavia. \gls{cpd} results of the \gls{mm}-tensor for $R$=$4$}
        \label{fig:pavia_emp_NNCPD_R4_Sparse_ASC}
    \end{minipage}
    
    \vspace{1mm}
    
    \begin{minipage}[b]{0.46\textwidth}
        \centering
        \begin{subfigure}[b]{\textwidth}
            \includegraphics[width=\textwidth]{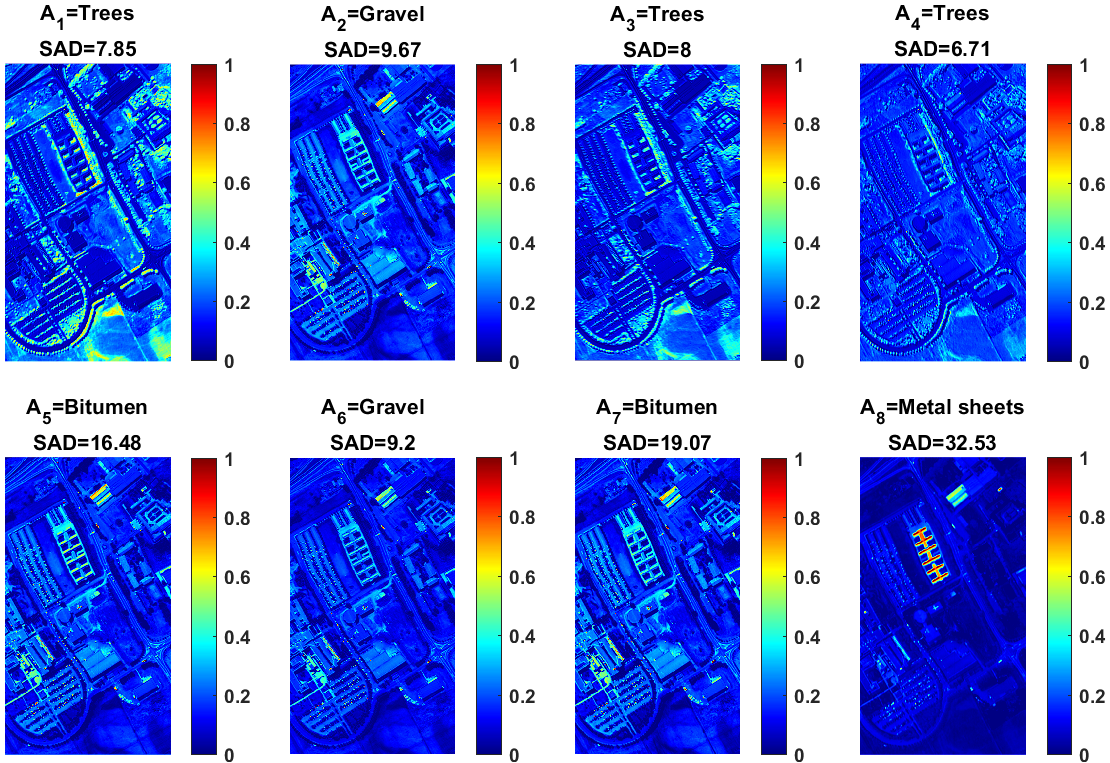}
            \caption{Components of $\matr{A}$}
            \label{subfig:pavia_pat_NNCPD_R8_Sparse_ASC_spatial_ensemble}
        \end{subfigure}
    
        \begin{subfigure}[b]{0.46\textwidth}
            \includegraphics[width=\textwidth]{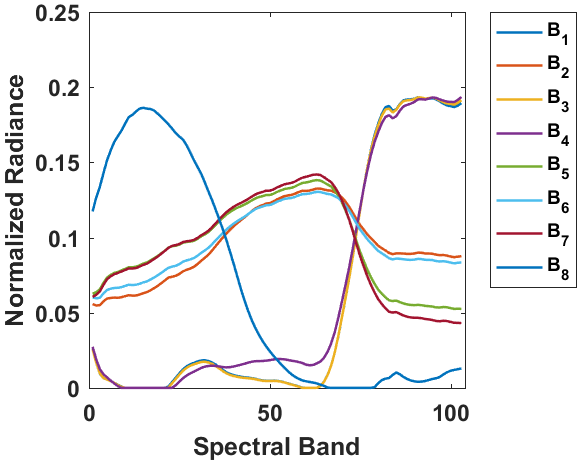}
            \caption{Components of $\matr{B}$}
            \label{subfig:pavia_pat_NNCPD_R8_Sparse_ASC_spectra_normalized}
        \end{subfigure}
        ~
        \begin{subfigure}[b]{0.46\textwidth}
            \includegraphics[width=\textwidth]{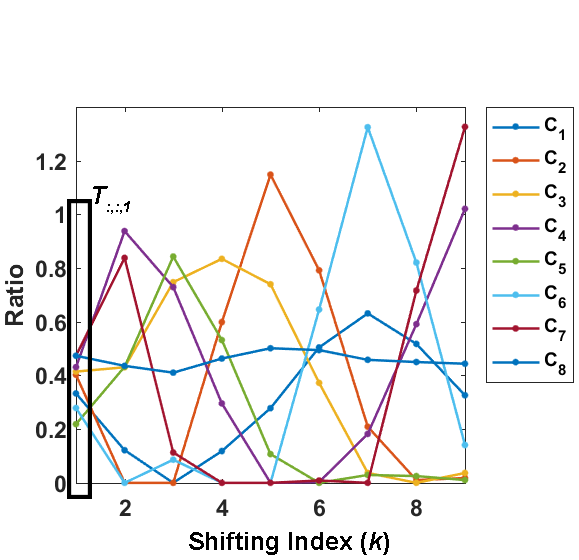}
            \caption{Components of $\matr{\Psi}$}
            \label{subfig:pavia_pat_NNCPD_R8_Sparse_ASC_morpho_normalized}
        \end{subfigure}
        \caption{Pavia. \gls{cpd} results of the Patch-tensor for $R$=$8$}
        \label{fig:pavia_pat_NNCPD_R8_Sparse_ASC}
    \end{minipage}
    ~~~~~~
    \begin{minipage}[b]{0.46\textwidth}
        \centering
        \begin{subfigure}[b]{\textwidth}
            \includegraphics[width=\textwidth]{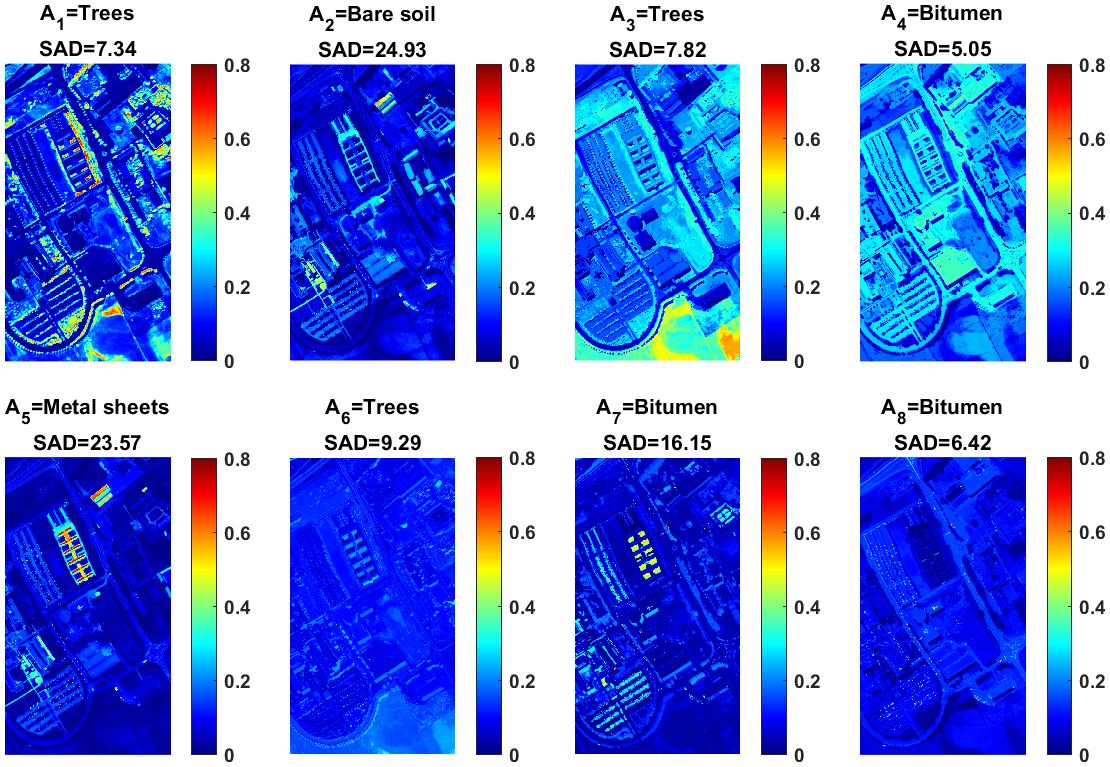}
            \caption{Components of $\matr{A}$}
            \label{subfig:pavia_NNCPD_Sparse_ASC_spatial_ensemble}
        \end{subfigure}
        
        \begin{subfigure}[b]{0.46\textwidth}
            \includegraphics[width=\textwidth]{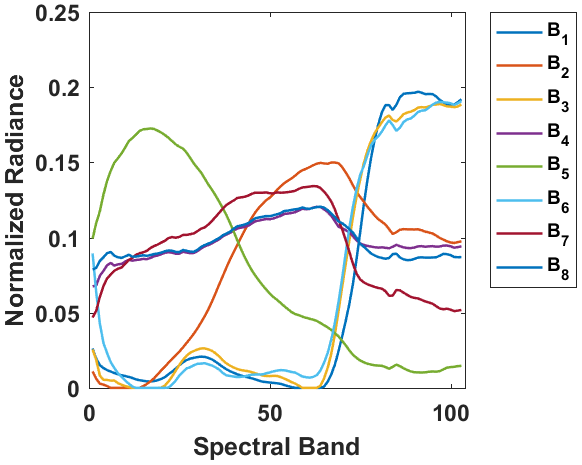}
            \caption{Components of $\matr{B}$}
            \label{subfig:pavia_NNCPD_Sparse_ASC_spectra_normalized}
        \end{subfigure}
        ~
        \begin{subfigure}[b]{0.46\textwidth}
            \includegraphics[width=\textwidth]{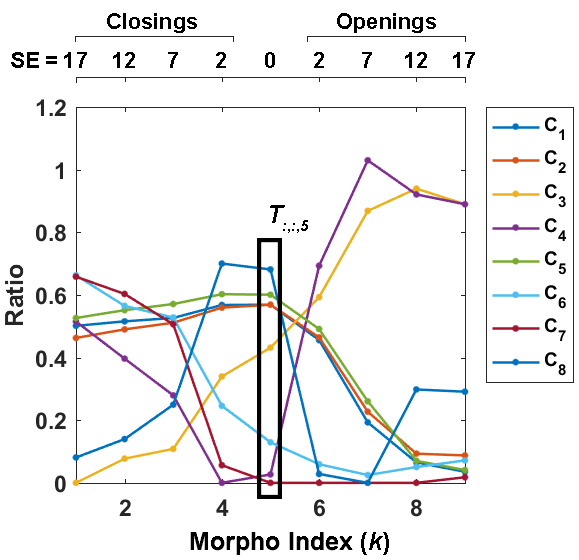}
            \caption{Components of $\matr{\Psi}$}
            \label{subfig:pavia_NNCPD_Sparse_ASC_morpho_normalized}
        \end{subfigure}
        \caption{Pavia. \gls{cpd} results of the \gls{mm}-tensor for $R$=$8$}
        \label{fig:pavia_NNCPD_noSparse_ASC}
    \end{minipage}
\end{figure*}

\subsubsection{Few latent components, \glsentryshort{elmm} and \glsentryshort{sv}}
\label{subsubsec:Unmixing_Experiments_Pavia_ELMM}

Here, we are interested in the property of \gls{cpd} dealing with \gls{sv}.
Since Patch-\gls{hsi} tensors have as an inherently low structure as that of \gls{nmf}, we start by considering $R = 4$ for patches and \gls{mm}. As explained in Section \ref{subsec:elmminterpretation}, such a low value highlights the \gls{sv} and \textit{spatial regularization} aspects of \gls{multihutd}.
We compile the discussion into two stages:
(a) Fig. \ref{fig:pavia_pat_NNCPD_R4_Sparse_ASC} representing Patch-\gls{cpd}, and
(b) Fig. \ref{fig:pavia_emp_NNCPD_R4_Sparse_ASC} representing \gls{mm}-\gls{cpd}.

\textit{2a)}
In Fig. \ref{fig:pavia_pat_NNCPD_R4_Sparse_ASC}, $\matr{B}_1$ and $\matr{B}_4$ are identified as \textit{Trees} and \textit{Metal Sheets} respectively, which reflects the areas highlighted in $\matr{A}_1$ and $\matr{A}_4$, while $\matr{B}_2$ and $\matr{B}_3$ can be seen as a bundle identified as \textit{Gravel}, which reflects the areas highlighted in $\matr{A}_2$ and $\matr{A}_3$ and where we already start to see \textit{replicated} components.
In fact, $\matr{A}_2$ and $\matr{A}_3$ are \textit{slightly-shifted} versions of each other, which is explained better in the following interpretation of $\matr{\Psi}_2$ and $\matr{\Psi}_3$.

Looking at Fig. \ref{subfig:pavia_pat_NNCPD_R4_Sparse_ASC_morpho_normalized}, which is the most interesting, one might intuitively expect to see horizontal curves since, quantitatively, the collective \gls{sv} is supposedly \textit{constant} in patches. However, while $\matr{\Psi}_1$ and $\matr{\Psi}_4$ look almost straight, $\matr{\Psi}_2$ and $\matr{\Psi}_3$ are not.
We notice here that where $\matr{\Psi}_2$ is high, $\matr{\Psi}_3$ is low and vice versa.
In part, this means that $\matr{A}_2$ represents the shifts where $k = \{2,3,4,5\}$, while $\matr{A}_3$ represents those where $k = \{6,7,8,9\}$.
In another part, the two columns fluctuate in a way that \textit{maintains} a constant \gls{sv} and \textit{balances out} their quantitative presence across the frontal slices.
As for $k = 1$, which represents the original \gls{hsi}, we notice that the scaling factors are almost equal, which means that the spectral vectors of the convex hull are \textit{equally present} in the \gls{hsi}, all of which shows that Patch-\gls{hsi} tensors do not account to physical spatial effects.

\textit{2b)}
In Fig. \ref{fig:pavia_emp_NNCPD_R4_Sparse_ASC}, $\matr{B}_1$, $\matr{B}_2$, and $\matr{B}_3$ are identified as \textit{Bare Soil}, \textit{Trees}, and \textit{Metal Sheets} respectively, which reflects the areas highlighted in $\matr{A}_1$, $\matr{A}_2$, and $\matr{A}_3$, all of which is similar to those obtained by patches.
As for $\{\matr{A}_4,\matr{B}_4\}$, while $\matr{B}_4$ and $\matr{B}_1$ can be seen as a spectral bundle, unlike patches, we notice that $\matr{A}_4$ highlights interesting \textit{shadow} areas (i.e., dark features), which clearly reflects the \textit{morphological awareness} incorporated into \gls{cpd} with \gls{mm}. The latter becomes more interesting with the following interpretation of $\matr{\Psi}$.

Looking at Fig. \ref{subfig:pavia_emp_NNCPD_R4_Sparse_ASC_morpho_normalized}, we observe three main patterns that can be associated to the chosen morphological parameters. First, $\matr{\Psi}_4$ corresponds to dark features (reflected by the shadows in $\matr{A}_4$) as it has higher values when $k$ corresponds to \gls{cbr}, then continues decreasing towards \gls{obr}. Second, $\matr{\Psi}_2$ and $\matr{\Psi}_3$ correspond to small features as they have higher values around the middle ($k = 5$) where the \glspl{se} are small, which is visually reflected through the small objects highlighted in $\matr{A}_2$ (trees) and $\matr{A}_3$ (metal sheets and vehicles). Third, $\matr{\Psi}_1$ is rather steady, which means that the spatial features shown in $\matr{A}_1$ are general.

As for $k = 5$, which represents the original \gls{hsi}, we notice that $\matr{\Psi}_2$ and $\matr{\Psi}_3$ have the highest scaling factors since they correspond to relatively bright objects of the scene, $\matr{\Psi}_1$ has a slightly lower factor since it corresponds to darker objects like asphalt roads, building roofs, parking lots, and bare soil areas, and $\matr{\Psi}_4$ has the lowest factors since it corresponds to dark shadows.
These relationships show the column- and row-wise significance of $\matr{\Psi}$ and how \gls{multihutd} can balance out the \glspl{sv} and simultaneously reconstruct the original \gls{hsi}.

\subsubsection{More latent components, \glsentryshort{elmm} and multi-feature separability}
\label{subsubsec:Unmixing_Experiments_Pavia_NCPD}

Here, we are interested in \gls{multihutd} when we have more degrees of freedom, where we dive deeper into the factors of patches and \gls{mm} for $R = 8$.
As explained in Section \ref{subsec:elmminterpretation}, such a higher value demonstrates the multi-feature separability of \gls{mm} and how patches only replicates its components.
We compile the discussion into two stages:
(a) Fig. \ref{fig:pavia_pat_NNCPD_R8_Sparse_ASC} representing Patch-\gls{cpd}, and
(b) Fig. \ref{fig:pavia_NNCPD_noSparse_ASC} representing \gls{mm}-\gls{cpd}.

\textit{3a)}
In Fig. \ref{fig:pavia_pat_NNCPD_R8_Sparse_ASC}, we end up with more replicas of the same $\matr{A}$ and $\matr{B}$ components obtained in Fig. \ref{fig:pavia_pat_NNCPD_R4_Sparse_ASC}, the bundles being associated to the column indices
$\{1,3,4\}$ detected as \textit{Trees},
and
$\{2,5,6,7\}$ detected as \textit{Gravel} and \textit{Bitumen}.
As for the plot of $\matr{\Psi}$, the same remarks of Fig. \ref{subfig:pavia_pat_NNCPD_R4_Sparse_ASC_morpho_normalized} about balancing the constant \glspl{sv} in patches apply on Fig. \ref{subfig:pavia_pat_NNCPD_R8_Sparse_ASC_morpho_normalized}, but since there are many replicas, the figure becomes hard to read.
Finally, we notice again that for $k = 1$, the scaling factors are almost equal.

\textit{3b)}
In Fig. \ref{fig:pavia_NNCPD_noSparse_ASC}, we notice that \gls{multihutd} is done based on spectral and morphological properties, where we observe three bundles: 
$\{1,3,6\}$,
$\{2,4,7,8\}$, and
$\{5\}$.
$\matr{B}_1$, $\matr{B}_3$, and $\matr{B}_6$ are identified as vegetation, which respectively reflects the areas highlighted in $\matr{A}_1$ (small vegetation areas like trees), $\matr{A}_3$ (big vegetation areas like meadow), and $\matr{A}_6$ (dark shadows on vegetation areas).
Therefore, while $\matr{B}_1$, $\matr{B}_3$, and $\matr{B}_6$ form a bundle, unlike patches, the corresponding \glspl{am} highlight interesting features accounting to the scale and brightness of vegetation objects, which we discuss in more depth with the interpretation of $\matr{\Psi}$ and which applies to the other components as well.
$\matr{B}_2$, $\matr{B}_4$, $\matr{B}_7$, and $\matr{B}_8$ are identified as \textit{Bitumen} and \textit{Gravel}, which respectively reflects the areas highlighted in $\matr{A}_2$ (small or short areas of roads and roofs), $\matr{A}_3$ (big connected areas of roads, roofs, and parking lots), $\matr{A}_7$ (dark shadows on parking lots and buildings), and $\matr{A}_8$ (tiny bright vehicles).
Finally, $\matr{B}_4$ is identified as \textit{Metal Sheets}, which are small.

\begin{figure*}[t]
    \centering
    
    \begin{minipage}[b]{0.46\textwidth}
        \centering
        \begin{subfigure}[b]{\textwidth}
            \includegraphics[width=\textwidth]{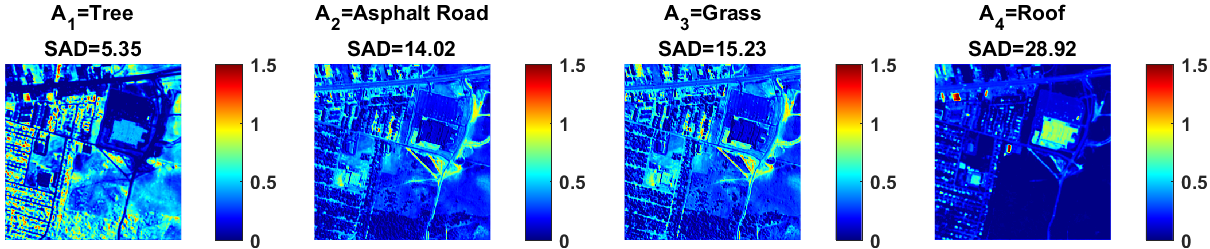}
            \caption{Components of $\matr{A}$}
            \label{subfig:urban_pat_NNCPD_R4_Sparse_ASC_spatial_ensemble}
        \end{subfigure}
        
        \begin{subfigure}[b]{0.46\textwidth}
            \includegraphics[width=\textwidth]{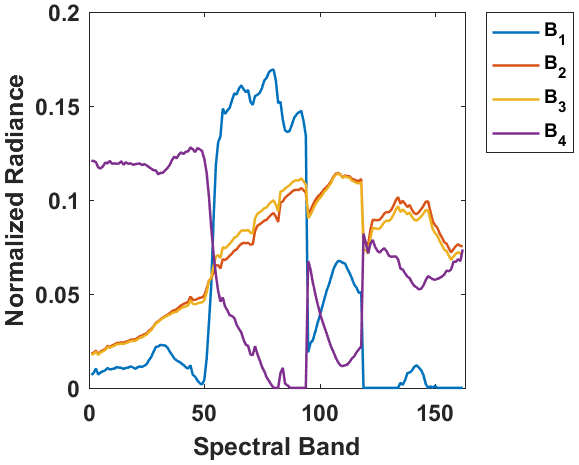}
            \caption{Components of $\matr{B}$}
            \label{subfig:urban_pat_NNCPD_R4_Sparse_ASC_spectra_normalized}
        \end{subfigure}
        ~
        \begin{subfigure}[b]{0.46\textwidth}
            \includegraphics[width=\textwidth]{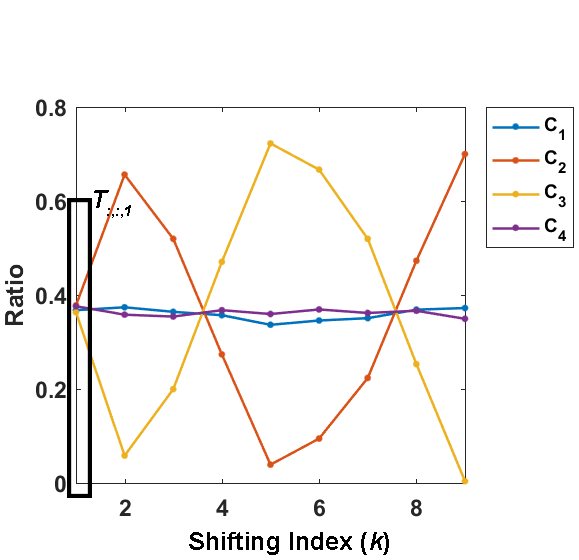}
            \caption{Components of $\matr{\Psi}$}
            \label{subfig:urban_pat_NNCPD_R4_Sparse_ASC_morpho_normalized}
        \end{subfigure}
        \caption{Urban. \gls{cpd} results of the Patch-tensor for $R$=$4$}
        \label{fig:urban_pat_NNCPD_R4_Sparse_ASC}
    \end{minipage}
    ~~~~~~
    \begin{minipage}[b]{0.46\textwidth}
        \centering
        \begin{subfigure}[b]{\textwidth}
            \includegraphics[width=\textwidth]{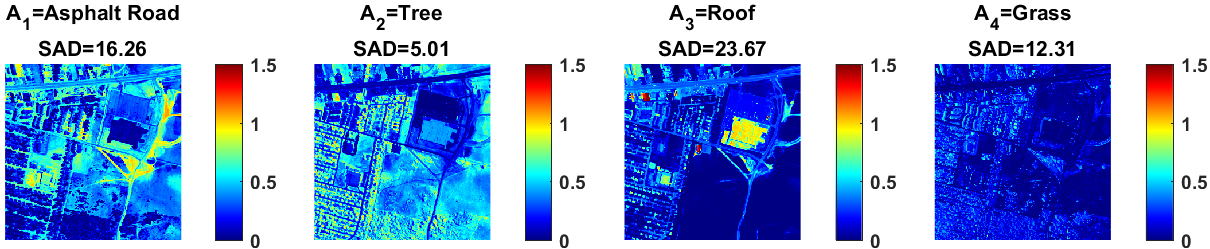}
            \caption{Components of $\matr{A}$}
            \label{subfig:urban_emp_NNCPD_R4_Sparse_ASC_spatial_ensemble}
        \end{subfigure}

        \begin{subfigure}[b]{0.46\textwidth}
            \includegraphics[width=\textwidth]{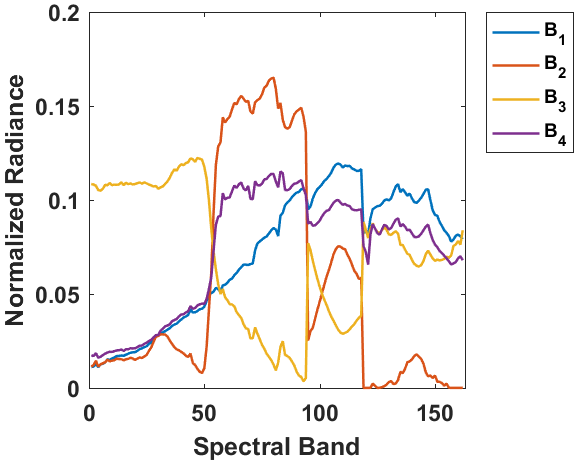}
            \caption{Components of $\matr{B}$}
            \label{subfig:urban_emp_NNCPD_R4_Sparse_ASC_spectra_normalized}
        \end{subfigure}
        ~
        \begin{subfigure}[b]{0.46\textwidth}
            \includegraphics[width=\textwidth]{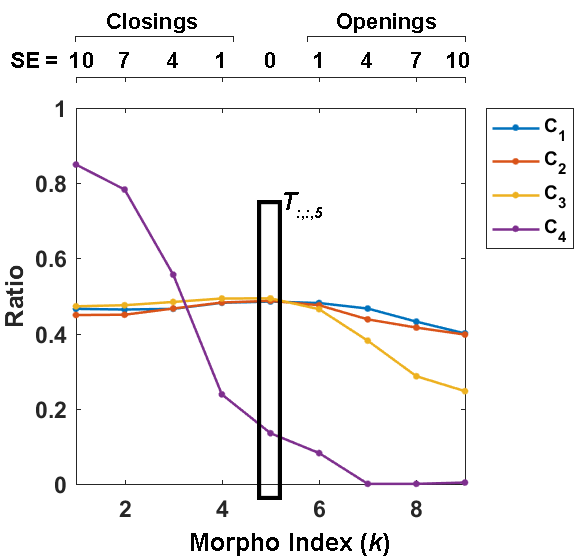}
            \caption{Components of $\matr{\Psi}$}
            \label{subfig:urban_emp_NNCPD_R4_Sparse_ASC_morpho_normalized}
        \end{subfigure}
        \caption{Urban. \gls{cpd} results of the \gls{mm}-tensor for $R$=$4$}
        \label{fig:urban_emp_NNCPD_R4_Sparse_ASC}
    \end{minipage}
\end{figure*}

Looking at Fig. \ref{subfig:pavia_NNCPD_Sparse_ASC_morpho_normalized}, we observe four main patterns that can be associated to the chosen morphological parameters:
First, $\matr{\Psi}_6$ and $\matr{\Psi}_7$ correspond to dark features (as observed in $\matr{A}_6$ and $\matr{A}_7$) as they have higher values when $k$ corresponds to \gls{cbr}, then continue decreasing towards \gls{obr}.
Second, $\matr{\Psi}_1$, $\matr{\Psi}_2$, and $\matr{\Psi}_5$ correspond to small features (as observed in $\matr{A}_1$, $\matr{A}_2$, and $\matr{A}_4$) as they have higher values around $k = 5$ where the \glspl{se} are small.
Third, $\matr{\Psi}_3$ and $\matr{\Psi}_4$ correspond to big features (as observed in $\matr{A}_3$ and $\matr{A}_4$) as they have higher values when $k$ corresponds to big \glspl{se} with \gls{obr}.
Fourth, $\matr{\Psi}_8$ corresponds to the tiny vehicles as it is the highest when $k$ corresponds to the smallest \gls{se}.
Finally, we talk about the original \gls{hsi} in the following (i.e., $\matr{T}_{:, :, 5}$).

As for $k = 5$, we notice that $\matr{\Psi}_1$, $\matr{\Psi}_2$, $\matr{\Psi}_5$, and $\matr{\Psi}_8$ have the highest scaling factors (relatively bright objects including the vehicles), $\matr{\Psi}_3$ and $\matr{\Psi}_4$ have lower factors (darker objects like asphalt roads, building roofs, parking lots, bare soil, and meadow areas), and $\matr{\Psi}_6$ and $\matr{\Psi}_7$ have the lowest factors (dark shadowy features).
These relationships showcase the separability of \gls{multihutd} when the third-mode has a significant physical meaning and when the number of latent components is set to be sufficiently high, which can also be interpreted in terms of \gls{elmm} and balancing the \gls{sv} factors.

\subsection{Results - Urban}
\label{subsec:Unmixing_Experiments_Urban}

In this section, we present the experiments of the Urban \gls{hsi} following the same order of Pavia.
Since we have the same observations, and in order to avoid repetition, we briefly go over the results.
But first, we note that the spatial and spectral references already come with the downloaded dataset, consisting of four endmembers, which we use as a spectral reference (shown in figure \ref{subfig:urban_spectralreference}) in the experiments, and four abundance maps (shown in figure \ref{subfig:urban_gt_abundancemaps}).
For the \gls{mm}-\gls{hsi} tensor, our \glspl{se} are disks with the successive radii: $\{1,4,7,10\}$ pixels.
Both Patch-\gls{hsi} and \gls{mm}-\gls{hsi} tensors then have $K = 9$ frontal slices and dimensions $94249\times 162\times 9$.
Finally, we also choose $R = 4$ and $R = 8$ for the number of latent components. 

\subsubsection{\glsentryshort{aoadmm}-\glsentryshort{asc}}

Table \ref{tab:urban_Algo} shows the \gls{rmse} results of \gls{mm}-\gls{cpd} between \gls{aoadmm}-\gls{asc} and Naive \gls{asc} \cite{VegaCFUDCC16:eusipco}, where again with \gls{aoadmm}-\gls{asc} we gain in \gls{rmse} with a small difference in the execution time.
\begin{table}[ht]
\begin{center}\begin{tabular}{c"c|c|c}
Algorithm & $R$ & \gls{rmse} $\%$ & Time (s) \\
\thickhline
Naive \gls{asc} \cite{VegaCFUDCC16:eusipco}  & $8$ & $7.88$ & \underline{$124$} \\
\gls{aoadmm}-\gls{asc} & $8$ & \underline{$6.87$} & $251$ \\
\end{tabular}
\caption{Urban. The results of \gls{aoadmm}-\gls{asc} and Naive \gls{asc} in terms of \gls{rmse} and execution time: $R$ is the number of latent components. The results of the minimum \gls{rmse} are shown.} \label{tab:urban_Algo}
\end{center}
\end{table}

\subsubsection{Few latent components, \glsentryshort{elmm} and \glsentryshort{sv}}
\label{subsubsec:Unmixing_Experiments_Urban_ELMM}

Here, we discuss the results for $R = 4$, where
(a)
Fig. \ref{fig:urban_pat_NNCPD_R4_Sparse_ASC} represents Patch-\gls{cpd} 
(b)
Fig. \ref{fig:urban_emp_NNCPD_R4_Sparse_ASC} represents
\gls{mm}-\gls{cpd}.
A fast look at the figures shows that we have the same observations as those of Pavia:

\textit{2a)}
In Fig. \ref{fig:urban_pat_NNCPD_R4_Sparse_ASC}, $\matr{B}_2$ and $\matr{B}_3$ form a bundle, and $\matr{A}_2$ and $\matr{A}_3$ are \textit{replicas} and represent \textit{Asphalt+Grass}. Moreover, in Fig. \ref{subfig:urban_pat_NNCPD_R4_Sparse_ASC_morpho_normalized}, we see the same patterns and fluctuations that were observed in Fig. \ref{subfig:pavia_pat_NNCPD_R4_Sparse_ASC_morpho_normalized} related to the constant \gls{sv} and its quantitative balance in patches, and the scaling factors are equal for $k = 1$.
On the other hand, $\{\matr{A}_1, \matr{B}_1\}$ and $\{\matr{A}_4, \matr{B}_4\}$ represent \textit{Tree+Grass} and \textit{Roof} respectively with steady $\matr{\Psi}_1$ and $\matr{\Psi}_4$ patterns.

\textit{2b)}
In Fig. \ref{fig:urban_emp_NNCPD_R4_Sparse_ASC}, while $\matr{B}_4$ and $\matr{B}_1$ form a bundle, we notice that $\matr{\Psi}_4$ has the same pattern observed in Fig. \ref{subfig:pavia_emp_NNCPD_R4_Sparse_ASC_morpho_normalized}, which corresponds to dark shadows and is reflected in $\matr{A}_4$, which highlights shadows of buildings and trees that fall on grass areas.
As for the other components, they can be interpreted similarly to those in the case of Pavia (including for $k = 5$), where $\{\matr{A}_1, \matr{B}_1, \matr{\Psi}_1\}$, $\{\matr{A}_2, \matr{B}_2, \matr{\Psi}_2\}$, and $\{\matr{A}_3, \matr{B}_3, \matr{\Psi}_3\}$ represent \textit{Asphalt+Grass}, \textit{Tree+Grass}, and \textit{Roof} respectively.

\begin{figure}[t]
    \centering
        \begin{subfigure}[b]{0.46\textwidth}
            \includegraphics[width=\textwidth]{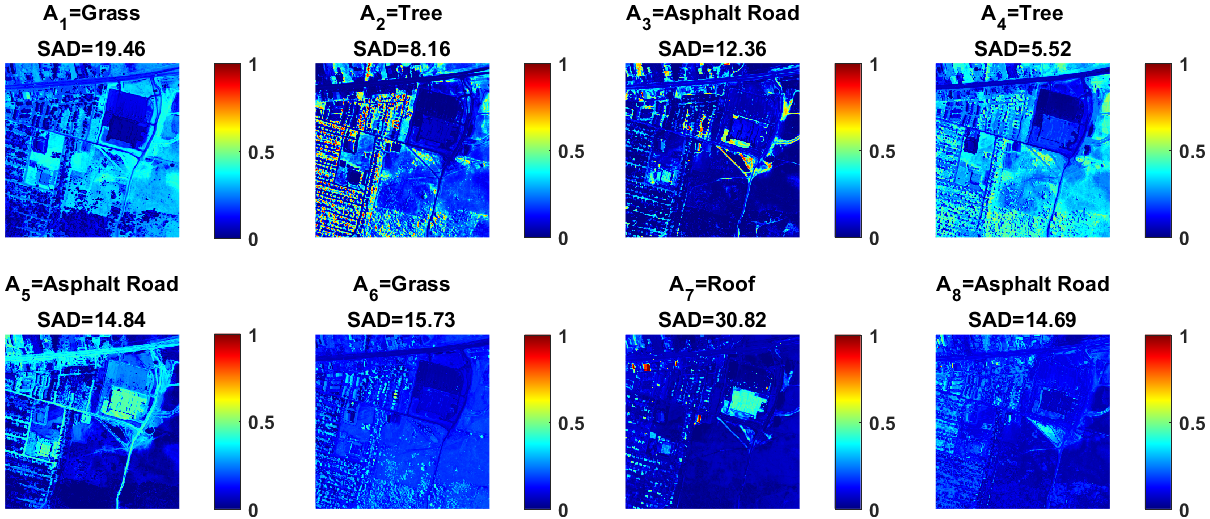}
            \caption{Components of $\matr{A}$}
            \label{subfig:urban_NNCPD_Sparse_ASC_spatial_ensemble}
        \end{subfigure}

        \begin{subfigure}[b]{0.2116\textwidth}
            \includegraphics[width=\textwidth]{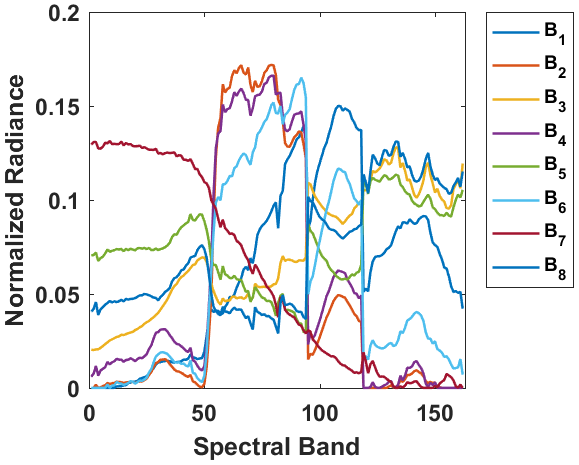}
            \caption{Components of $\matr{B}$}
            \label{subfig:urban_NNCPD_Sparse_ASC_spectra_normalized}
        \end{subfigure}
        ~
        \begin{subfigure}[b]{0.2116\textwidth}
            \includegraphics[width=\textwidth]{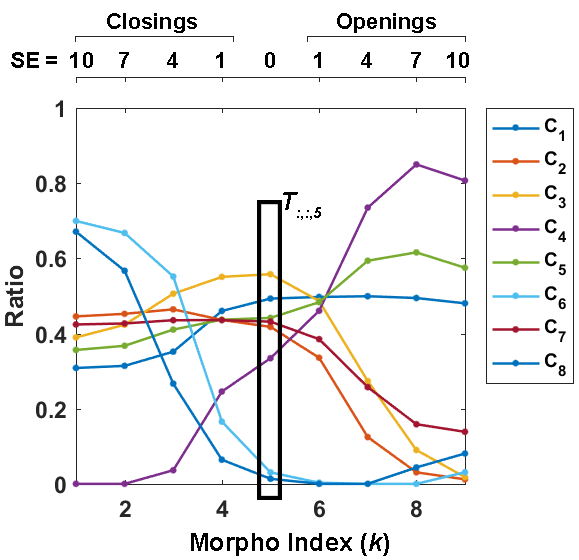}
            \caption{Components of $\matr{\Psi}$}
            \label{subfig:urban_NNCPD_Sparse_ASC_morpho_normalized}
        \end{subfigure}
    \caption{Urban. \gls{cpd} results of the \gls{mm}-tensor for $R$=$8$}
    \label{fig:urban_NNCPD_Sparse_ASC}
\end{figure}

\subsubsection{More latent components, \glsentryshort{elmm} and multi-feature separability}
\label{subsubsec:Unmixing_Experiments_Urban_NCPD}

Here, we discuss the results for $R = 8$, where
Fig. \ref{fig:urban_NNCPD_Sparse_ASC} represents \gls{mm}-\gls{cpd}.
We skip the case of Patch-\gls{cpd} in order to avoid repetition, where we simply end up with more \textit{replicas} of the components of figures \ref{subfig:urban_pat_NNCPD_R4_Sparse_ASC_spatial_ensemble} and \ref{subfig:urban_pat_NNCPD_R4_Sparse_ASC_spectra_normalized}.
In Fig. \ref{fig:urban_NNCPD_Sparse_ASC}, we are interested in the features of the \glspl{am} that do not appear in Patch-\gls{cpd}
as the comments on the spectral and morphological patterns are the same as those of Pavia; where
the plots reflect the qualitative features that appear in the respective \glspl{am}.

We observe three column sets forming three spectral bundles:
$\{1, 2, 6\}$,
$\{3, 5, 7, 8\}$, and
$\{4\}$.
First, $\matr{A}_1$, $\matr{A}_2$, $\matr{A}_4$, and $\matr{A}_6$ were identified as \textit{Vegetation}:
$\matr{A}_1$ highlights grass fields, which is close to the \textit{Grass} reference and does not appear in Patch-\gls{cpd}.
$\matr{A}_2$ and $\matr{A}_4$ highlight small and big areas, and together they correspond to the \textit{Tree} reference.
$\matr{A}_6$ highlights dark shadows (which is reflected in $\matr{\Psi}_6$).
Second, $\matr{A}_3$, $\matr{A}_5$, and $\matr{A}_8$ are identified as \textit{Asphalt Road}:
$\matr{A}_8$ seems to correspond to dark features (refer to $\matr{\Psi}_8$).
$\matr{A}_3$ highlights small roads such as dirt and narrow streets, while $\matr{A}_5$ highlights large roads like the main and connected roads,
which are clearly highlighted unlike the case of patches.
Third, $\matr{A}_7$ is identified as \textit{Roof} and highlights both small and large building roofs.

\section{Conclusion}
\label{sec:Conclusion}

In this paper, we proposed a methodological framework for \gls{multihutd} based on \gls{cpd} and the \gls{aoadmm}-\gls{asc} algorithm, where the samples (pixels) represent a convex combination of the sources. We also established a unified framework for the interpretability of \gls{multihutd} into ``multilinear'' subspaces which involved mathematical, physical, and graphical representations of the \gls{cpd} model with \gls{asc}, \gls{elmm} and \gls{sv}. Finally, we proposed to include \gls{mm} as spatial features in a spectral-spatial \gls{hu} and dived further into the case of neighborhood patches, where \gls{mm} incorporates physically meaningful features into the data tensor. Through the comparison between the two third-mode examples, we provided in-depth insights on the interpretability of \gls{multihutd} including the physical significance of the factor matrices and the input rank.
To conclude, we summarize some key properties of \gls{multihutd} as follows:
\begin{itemize}
	\item Multi-feature hyperspectral data is useful for low-rank latent variable analysis, such as unmixing.
	
	\item Having multiple modalities of features allows to exploit more information on the scene, relaxing the dependency on the high-rank spatial structures while conserving enough context of the scene.
	
	\item Having multiple modalities of features with \gls{cpd} acts as an implicit prior on the scene. The \gls{multihutd} framework is then equivalent to performing a coupled matrix decomposition on each of the tensor slices where the abundances matrix $\matr{A}$ is the common factor.
	
	\item Multi-feature unmixing through low-rank tensor decomposition factorizes the pixel and spectral information and implicitly models the spectral variabilities of the scene.
\end{itemize}

In the future, we plan to explore \gls{btd} which allows some flexibility with the tensor structure and can be seen as an extension to Spectral Bundles for \gls{sv} \cite{borsoi2021spectral}, but also comes with many challenges such as the rank and the interpretation of the subspaces. Moreover, areas of \gls{bss} other than \gls{hu} may be explored.
Finally, it is worth mentioning that some deep learning approaches are being considered for \gls{hu} (which still suffer from the increasing and flexible dimensionality of \glspl{hsi} and the difficulty of finding data sets for training especially in a blind framework).
However, by developing our methodological study of tensor-based unmixing and pushing for interpretability, this framework can help interpretability in data driven methods based on tensor decomposition \cite{GattDKFJ21:esa, Gatt20:thesis, BataSGSF22:mlwa}.

\section{Acknowledgments}

We would like to thank the associate editor and the reviewers for their valuable comments and for enriching the content and the context of this work.

\appendices

\section{Acronyms}
\label{appendix:acronyms}


\printglossaries


\section{Sparse-\glsentryshort{nmf} results on the matricized \glsentryshort{hsi}}
\label{appendix:NMF_results}

In this appendix, we include the results obtained by applying sparse \gls{nmf} (with \gls{asc}) \cite{YangZXDYZ10:tgrs}, which partly inspired this work.
We note that these results cannot be compared with those of tensor decomposition in terms of \gls{rmse} and execution time due to the following reasons:
\begin{itemize}
	\item In terms of \gls{rmse}, on the one hand, we reconstruct a multi-feature \gls{hsi} tensor, while on the other hand, we reconstruct a \gls{hsi} matrix which does not apply in \textit{multi-feature analysis}, so the reconstructed data represent different types of information.
	
	\item
	In terms of execution time, \gls{nmf} typically has shorter execution times than tensor-based methods due to the added complexity. However, both tools are fundamentally different and can not be used for the same multi-linear application.
\end{itemize}
With that said, the obtained results serve only as a qualitative baseline or reference for the \textit{abundance maps} and \textit{spectral components} of decomposing the two \gls{hsi} datasets, which can indeed be compared with those obtained in the case of \gls{mm}-\gls{hsi} and Patch-\gls{hsi} tensors.

\begin{figure}[t]
	\centering
	
	\begin{minipage}[b]{\linewidth}
		\centering
		\begin{subfigure}[b]{\linewidth}
			\includegraphics[width=\textwidth]{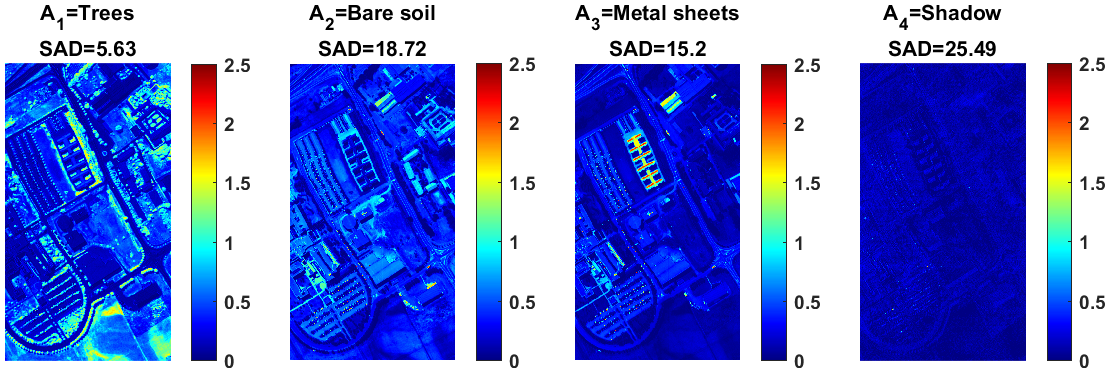}
			\caption{Components of $\matr{A}$}
			\label{subfig:pavia_NMF_ASC_spatial_ensemble}
		\end{subfigure}
		
		\begin{subfigure}[b]{0.48\textwidth}
			\includegraphics[width=\textwidth]{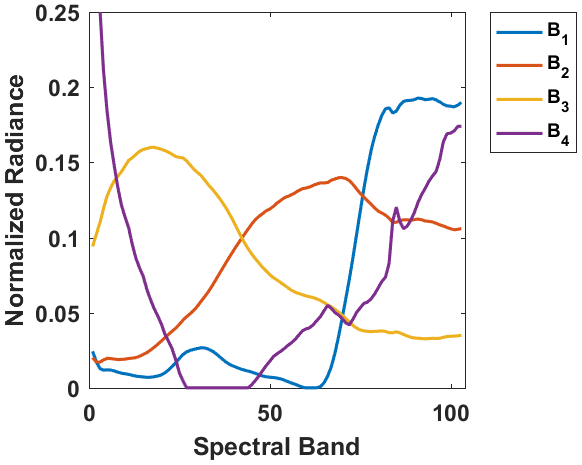}
			\caption{Components of $\matr{B}$}
			\label{subfig:pavia_NMF_ASC_spectra_normalized}
		\end{subfigure}
		
		\caption{Pavia. \gls{nmf} results of the \gls{hsi} matrix for $R=4$}
		\label{fig:pavia_NMF_ASC}
	\end{minipage}
	
	\vspace{3mm}
	
	\begin{minipage}[b]{\linewidth}
		\centering
		\begin{subfigure}[b]{\linewidth}
			\includegraphics[width=\textwidth]{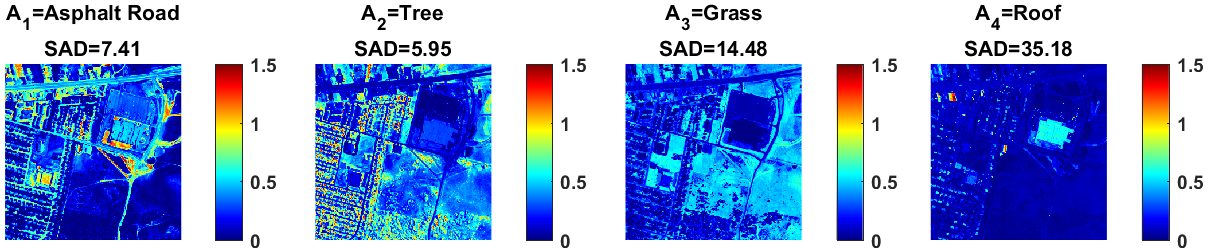}
			\caption{Components of $\matr{A}$}
			\label{subfig:urban_NMF_ASC_spatial_ensemble}
		\end{subfigure}
		
		\begin{subfigure}[b]{0.48\textwidth}
			\includegraphics[width=\textwidth]{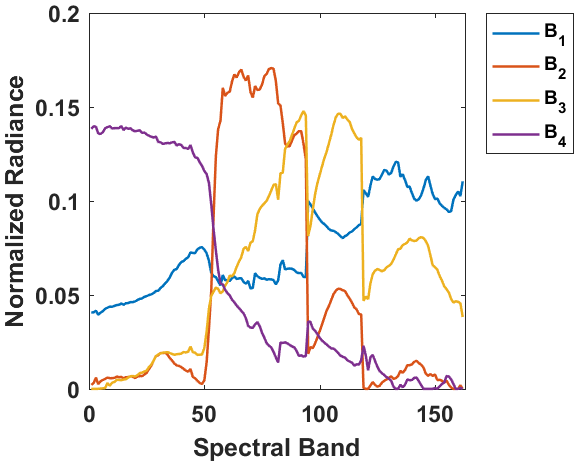}
			\caption{Components of $\matr{B}$}
			\label{subfig:urban_NMF_ASC_spectra_normalized}
		\end{subfigure}
		
		\caption{Urban. \gls{nmf} results of the \gls{hsi} matrix for $R=4$}
		\label{fig:urban_NMF}
	\end{minipage}

\end{figure}

\subsection{Pavia dataset}
Fig. \ref{fig:pavia_NMF_ASC} shows the results obtained for the dataset of Pavia University.
Each abundance map is shown with the class that is assigned to it based on the minimum \gls{sad} value, which is reported as well.

First, we look at the first three components, \textit{Trees}, \textit{Bare Soil}, and \textit{Metal Sheets}.
Their spectral signatures $\{\matr{B}_1, \matr{B}_2, \matr{B}_3\}$ shown in Fig. \ref{subfig:pavia_NMF_ASC_spectra_normalized} look very similar to those of the reference, but the corresponding \gls{sad} values are relatively bad (high) compared to those obtained using tensor decomposition in Fig. \ref{fig:pavia_pat_NNCPD_R4_Sparse_ASC} and \ref{fig:pavia_emp_NNCPD_R4_Sparse_ASC}.

The abundance maps $\{\matr{A}_1, \matr{A}_2, \matr{A}_3\}$ of Fig. \ref{subfig:pavia_NMF_ASC_spatial_ensemble} show highlighted elements belonging to \textit{Trees}, \textit{Bare Soil}, and \textit{Metal Sheets} respectively.
However, we notice that other areas of the scene belonging to these categories are barely or faintly highlighted (e.g., asphalt road, brick parking lots, other soil areas), which is due to the insufficiency of \gls{lmm} to model their variabilities.

Regarding the \textit{Shadow} component, the spectral signature $\matr{B}_4$ looks slightly similar to the reference, but the shadows in the map are barely visible due to their relatively very low brightness $\matr{A}_4$.

\subsection{Urban dataset}
Fig. \ref{fig:urban_NMF} shows the abundance maps and spectral signatures of $\matr{A}$ and $\matr{B}$ respectively.
We obtain four components with relatively low (good) \gls{sad} values and good abundance maps similarity with respect to the reference. Finally, we also note that some dark areas like asphalt roads are not highlighted.

\section{Synthetic \glsentryshort{hsi} example}
\label{appendix:synthetic_hsi}

In the following, we simulate a time-series \gls{hsi} tensor through a synthetic example. The noiseless tensor is reconstructed from its building blocks, i.e., the factor matrices, which are considered here as the noiseless ground-truths.
This simplified example serves as a demonstration of the performance of \gls{multihutd} as we increase the \gls{snr}.

\subsection{Description of the synthetic dataset}
\begin{figure}
	
	\begin{minipage}[t]{\linewidth}
		\centering
		\includegraphics[width=\columnwidth]{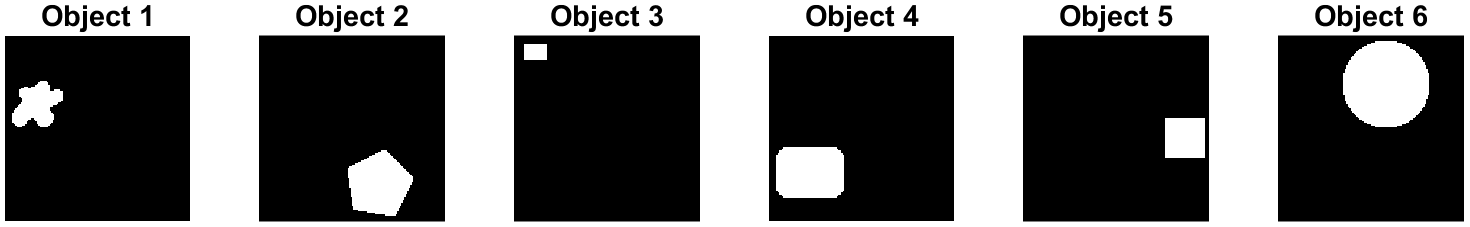}
		\caption{
			The simulated spatial objects of the synthetic HSI, denoted as Objects ``1'' through ``6''.
		}
		\label{subfig:Toy_spatial_objects}
	\end{minipage}
	
	\vspace{4mm}
	
	\begin{minipage}[t]{\linewidth}
		\centering
		\begin{subfigure}[t]{\columnwidth}
			\centering
			\includegraphics[width=0.22\textwidth]{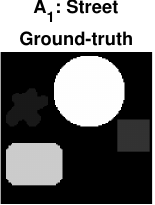}
			\hspace{3mm}
			\includegraphics[width=0.22\textwidth]{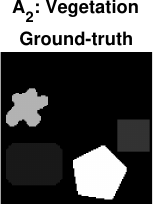}
			\hspace{3mm}
			\includegraphics[width=0.22\textwidth]{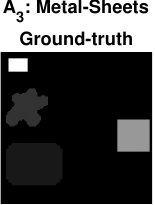}
			
			\caption{Ground-truth of $\matr{A}$}
			\label{subfig:toy_tensor_GT_A}
		\end{subfigure}
		
		\vspace{2mm}
		
		\begin{subfigure}[t]{0.4232\textwidth}
			\includegraphics[width=\textwidth]{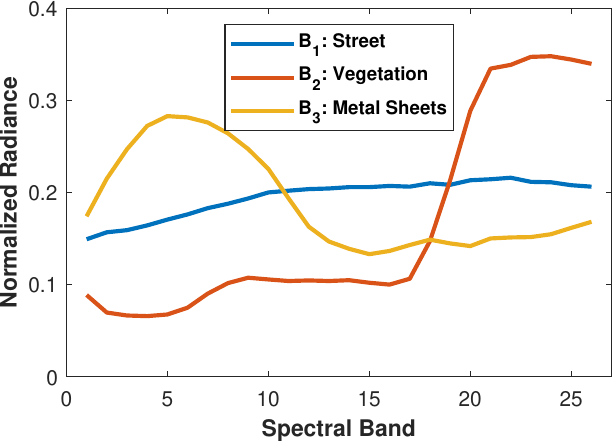}
			\caption{Ground-truth of $\matr{B}$}
			\label{subfig:toy_tensor_GT_B}
		\end{subfigure}
		\,
		\begin{subfigure}[t]{0.4232\textwidth}
			\includegraphics[width=\textwidth]{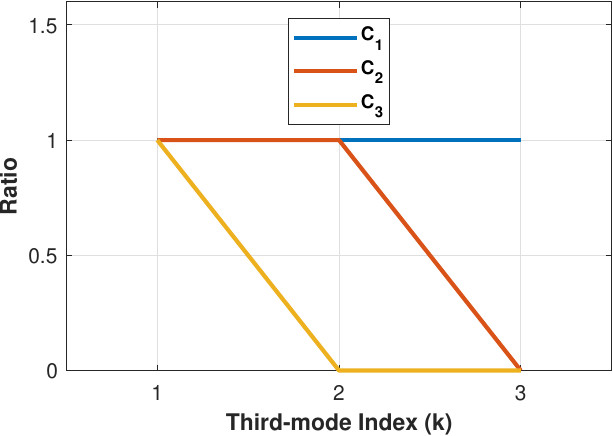}
			\caption{Ground-truth of $\matr{C}$}
			\label{subfig:toy_tensor_GT_C}
		\end{subfigure}
		
		\caption{
			Ground-truth components of the synthetic tensor $\tens{T}$.
		}
		\label{subfig:toy_tensor_GT}
	\end{minipage}
	
	\vspace{4mm}
	
	\begin{minipage}[t]{\linewidth}
		\centering
		\begin{subfigure}[t]{0.24\textwidth}
			\includegraphics[width=\textwidth]{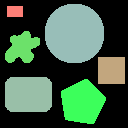}
			\caption{Noiseless}
			\label{subfig:Toy_color_noiseless}
		\end{subfigure}
		\hspace{3mm}
		\begin{subfigure}[t]{0.24\textwidth}
			\includegraphics[width=\textwidth]{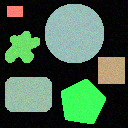}
			\caption{$\sigma^2=10^{-4}$}
			\label{subfig:Toy_color_noisy_var_10-4}
		\end{subfigure}
		\hspace{3mm}
		\begin{subfigure}[t]{0.24\textwidth}
			\includegraphics[width=\textwidth]{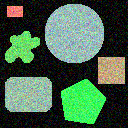}
			\caption{$\sigma^2=10^{-3}$}
			\label{subfig:Toy_color_noisy_var_10-3}
		\end{subfigure}
		
		\vspace{2mm}
		
		\begin{subfigure}[t]{0.24\textwidth}
			\includegraphics[width=\textwidth]{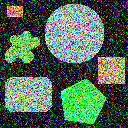}
			\caption{$\sigma^2=10^{-2}$}
			\label{subfig:Toy_color_noisy_var_10-2}
		\end{subfigure}
		\hspace{3mm}
		\begin{subfigure}[t]{0.24\textwidth}
			\includegraphics[width=\textwidth]{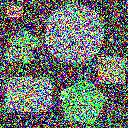}
			\caption{$\sigma^2=10^{-1}$}
			\label{subfig:Toy_color_noisy_var_10-1}
		\end{subfigure}
		
		\caption{
			The synthetic HSI scene at a single time stamp $k=1$ in False colors, with Gaussian noise of variance $\sigma^2$ ranging from $0$ (being noiseless) to $10^{-1}$ (being very noisy).
		}
		\label{subfig:Toy_color}
	\end{minipage}
	
\end{figure}

The synthetic \gls{hsi} cube $\tens{M}$ of a signle time-stamp has dimensions $128\times 128 \times 26$, whose matricized version is denoted by $\matr{M}$ of dimensions ${16384} \times {26}$. In particular:
\begin{itemize}
	\item The spatial scene is composed of six objects that vary in size, as shown in \figurename~\ref{subfig:Toy_spatial_objects}, denoted by ``Object 1'' through ``Object 6''. The scene is shown in \figurename~\ref{subfig:Toy_color_noiseless} in false colors.
    
	\item Three independent spectral signatures are simulated from the \gls{hsi} of Pavia University, corresponding to \texttt{Street}, \texttt{Vegetation}, and \texttt{Metal Sheets} which are shown in \figurename~\ref{subfig:toy_tensor_GT_B}.
	
	\item The objects are assigned linear mixtures of the three simulated spectra as shown in \tablename~\ref{tab:objects_spectral_mixtures}.
	This means that the ground-truth of the abundances satisfies the \gls{asc}.
\end{itemize}
That said, $\matr{M}$ has rank $3$ where $\matr{M} = \matr{A} \matr{B}^{\T}$, such that $\matr{A} \in \RR{16384}{3}$ and $\matr{B} \in \RR{26}{3}$ are two factor matrices. The spatial and spectral ground-truths, of $\matr{A}$ and $\matr{B}$ respectively, are shown in \figurename~\ref{subfig:toy_tensor_GT_A} and \ref{subfig:toy_tensor_GT_B}.
\begin{table}[h]
	\centering
	\begin{tabular}{c||c|c|c|c|c|c}
		Object number & $1$ & $2$ & $3$ & $4$ & $5$ & $6$ \\
		\hline
		\hline
		\texttt{Street $(\%)$} & $10$ & $0$ & $0$ & $\textbf{80}$ & $20$ & $\textbf{100}$ \\
		\hline
		\texttt{Vegetation $(\%)$} & $\textbf{70}$ & $\textbf{100}$ & $0$ & $10$ & $20$ & $0$ \\
		\hline
		\texttt{Metal Sheets $(\%)$} & $20$ & $0$ & $\textbf{100}$ & $10$ & $\textbf{60}$ & $0$ \\
		\hline
		\hline
		\texttt{Total $(\%)$} & $100$ & $100$ & $100$ & $100$ & $100$ & $100$ \\
	\end{tabular}
	\caption{Spectral mixture of each object based on the endmembers, \texttt{Street}, \texttt{Vegetation}, and \texttt{Metal Sheets}.}
	\label{tab:objects_spectral_mixtures}
\end{table}

As for the full time-series \gls{hsi} tensor, it is composed of three stamps where the objects of the scene change in color or disappear in time. In principle, this corresponds to a time-series tensor $\tens{D}$ of dimensions $128 \times 128 \times 26 \times 3$. After reordering the pixel modalities in lexicographic order, we would obtain a tensor $\tens{T}$ of dimensions $16384 \times 26 \times 3$.
We synthesize the time-series \gls{hsi} tensor $\tens{T}$ of dimensions $\RRR{16384}{26}{3}$ from the product with an additional matrix $\matr{C} \in \RR{3}{3}$ such that:
\begin{equation}
	\tens{T} = \tens{I} \con_1 \matr{A} \con_2 \matr{B} \con_3 \matr{C}
\end{equation}
where $\tens{I} \in \RRR{3}{3}{3}$ is a diagonal tensor of ones, and $\matr{C}$ is described as follows:
\begin{equation}
	\label{eq:time_series_ground_truth}
	\matr{C} = 
	\begin{bmatrix}
		1 & 1 & 1 \\
		1 & 1 & 0 \\
		1 & 0 & 0
	\end{bmatrix}
	<=>
	\begin{pmatrix}
		k = 1 \\
		k = 2 \\
		k = 3
	\end{pmatrix}
\end{equation}
where $k$ is the index spanning the third modality, which is that of time stamps. $C$ is considered the ground-truth of the temporal signatures, and its columns are plotted in \figurename~\ref{subfig:toy_tensor_GT_C}.

In order to assess the performance of the proposed framework in the presence of noise, Gaussian noise was added on the tensor $\tens{D}$ with varying levels of noise, where the variance $\sigma^2 \in \{0, 10^{-4}, 10^{-3}, 10^{-2}, 10^{-1}\}$, as described in \figurename~\ref{subfig:Toy_color}.
After adding noise, the tensor is reshaped back to $\tens{T}$.

\begin{figure}[t]
	\centering
	\begin{subfigure}[t]{\columnwidth}
		\centering
		\includegraphics[width=0.22\textwidth]{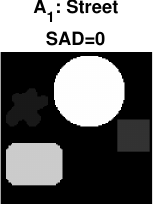}
		\hspace{3mm}
		\includegraphics[width=0.22\textwidth]{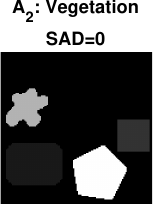}
		\hspace{3mm}
		\includegraphics[width=0.22\textwidth]{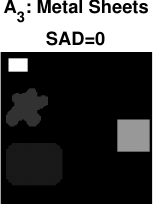}
		
		\caption{Recovered columns of $\matr{A}$}
		\label{subfig:toy_tensor_results_A_var_0}
	\end{subfigure}
	
	\vspace{2mm}
	
	\begin{subfigure}[t]{0.2116\textwidth}
		\includegraphics[width=\textwidth]{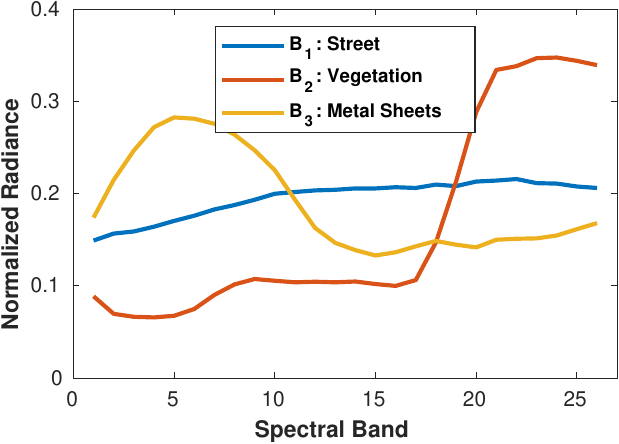}
		\caption{Recovered columns of $\matr{B}$}
		\label{subfig:toy_tensor_results_B_var_0}
	\end{subfigure}
	\,
	\begin{subfigure}[t]{0.2116\textwidth}
		\includegraphics[width=\textwidth]{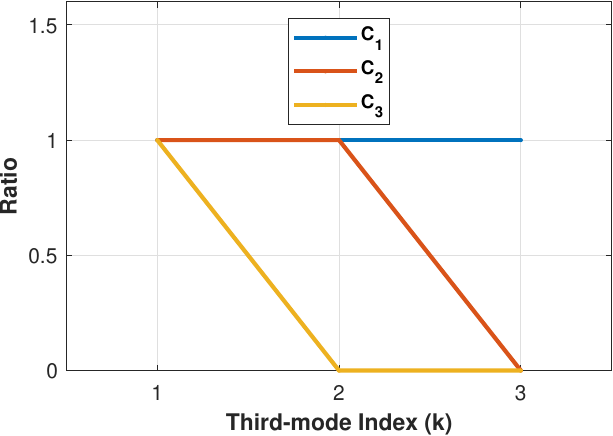}
		\caption{Recovered columns of $\matr{C}$}
		\label{subfig:toy_tensor_results_C_var_0}
	\end{subfigure}
	
	\caption{
		Synthetic \gls{hsi}. Results of decomposing the \gls{hsi} tensor $\tens{T}$ without noise, where we also have $\textrm{RMSE} = 0$.
	}
	\label{subfig:toy_tensor_results_var_0}
\end{figure}

\begin{figure}[t]
	\centering
	\begin{subfigure}[t]{\columnwidth}
		\centering
		\includegraphics[width=0.22\textwidth]{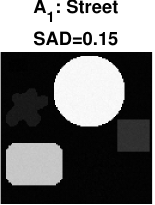}
		\hspace{3mm}
		\includegraphics[width=0.22\textwidth]{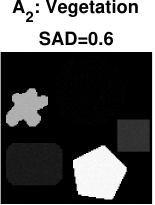}
		\hspace{3mm}
		\includegraphics[width=0.22\textwidth]{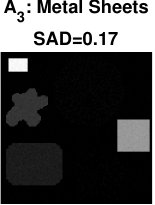}
		
		\caption{Recovered columns of $\matr{A}$}
		\label{subfig:toy_tensor_results_A_var_10-4}
	\end{subfigure}
	
	\vspace{2mm}
	
	\begin{subfigure}[t]{0.2116\textwidth}
		\includegraphics[width=\textwidth]{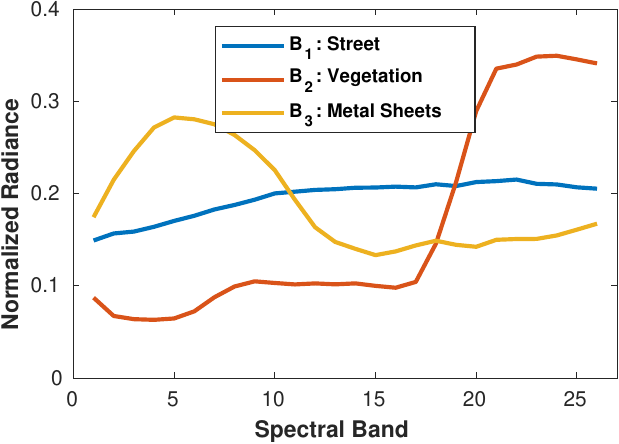}
		\caption{Recovered columns of $\matr{B}$}
		\label{subfig:toy_tensor_results_B_var_10-4}
	\end{subfigure}
	\,
	\begin{subfigure}[t]{0.2116\textwidth}
		\includegraphics[width=\textwidth]{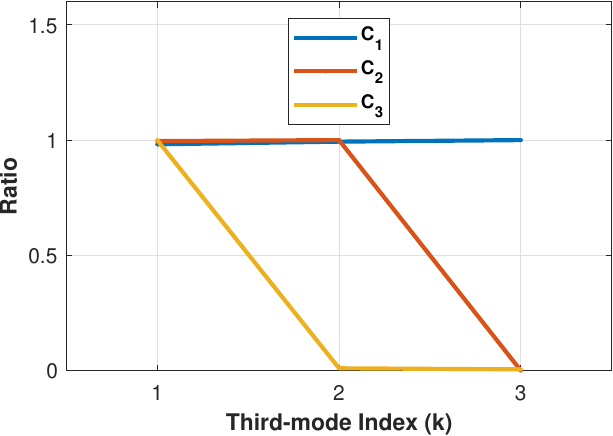}
		\caption{Recovered columns of $\matr{C}$}
		\label{subfig:toy_tensor_results_C_var_10-4}
	\end{subfigure}
	
	\caption{
		Synthetic \gls{hsi}. Results of decomposing the \gls{hsi} tensor $\tens{T}$ with Gaussian noise of variance $10^{-4}$.
	}
	\label{subfig:toy_tensor_results_var_10-4}
\end{figure}

\begin{figure}[t]
	\centering
	\begin{subfigure}[t]{\columnwidth}
		\centering
		\includegraphics[width=0.22\textwidth]{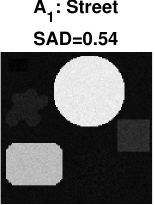}
		\hspace{3mm}
		\includegraphics[width=0.22\textwidth]{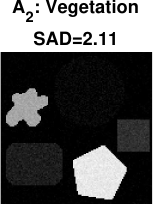}
		\hspace{3mm}
		\includegraphics[width=0.22\textwidth]{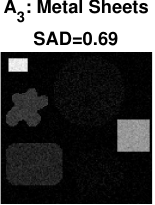}
		
		\caption{Recovered columns of $\matr{A}$}
		\label{subfig:toy_tensor_results_A_var_10-3}
	\end{subfigure}
	
	\vspace{2mm}
	
	\begin{subfigure}[t]{0.2116\textwidth}
		\includegraphics[width=\textwidth]{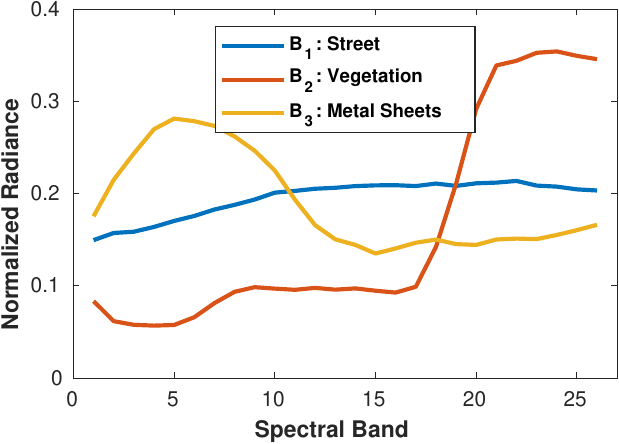}
		\caption{Recovered columns of $\matr{B}$}
		\label{subfig:toy_tensor_results_B_var_10-3}
	\end{subfigure}
	\,
	\begin{subfigure}[t]{0.2116\textwidth}
		\includegraphics[width=\textwidth]{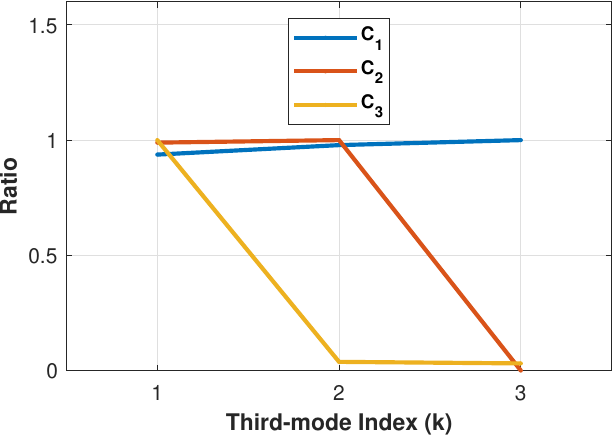}
		\caption{Recovered columns of $\matr{C}$}
		\label{subfig:toy_tensor_results_C_var_10-3}
	\end{subfigure}
	
	\caption{
		Synthetic \gls{hsi}. Results of decomposing the \gls{hsi} tensor $\tens{T}$ with Gaussian noise of variance $10^{-3}$.
	}
	\label{subfig:toy_tensor_results_var_10-3}
\end{figure}

\begin{figure}[t]
	\centering
	\begin{subfigure}[t]{\columnwidth}
		\centering
		\includegraphics[width=0.22\textwidth]{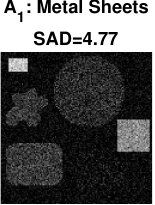}
		\hspace{3mm}
		\includegraphics[width=0.22\textwidth]{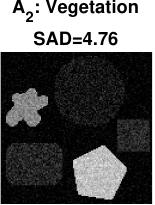}
		\hspace{3mm}
		\includegraphics[width=0.22\textwidth]{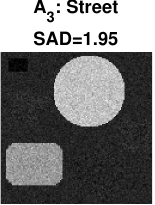}
		
		\caption{Recovered columns of $\matr{A}$}
		\label{subfig:toy_tensor_results_A_var_10-2}
	\end{subfigure}
	
	\vspace{2mm}
	
	\begin{subfigure}[t]{0.2116\textwidth}
		\includegraphics[width=\textwidth]{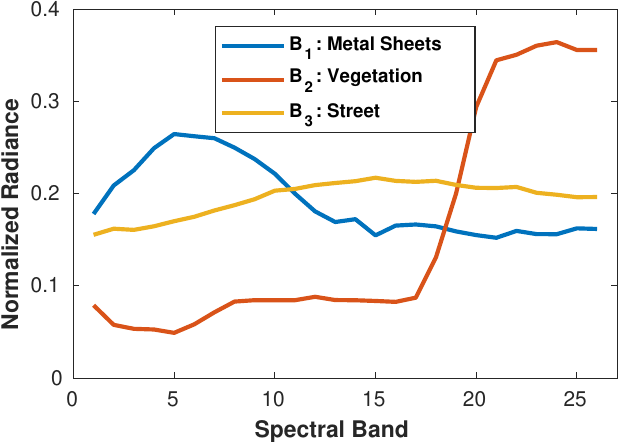}
		\caption{Recovered columns of $\matr{B}$}
		\label{subfig:toy_tensor_results_B_var_10-2}
	\end{subfigure}
	\,
	\begin{subfigure}[t]{0.2116\textwidth}
		\includegraphics[width=\textwidth]{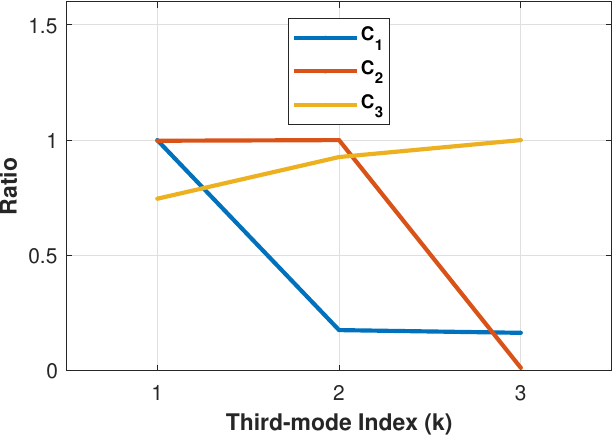}
		\caption{Recovered columns of $\matr{C}$}
		\label{subfig:toy_tensor_results_C_var_10-2}
	\end{subfigure}
	
	\caption{
		Synthetic \gls{hsi}. Results of decomposing the \gls{hsi} tensor $\tens{T}$ with Gaussian noise of variance $10^{-2}$.
	}
	\label{subfig:toy_tensor_results_var_10-2}
\end{figure}

\begin{figure}[t]
	\centering
	\begin{subfigure}[t]{\columnwidth}
		\centering
		\includegraphics[width=0.22\textwidth]{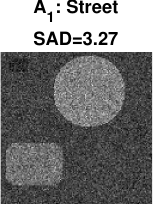}
		\hspace{3mm}
		\includegraphics[width=0.22\textwidth]{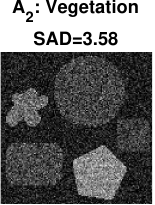}
		\hspace{3mm}
		\includegraphics[width=0.22\textwidth]{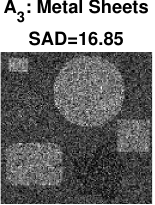}
		
		\caption{Recovered columns of $\matr{A}$}
		\label{subfig:toy_tensor_results_A_var_10-1}
	\end{subfigure}
	
	\vspace{2mm}
	
	\begin{subfigure}[t]{0.2116\textwidth}
		\includegraphics[width=\textwidth]{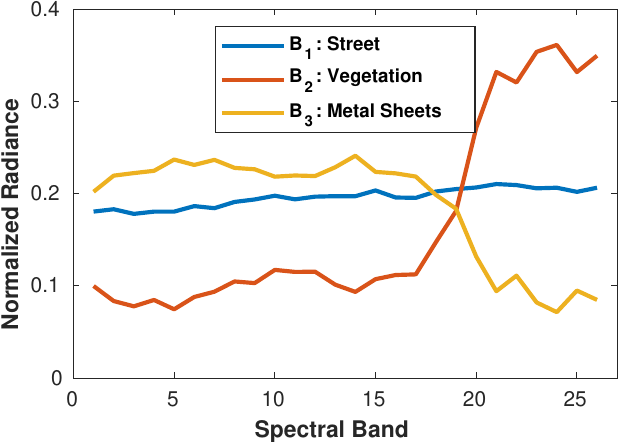}
		\caption{Recovered columns of $\matr{B}$}
		\label{subfig:toy_tensor_results_B_var_10-1}
	\end{subfigure}
	\,
	\begin{subfigure}[t]{0.2116\textwidth}
		\includegraphics[width=\textwidth]{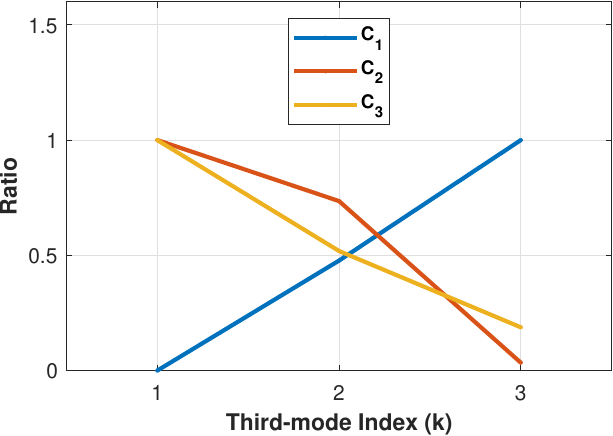}
		\caption{Recovered columns of $\matr{C}$}
		\label{subfig:toy_tensor_results_C_var_10-1}
	\end{subfigure}
	
	\caption{
		Synthetic \gls{hsi}. Results of decomposing the \gls{hsi} tensor $\tens{T}$ with Gaussian noise of variance $10^{-1}$.
	}
	\label{subfig:toy_tensor_results_var_10-1}
\end{figure}

\begin{figure}
	\centering
	\includegraphics[width=0.8\columnwidth]{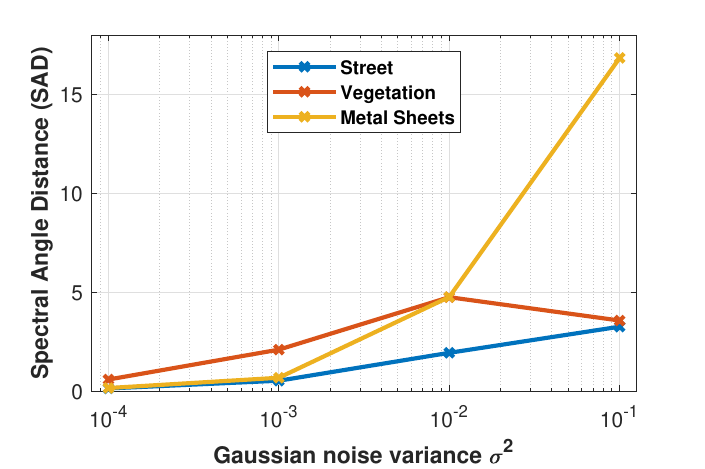}
	\caption{
		Synthetic \gls{hsi}. Evolution of \gls{sad} index of the estimated spectra compared to the spectral ground-truth.
	}
	\label{fig:sad_noise_variance_evolution}
\end{figure}

\subsection{Results of unmixing without adding noise}
\figurename~\ref{subfig:toy_tensor_results_var_0} shows the components of the factor matrices after decomposing the synthetic time-series \gls{hsi} tensor described using CP decomposition.
The components of the factor matrices are perfectly recovered thanks to the CP uniqueness of the data tensor \cite{Krus77:laa}.
Moreover, it is worth noting that the \gls{rmse} between the original data tensor and the reconstructed one is $0$, which means perfect reconstruction.

\subsection{Results of unmixing with varying levels of noise}
In this case, we discard the reconstructability of the tensor itself and focus on the factor matrices, of which we have the abundance matrix $\matr{A}$.
\figurename~\ref{subfig:toy_tensor_results_var_10-4},
\figurename~\ref{subfig:toy_tensor_results_var_10-3},
\figurename~\ref{subfig:toy_tensor_results_var_10-2},
and
\figurename~\ref{subfig:toy_tensor_results_var_10-1}
show the results of decomposing the tensor under varying levels of Gaussian noise, with variances of $10^{-4}$, $10^{-3}$, $10^{-2}$, and $10^{-1}$ respectively.
Moreover, \figurename~\ref{fig:sad_noise_variance_evolution} shows the evolution of the \gls{sad} index of the estimated endmembers as the noise level increases.
We note that the experiments are done without applying any spatial denoising.

The components of the factor matrices, including the abundance matrix $\matr{A}$ whose rows sum to one, are quite recoverable up to a level of noise of variance $\sigma^2 = 10^{-3}$. In the case of $\sigma^2 = 10^{-2}$, the components are still recoverable even though the tensor looks quite noisy in \figurename~\ref{subfig:Toy_color_noisy_var_10-2}. In the case of $\sigma^2 = 10^{-1}$, which is very noisy that some of the objects are indistinguishable in \figurename~\ref{subfig:Toy_color_noisy_var_10-1}, the reconstructed spectra are still fairly close to their ground-truths and the spatial structures in the factors of the abundance matrix can still be recognized.
In \figurename~\ref{fig:sad_noise_variance_evolution}, we can see that the \gls{sad} index generally increases with the level of noise. However, the values remain quite small, i.e., less than $5^{\circ}$, indicating the recoverability of the estimated endmembers even under such high levels of noise. The only exception in this case is that of \textit{Metal Sheets} under a variance of $10^{-1}$ which results in a \gls{sad} index of about $16^{\circ}$; this can be due to the small size of \textit{Metal Sheets} objects, which makes them more susceptible to noise.

Moreover, we note that with the addition of noise, the tensor $\tens{T}$ which was synthesized to be of rank $R=3$ becomes full rank. Decomposing the noisy tensor with a low rank can be roughly seen as a denoising procedure since it forces the projection of the data onto a lower-rank multi-linear latent subspace of rank $R=3$.
However, spatial denoising is still needed in order to recover a better representation of the spatial components, but said application is out of the scope of this paper.

Finally, this is a simple, controlled, and minimalistic example that serves as an intuition for more complex structures where the situation is completely blind, such as in real \gls{hsi} tensors.

\section{Computational Complexity of AO-ADMM-ASC}
\label{appendix:computational_complexity_aoadmmasc}

First, we refer to paper~\cite{HuanSL16:tsp} for the detailed explanations concerning the CPD by using AO-ADMM.
Let us consider a third-order tensor $\tens{T} \in \mathbb{R}^{I_1 \times I_2 \times I_3}$ (where $I_1 >>> I_2 I_3$) with a low rank $R$, and we consider the complexity as per \gls{admm} iterations.
Also, it is important to note that  matrices $\Tilde{\matr{W}}$ and $\matr{T}$ are independent from the inner-ADMM updates, so they can be used only once to compute the products $\Tilde{\matr{W}}^{\T} \Tilde{\matr{W}}$ and $\Tilde{\matr{W}}^{\T}\matr{T}$, whose values can be cached before the ADMM update allowing to save a lot of repetitive computations.
Now, we split the problem into three steps:
\begin{enumerate}
    \item Unconstrained CPD:
    In this case, the complexity of the algorithm is dominated only by the updates of the factor matrices. Hence, the complexity is $\tens{O}(I_d R^2)$ $\forall d \in \{1, 2, 3\}$.
    
    \item Nonnegative CPD, which is relevant for the updates of each of the factor matrices:
    Nonnegativity requires only element-wise projection, i.e., a complexity of $\tens{O}(I_d R)$ $\forall d \in \{1, 2, 3\}$, which is negligible compared to $\tens{O}(I_d R^2)$. Hence, the complexity is still dominated by $\tens{O}(I_d R^2)$.
    
    \item Nonnegative CPD with sparsity and ASC, which is only relevant for the update of $\matr{A}$:
    \textit{Sparsity} is like nonnegativity as it boils down to an element-wise subtraction with complexity $\tens{O}(I_1 R)$, which is negligible compared to $\tens{O}(I_1 R^2)$.
    As for \textit{ASC}, it requires two updates:
    \begin{itemize}
        \item $b_{J+1,r}=\delta \psi_{K,r}^{-1}$ $\forall r\in\{1,\dots,R\}$, i.e., a complexity of $\tens{O}(R)$, which is negligible.

        \item $t_{i,J+1,k} = \sum_{r=1}^R a_{i,r} b_{J+1,r} \psi_{k,r}$ $\forall i\in\{1,\dots,I\}$ and $\forall k\in\{1,\dots,K-1\}$, i.e., a complexity of $\tens{O}(I_1 I_3 R)$.
    \end{itemize}
\end{enumerate}
Considering that MultiHU-TD admits a \textit{low-rank} CP decomposition, $R$ is usually small, and in most of the cases we would have $R < I_3$ (or at least very close). In which case, the complexity is dominated by $\tens{O}(I_1 I_3 R)$, which changes linearly with either the number of pixels $I_1$, the third-mode features $I_3$, or the latent components $R$.




\clearpage

%

\begin{IEEEbiography}[{\includegraphics[width=1in,height=1.25in,clip,keepaspectratio]{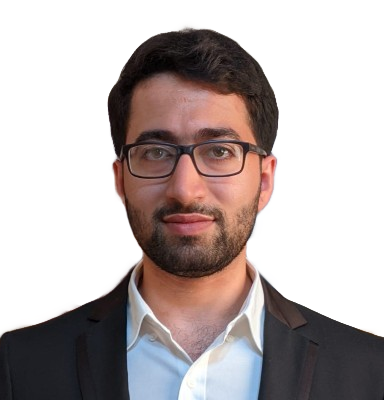}}]{Mohamad Jouni}
    (S'14, GS'17, M'23) received the B.Eng. degree in Computer and Communications Engineering from the Lebanese University, Beirut, Lebanon, in 2016, and the M.Sc. and Ph.D. degrees in Signal and Image Processing from Grenoble Institute of Technology and the University of Grenoble Alpes, Grenoble, France, in 2017 and 2021 respectively.
    Since 2021, he has been a Postdoctoral Researcher at Grenoble Institute of Technology, Grenoble, France.
    In 2019 and 2023, he was a visiting researcher for 10 and 6 weeks, respectively, at Tokyo Institute of Technology, Tokyo, Japan.
    His interests include computational imaging, tensor algebra, hybrid AI methods, applications of multimodal and hyperspectral data analysis, and technology transfer.
\end{IEEEbiography}

\begin{IEEEbiography}[{\includegraphics[width=1in,height=1.25in,clip,keepaspectratio]{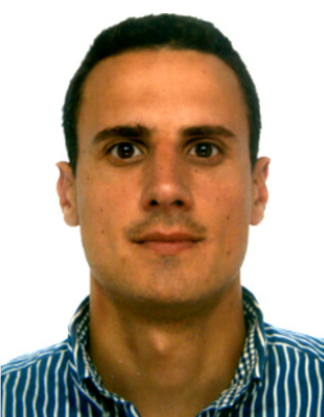}}]{Mauro Dalla Mura}
    (S'08, M'11, SM'18) received the B.Sc. and M.Sc. degrees in Telecommunication Engineering from the University of Trento, Italy in 2005 and 2007, respectively.
    He obtained in 2011 a joint Ph.D. degree in Information and Communication Technologies (Telecommunications Area) from the University of Trento, Italy and in Electrical and Computer Engineering from the University of Iceland, Iceland.
    In 2011 he was a Research fellow at Fondazione Bruno Kessler, Trento, Italy, conducting research on computer vision.
    He is currently an Assistant Professor at Grenoble Institute of Technology (Grenoble INP), France since 2012. He is conducting his research at the Grenoble Images Speech Signals and Automatics Laboratory (GIPSA-Lab).  He is a Junior member of the Institut Universitaire de France (2021-2026).
    Dr. Dalla~Mura has been appointed "Specially Appointed Associate Professor" at the School of Computing, Tokyo Institute of Technology, Japan for 2019-2022.
    His main research activities are in the fields of remote sensing, computational imaging, image and signal processing.
    Dr. Dalla~Mura was the recipient of the IEEE GRSS Second Prize in the Student Paper Competition of the 2011 IEEE IGARSS 2011 and co-recipient of the Best Paper Award of the International Journal of Image and Data Fusion for the year 2012-2013 and the Symposium Paper Award for IEEE IGARSS 2014.
    Dr. Dalla~Mura was the IEEE GRSS Chapter's Committee Chair for 2020-2021. He was President of the IEEE GRSS French Chapter 2016-2020 (he previously served as Secretary 2013-2016). In 2017 the IEEE GRSS French Chapter was the recipient of the IEEE GRSS Chapter Award and the ``Chapter of the year 2017'' from the IEEE French Section.
    He is on the Editorial Board of the IEEE Journal of Selected Topics in Applied Earth Observations and Remote Sensing (J-STARS) since 2016.
\end{IEEEbiography}


\begin{IEEEbiography}[{\includegraphics[width=1in,height=1.25in,clip,keepaspectratio]{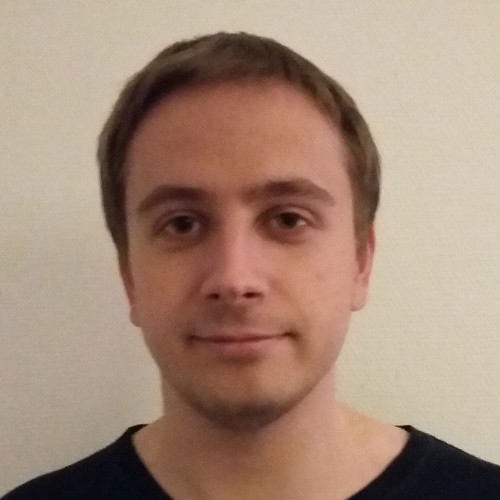}}]{Lucas Drumetz}
    (Member, IEEE) received the M.Eng. degree form Grenoble INP in 2013, and the Ph.D degree in 2016 in image and signal processing from Université Grenoble Alpes, for works carried out at GIPSA-lab, Grenoble, France. This work has been awarded the PhD award of the University of Grenoble Alpes in 2017. In 2017, he was a Visiting Assistant Professor at the Department of Mathematics at the University of California, Los Angeles (UCLA). In 2017, he was also a visiting researcher for 10 weeks at the RCAST laboratory at the University of Tokyo, Japan. Since He has been an Associate Professor at IMT Atlantique, in the Mathematical and Electrical Engineering department since 2018. He is part of the OSE (Observations, Signal and Environment) team of UMR CNRS 6285 Lab-STICC. His research interests include inverse problems and machine learning for remote sensing applications, signal and image processing, and optimization techniques.
\end{IEEEbiography}

\begin{IEEEbiography}[{\includegraphics[width=1in,height=1.25in,clip,keepaspectratio]{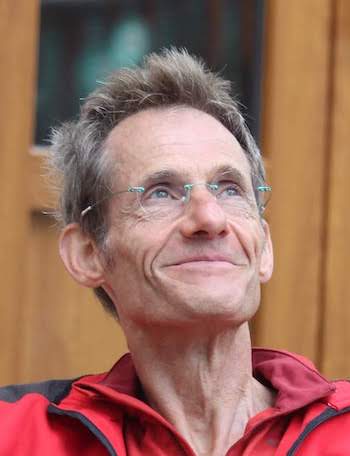}}]{Pierre Comon}
    (M’87 - SM’95 - F’07) received the Graduate degree in 1982, and the Doctorate degree in 1985, both from the University of Grenoble, France. He received the Habilitation to Lead Researches in 1995, from the University of Nice, France. He has been for nearly 13 years in industry, first with Crouzet-Sextant, Valence, France, between 1982 and 1985, and then with Thomson Marconi, Sophia Antipolis, France, between 1988 and 1997. He was with the ISL laboratory, Stanford University, CA, USA, in 1987. He joined in 1997 the Eurecom Institute, Sophia Antipolis, France. He is a Research Director with CNRS since 1998, first with the Laboratory I3S, Sophia Antipolis, France, until 2012, and then with Gipsa-Lab, Grenoble, France. He has been the Director of Labex Persyval, Grenoble, until 2022. 
    His research interests include high-order statistics (HOS), blind techniques, statistical signal and array processing, tensor decompositions, multi-way factor analysis, and data science. 
    He was an Associate Editor for the IEEE Transactions on Signal Processing from 1995 to 1998, and a member of the French National Committee of Scientific
    Research from 1995 to 2000. He was the Coordinator of the European Basic Research Working Group on HOS, ATHOS, from 1992 to 1995. Between 1992 and 1998, he was a member of the Technical and Scientific Council of the Thomson Group. Between 2001 and 2004, he acted as a Launching Associate Editor with the IEEE Transactions on Ciruits and Systems I, in the area of Blind Techniques. He has also been a member of the editorial board of the Elsevier journal Signal Processing from 2006 to 2011, and member of several IEEE Technical Committees. He was in the Editorial Board of the SIAM Journal on Matrix Analysis and Applications from 2011 to 2017. He received several prizes, including the Silver medal of CNRS in 2018. Dr Comon is also a Fellow of Eurasip and SIAM.
\end{IEEEbiography}




\end{document}